\documentclass{aa}

\usepackage[utf8]{inputenx}
\usepackage{graphicx}
\usepackage{txfonts}

\begin{document}

\title{A $\sim$15\,kpc outflow cone piercing through the halo of the blue
  compact metal-poor galaxy SBS\,0335-052E} \titlerunning{Outflow cone of
  SBS\,0335-052E}

\author{
  E.~C.~Herenz\inst{\ref{esochile},\ref{leiden}}
  \and
  J.~Inoue\inst{\ref{stpaul}}
  \and
  H.~Salas\inst{\ref{aip}} % add correct institute
  \and
  B.~Koenigs\inst{\ref{stpaul}}
  \and
  C.~Moya-Sierralta\inst{\ref{pontifica}}
  \and
  J.~M.~Cannon\inst{\ref{stpaul}}
  \and
  M.~Hayes\inst{\ref{stockholm}}
  \and
  P.~Papaderos\inst{\ref{porto}}
  \and
  G.~\"{O}stlin\inst{\ref{stockholm}}
  \and
  A.~Bik\inst{\ref{stockholm}}
  \and
  A.~Le~Reste\inst{\ref{stockholm}}
  \and
  H.~Kusakabe\inst{\ref{geneve}}
  \and
  A. Monreal-Ibero\inst{\ref{leiden}}
  \and
  J. Puschnig\inst{\ref{bonn}}
}

\institute{
  European Southern Observatory,
  Av. Alonso de C\'ordova 3107,
  763 0355 Vitacura,
  Santiago, Chile \\ e-mail: \url{eherenz@eso.org}
  \label{esochile}\and
  Leiden Observatory, Leiden University, Niels Bohrweg 2, NL 2333 CA Leiden, The
  Netherlands
  \label{leiden}\and
  Department of Physics \&
  Astronomy, Macalester College, 1600 Grand Avenue, Saint Paul, MN
  55105, USA
  \label{stpaul}\and
  Leibniz-Institut f\"{u}r Astrophysik Potsdam (AIP),
  An der Sternware 16, 14482 Potsdam, Germany
  \label{aip}\and  
  Pontificia Universidad Cat\'{o}lica de Chile - Instituto de Astrof\'{i}sica,
  Av. Vicu\~{n}a Mackenna 4860, 897 0117 Macul, Santiago, \\ Chile  
  \label{pontifica}\and
  Department of Astronomy, Stockholm University,
  AlbaNova University Centre, SE-106 91,
  Stockholm, Sweden
  \label{stockholm}
  \and
  Instituto de Astrof\'{i}sica e Ci\^{e}ncias do Espaço - Centro de
  Astrof\'isica da Universidade do Porto, Rua das Estrelas, 4150-762
  Porto, Portugal
  \label{porto}
  \and
  Observatoire de Gen\`{e}ve, Universit\'{e} de Gen\`{e}ve, Chemin Pegasi 51,
  1290 Versoix,  Switzerland
  \label{geneve}\and
  Universit\"{a}t Bonn, Argelander-Institut für Astronomie, Auf dem H\"{u}gel
  71, D-53121 Bonn, Germany
  \label{bonn}
}

\abstract{% Context
  Outflows from low-mass star-forming galaxies are a fundamental ingredient for
  models of galaxy evolution and cosmology.  Despite seemingly favourable
  conditions for outflow formation in compact starbursting galaxies, convincing
  observational evidence for kiloparsec-scale outflows in such systems is scarce.
}{
  % Aims
  The onset of kiloparsec-scale ionised filaments in the halo of the metal-poor
  compact dwarf SBS 0335-052E was previously not linked to an outflow.  In this paper
  we investigate whether these filaments provide evidence for an outflow.  }{
  % Methods
  We obtained new VLT/MUSE WFM and deep NRAO/VLA B-configuration 21cm data of
  the galaxy.  The MUSE data provide morphology, kinematics, and emission line
  ratios of H$\beta$/H$\alpha$ and [\ion{O}{iii}]$\lambda5007$/H$\alpha$ of the low
  surface-brightness filaments, while the VLA data deliver morphology and
  kinematics of the neutral gas in and around the system.  Both datasets are
  used in concert for comparisons between the ionised and the neutral phase.  }{
  % Results
  We report the prolongation of a lacy filamentary ionised structure up to a
  projected distance of 16\,kpc at
  $\mathrm{SB}_\mathrm{H\alpha} =
  1.5\times10^{-18}$erg\,s\,cm$^{-2}$\,arcsec$^{-2}$.  The filaments exhibit
  unusual low H$\alpha$/H$\beta \approx 2.4$ and low
  [\ion{O}{iii}]/H$\alpha \sim 0.4 - 0.6$ typical of diffuse ionised gas.  They
  are spectrally narrow ($\sim 20$\,km\,s$^{-1}$) and exhibit no velocity
  sub-structure.  The filaments extend outwards from the elongated \ion{H}{I}
  halo.  On small scales, the $N_\mathrm{HI}$ peak is offset from the main
  star-forming sites.  The morphology and kinematics of \ion{H}{I} and \ion{H}{II}
  reveal how star-formation-driven feedback interacts differently with the
  ionised and the neutral phase.
}{
  % Conclusions
  We reason that the filaments are a large-scale manifestation of star-formation-  driven feedback, namely limb-brightened edges of a giant outflow cone that
  protrudes through the halo of this gas-rich system.  A simple toy model of
  such a conical structure is found to be commensurable with the observations. }


\maketitle

\defcitealias{Herenz2017b}{H17}

\section{Introduction}
\label{sec:int}

The response of interstellar gas to energy and momentum deposition from
supernovae and stellar winds is the growth of a hot bubble surrounded by a dense
shell.  Star-forming populations can inject energy and momentum long enough to
sweep the shell up and out to lower density gas, where Rayleigh-Taylor
instabilities eventually deform it before breaking it up.  The hot gas then
vents into the circum-galactic halo of the galaxy as a wind that also entrains
warm and cold gas.  If the wind is powerful enough, the gas may escape the
gravitational potential of a galaxy and thereby enrich the intergalactic medium (IGM)
with metals.  The gas from less powerful outflows then rains back on the galaxy
and is available to form the next generation of stars.  Outflows and winds -- next
to inflows of fresh gas -- are a cornerstone of galaxy formation models and are
vital for regulating cosmic chemical evolution (see reviews by
\citealt{Veilleux2005,Veilleux2020,Collins2022}, as well as
\citealt{Schneider2018}, \citealt{Nelson2019}, \citealt{Mitchell2020},
\citealt{Schneider2020}, and \citealt{Pandya2021} for state-of-the-art computer
models).

Observationally, signatures of galactic outflows in star-forming galaxies are
ubiquitous, both in the nearby \citep[e.g.][]{Heckman2011,Chisholm2016} and in
the high-redshift universe \citep[review by][]{Erb2015}.  Galaxies in the local
Universe allow for detailed panchromatic mappings of outflow phenomena, from the
highest energies (cosmic rays and hot gas) to the longest wavelengths possible
(cold and molecular gas) -- especially if the outflow occurs `edge-on'.  The
most detailed observational studies that analyse multiple phases of outflows
simultaneously are performed on nearby, more evolved, and massive systems.  These
observations nourish our understanding and provide robust constraints of how
mass, energy, and momentum from star formation couple with the different phases
of the interstellar-medium (ISM) to drive outflows\footnote{ Proto-typical example galaxies with well-mapped outflows are M\,82 \citep[e.g.,][]{Bland1988,Leroy2015,Lokhorst2022}
  and Arp\,220 \citep[e.g.][]{Perna2020}; see \cite{Veilleux2005} and
  \cite{Veilleux2020} for more examples.  We also mention in passing that the
  centre of our galaxy also powers an outflow that can be studied in exquisite
  detail \citep[e.g.][]{Hsieh2016,Ponti2021}.}.  However, from a cosmological
perspective, similar constraints are especially needed for the abundant and, in
the early universe, dominating population of low-mass star-forming galaxies.
Currently, cosmological models have to implement those processes in the form of
phenomenological `sub-grid physics' recipes that are tuned to match observed
galaxy population statistics, especially luminosity- and stellar-mass functions
\citep{Somerville2015}.

Understanding the effects of feedback and winds in low-mass systems is
especially required to grasp the large-scale physics of the universe during the
cosmological phase transition known as the Epoch of Reionisation at $z \gtrsim 6$.
This is because dense shells in the pre-fragmentation stage are more opaque to
hydrogen ionising radiation than perforated bubbles at a later stage
\citep[e.g.][]{Fujita2003}. 
Theoretical modelling requires sustained momentum and energy injection to drive
a wind and thereby significant Lyman continuum (LyC) photons out of low-mass galaxies
\citep{Kimm2014,Paardekooper2015}.  Detailed `zoom-in' simulations of
individual dwarf star-forming galaxies suggest that LyC escape is highly
variable and highly anisotropic \citep{Trebitsch2017}.  The angular dependence
is modulated by high optical depth sight-lines along cold-flow accretion
filaments \citep[][]{Park2021} and low optical depth sight-lines that follow the
direction of the outflow phenomena \citep[][]{Trebitsch2017}.  Such anisotropic
beamed LyC may have consequences for the mechanisms by which \ion{H}{II} zones
percolate during reionisation \citep{Paardekooper2015,Furlanetto2016}.

While a pan-chromatic approach is required for a detailed understanding of the
multiple phases of the outflowing material, narrow band imaging and kinematic
analysis of the classical principal emission lines can expose and characterise
feedback, outflow, and wind phenomena just from the $T\sim10^4$K phase
\citep[e.g.][]{Calzetti2004,Westmoquette2008,Zastrow2011,Lee2016,McQuinn2019}.
The increased sensitivity for low surface-brightness line emission offered by
modern integral-field spectrographs \citep[IFSs;][]{Bacon2017a} allows for
unparalleled discoveries in this respect.

Observational evidence for the $T\sim10^4$K phase of feedback-driven winds in
nearby dwarf galaxies and potential early-universe analogues have been amassed
for decades from such imaging and spectral analyses
\citep[e.g.][]{Marlowe1995,McQuinn2015,McQuinn2018,Collins2022}.  Combined
\ion{H}{I} -- \ion{H}{ii} morpho-kinematical investigations of those targets are
feasible as well.  The studies by
\cite{vanEymeren2009,vanEymeren2009b,vanEymeren2010} are noteworthy as they compare 21cm HI and
HII kinematics from H$\alpha$ in five nearby irregular dwarf galaxies
(NGC\,4861, NGC\,2366, NGC\,5408, and IC\,4622).  In these galaxies, multiple localised
velocity offsets between the ionised and the neutral phase are found and are
interpreted as being driven feedback (see also \citealt{Bomans2001,Bomans2007} for
reviews and X-ray follow-up).  This idea is corroborated by filamentary features
seen in H$\alpha$ that extend towards regions of lower neutral columns, as well
as the detection of partially broken super-shells in H$\alpha$.  However, in the
\citeauthor{vanEymeren2009} studies, feedback was never found to be strong enough
to drive outflows into the IGM.  A similar conclusion was
reached by \cite{Westmoquette2008} for the dwarf starburst NGC 1569, where the
hot wind fluid is still presumed to be confined by super shells that extend up
to 1\,kpc outwards from the super-star clusters in this system.
\cite{Roychowdhury2012} compared \ion{H}{I} with \ion{H}{II} in three compact
dwarf galaxies but found only tentative evidence for outflows.  More recently,
\cite{McQuinn2019} analysed H$\alpha$ and 21cm morphologies simultaneously in a
sample of 12 nearby dwarfs.  They traced potential galactic winds via low
surface-brightness H$\alpha$ emission isophotes
($\gtrsim 3 \times 10^{-18}$erg\,s$^{-1}$cm$^{-2}$arcsec$^{-2}$) that extend
beyond regions of their limiting neutral column
($N_\mathrm{HI} \sim 10^{20}$\,cm$^{-2}$ or 1 M$_\odot$\,pc$^{-2}$) gas; however,
even the most convincing winds found in this study were not deemed strong enough
to have material escaping into the IGM.

More compelling early universe analogues for such observational studies are
small, low-mass, and low-metallicity galaxies that exhibit high specific
star-formation rates -- so-called blue compact dwarf galaxies
(\citealt{dePaz2003}, see also the review by \citealt{Thuan2008} and the recent, more
general review on dwarf galaxies by \citealt{Henkel2022}).  Their shallow
gravitational potential and the spatially and temporally concentrated energy and
momentum injection from the starbursting population make them prime candidates for powering large-scale outflows \citep{MacLow1999}.  And indeed, the unprecedented sensitivity
for mapping low-surface brightness emission lines
offered by the \emph{Multi   Unit Spectroscopic Explorer} (MUSE) at ESOs Very Large Telescope (VLT) UT4
\citep[][]{Bacon2014} unveiled spectacular outflows around the blue compact
dwarfs Haro 11 \citep{Menacho2019}, Haro 14 \citep{Cairos2022},
and ESO\,0338-IG04 \citep{Bik2015,Bik2018}, as well as for the nearby HII galaxy
Heinze 2-10 \citep{Cresci2017}.

In this paper we report new observational results on \ion{H}{I} - \ion{H}{II}
morphology and kinematics from the interstellar and circumgalactic medium of the
observationally well-studied blue compact dwarf galaxy SBS\,0335-052E
\citep[e.g.][and references
therein]{Izotov1990,Thuan1997,Izotov2001,Papaderos1998,Johnson2009,Adamo2010,Hunt2014,Kehrig2018,Wofford2021}.
Numerous properties render this galaxy a unique analogue of early universe
galaxies.  With an absolute UV magnitude of $M_\mathrm{UV} = -16.8$, it is
significantly fainter than the characteristic absolute UV magnitude of the
$z \gtrsim 6$ luminosity functions
\citep[$M_\mathrm{UV}^\star < -20.6$;][]{Bouwens2015a}.  With
$12 + \log(\mathrm{O/H}) \sim 7.25$ \citep{Izotov2009}, it is one of the most
metal-poor galaxies in the local Universe.  This low-mass galaxy
\citep[$M_\star = 6 \times 10^6$\,M$_\odot$;][]{Reines2008} is also extremely
compact; the starburst is taking place within 500 pc, where six super-star
clusters with diameters $\lesssim 60$\,pc form stars at
$\sim$0.7\,M$_\odot$\,yr$^{-1}$ or at specific star-formation rates up to
20\,M$_\odot$\,yr$^{-1}$kpc$^{-2}$.
% age gradient / propagating star formation!!!

Our previous analysis of VLT/MUSE integral-field spectroscopic data from this
galaxy unveiled two prominent ionised filaments in its halo \cite[][hereafter
H17]{Herenz2017b}.  Here we present an analysis of new MUSE data in concert with
new \textit{Karl G. Jansky} Very Large Array (VLA) B-configuration observations
that show that these filaments are likely related to an unprecedented large
conical outflow structure that pinches deep into the halo of this compact
starburst galaxy (Sect.~\ref{sec:disc}).  Prior to this, we summarise
observations and data reduction in Sect.~\ref{sec:obs}; also, the morphological and
kinematic analyses of neutral and ionised gas are detailed in
Sect.~\ref{sec:res}. In Sect.~\ref{sec:sc} we summarise and conclude.  For
distance conversions we assume a flat $\Lambda$ cold dark matter Universe with $\Omega_m = 0.3$
and $H_0 = 70$\,km\,s$^{-1}$Mpc$^{-1}$, and we adopt $z = 0.01352$ as the
cosmological redshift for SBS\,0335-052E \citep{Moiseev2010}.  This translates
to a luminosity distance of 58.42\,Mpc and an angular scale of
276\,pc/\arcsec{}.

\section{Observations and data reduction}
\label{sec:obs}

\subsection{VLT/MUSE}
\label{sec:muse}

For the present analysis we combine data from two ESO/VLT MUSE programmes:
Programme 096.B-0690 (PI: Hayes) with data taken on November 16th and 17th in
2015 (clear to photometric sky, DIMM seeing $\approx$0.7\arcsec{} -
0.9\arcsec{}) and Programme 0104.B-0834 (PI: Herenz) with data taken on October
2nd and 3rd in 2020 (photometric sky, DIMM seeing 1.1\arcsec{} - 1.4\arcsec{}).
Both programmes were observed in wide-field mode (1\arcmin{}$\times$1\arcmin{}
field of view) with the blue cut-off filter removed (extended mode, wavelength
range 465\,nm -- 930\,nm) and without adaptive optics.  Programme 096.B-0690 
centred the field of view on SBS\,0335-052E (RA:
03$^\text{h}$37$^\text{m}$44.075$^\text{s}$ Dec:
$-$05\degr{}02\arcmin{}39.5\arcsec{}).  These data lead to the discovery of
ionised filaments in the halo of SBS\,0335-052E \citepalias{Herenz2017b}, and
were also analysed by \cite{Kehrig2018} and \cite{Wofford2021} to understand the
properties of the ionising sources of this galaxy.  In order to map the extent
of the filament we obtained new MUSE data in programme 0104.B-0834 with a
pointing centred north-west of the galaxy (RA:
03$^\text{h}$37$^\text{m}$41.9167$^\text{s}$ Dec:
$-$05\degr{}01\arcmin{}57.2\arcsec{}'). This pointing overlaps partly with the
discovery data.  The total open shutter times are 5680\,s (8 exposures a 710\,s)
and 5576\,s (8 exposures a 697\,s) for the central- and offset pointing,
respectively.

We reduced the data with version 2.8.3 of the MUSE data processing pipeline
\citep{Weilbacher2020}.  The reduction of the 2015 data was already detailed in
\citet[][Sect. 3.2]{Wofford2021} and we apply here the same recipes to our new
2020 data.  These recipes provide us first with sky-subtracted, flux calibrated,
and astrometrically roughly calibrated pixel tables for each exposure.  As
explained in \citetalias{Herenz2017b}, the spatially extended line emission in
the data requires a modification of the default pipeline sky subtraction
procedure; significant flux from the extended principal \ion{H}{ii} emission
around SBS\,0335-052E would be removed without this modification.  The most
prominent oversubtracted emission shows up as spikes at the observed wavelengths
of H$\alpha$ and [\ion{O}{III}] $\lambda$5007 in the sky continuum spectra that
the pipeline determines after the best-fit sky emission line spectra have been
subtracted \citep[][Sect.~3.9.1]{Weilbacher2020}.  We thus altered the output
sky continuum spectra from the initial run of the sky subtraction routine by
replacing the values around principal emission lines via linear interpolation
from the surrounding spectral bins.  The corrected sky continuum spectra and the
unaltered sky emission line spectra from the first run were then used as input
in a second run of the sky subtraction procedure.  This removed the visible
oversubtraction artefacts at H$\alpha$ and [\ion{O}{III}] $\lambda$5007 in the
initial reduction, and to be secure we also repeated the procedure for the lines
[\ion{O}{III}] $\lambda$4363, \ion{He}{II} $\lambda$4686, \ion{He}{I}
$\lambda$5876, [\ion{S}{ii}] $\lambda\lambda$6717,6731, and [\ion{O}{I}]
$\lambda$6300, even if such strong spikes were apparent in the continuum sky
spectrum from the initial run of the sky subtraction.  Lastly, the individual
exposure pixel tables are resampled onto a 3-dimensional
473$\times$532$\times$3802 grid (datacube); the first two axes sample the
spatial domain with 0.2\arcsec{}$\times$0.2\arcsec{} spectral pixels (spaxels)
parallel to RA and Dec and the third axis samples wavelength linearly from
4599.96\AA{} to 9351.21\AA{} with 1.25\AA{} bins.  The final coordinate
transformation between datacube spaxels and (RA,Dec)-coordinates is registered
by cross-correlating the Pan-STARRS \citep{Chambers2016, Magnier2020}
\textit{r}-band image against a synthetic \textit{r}-band image created from the
datacube.

\subsection{VLA}
\label{sec:vla}

We obtained new HI 21cm observations with the Karl G. Jansky Very Large Array
(VLA) in its B-configuration.  The observations (project number 17B-234; PI:
Herenz) made use of the L-Band receiver centred on the redshifted 21cm frequency
of the galaxy.  The programme was executed in 10$\times$3h blocks from Sep 30th
to Oct 20th, 2017. 

All raw VLA data are calibrated following standard prescriptions using the
National Radio Astronomy Observatory (NRAO) {\tt casa} software package (version
5.6.1-8; \citealt{McMullin2007,Emonts2020}).  Radio-frequency interference is
manually identified and excised from the raw data.  These data are then
calibrated, and once a satisfactory calibration is achieved, the data are
continuum subtracted and weighted using CASA's {\tt uvcontsub} and {\tt statwt}
functions. ``Satisfactory" calibration constitutes an RMS noise value in
source-free channels of the cube that is within 10\% of the theoretically
optimal RMS value given the duration of the observation, the configuration of
the VLA, and the spectral properties of the datacube.

Using {\tt tclean} and the masking algorithm {\tt Auto-Multithresh}
\citep[AMT;][]{Kepley2020}, we generate an intermediate image cube using {\tt
  casa} version 6.3.0.48. From this intermediate cube, we extract the
AMT-generated mask. We chose AMT for mask generation as it produced high quality
data products with reproducibility and speed. All image cubes are created using
Briggs weighting with a robust parameter of 0.5 for a compromise between angular
resolution and
sensitivity. 

We set the AMT noise threshold to 3$\sigma$. In doing so, AMT finds
and masks all emission associated with our target sources, in addition
to noise peaks that rise above this threshold. To address this, we
then examine the AMT-generated mask interactively and extract only the
regions that include source emission. Through this process of
automated mask generation and interactive selection, we end up with a
final, robustly generated mask that solely and fully masks the target
sources.  With this mask in hand, we then initiate a final {\tt
  tclean} with the same parameters as used for the mask-generation
clean, but now providing the source-only mask generated in the
previous step.  We clean down to 0.5$\sigma$, as this removes the need
for residual flux rescaling.  The resulting final image cube covers an
area of 22.8 arcmin$^2$ sampled at 1\arcsec{}/pixel, and the channel
width is 10.26\,km\,s$^{-1}$.  The typical rms-noise per channel is
2.5$\times10^{-4}$ Jy\,beam$^{-1}$.  The beam is
6.1\arcsec{}$\times$4.6\arcsec{} and it is oriented 23.6$^\circ$ west
of north.  The positional accuracy of the VLA HI data, which is based
on milli-arcsecond precision positions of quasars in the VLA
calibrator manual\footnote{
  \url{https://science.nrao.edu/facilities/vla/observing/callist}},
can be assumed to be typically 10\% of the synthesised beam size.

We also create another datacube from the B-configuration data, that
provides us with a different spatial resolution.  This cube is created
by applying a 7k$\lambda$ UV-taper in the imaging process.  This
UV-tapering downweights the longest baselines using a Gaussian
function in the \textit{uv}-plane (with the FWHM measured in units of
wavelengths), producing a cube at a spatial resolution that is roughly
equivalent to VLA C-configuration observations (beam
15.9\arcsec{}$\times$14.7\arcsec{} at 66.4$^\circ$ east of north).
The datacube differs from the untapered cube by an adjusted spatial
sampling of 3\arcsec{}/pixel and the RMS-noise per channel is
4$\times10^{-4}$ Jy\,beam$^{-1}$.  The 7k$\lambda$-tapered data are
less sensitive than the untapered data, but this is compensated by the
larger beam that enables the mapping of lower columns over larger areas.

\section{Analysis and results}
\label{sec:res}

\subsection{Detection and morphology of extended \ion{H}{II} and \ion{H}{I} emission}
\label{sec:detect-morph-extend}

\begin{figure*}[]
  \centering
  \includegraphics[width=\textwidth,trim=3 3 3
  0,clip=true]{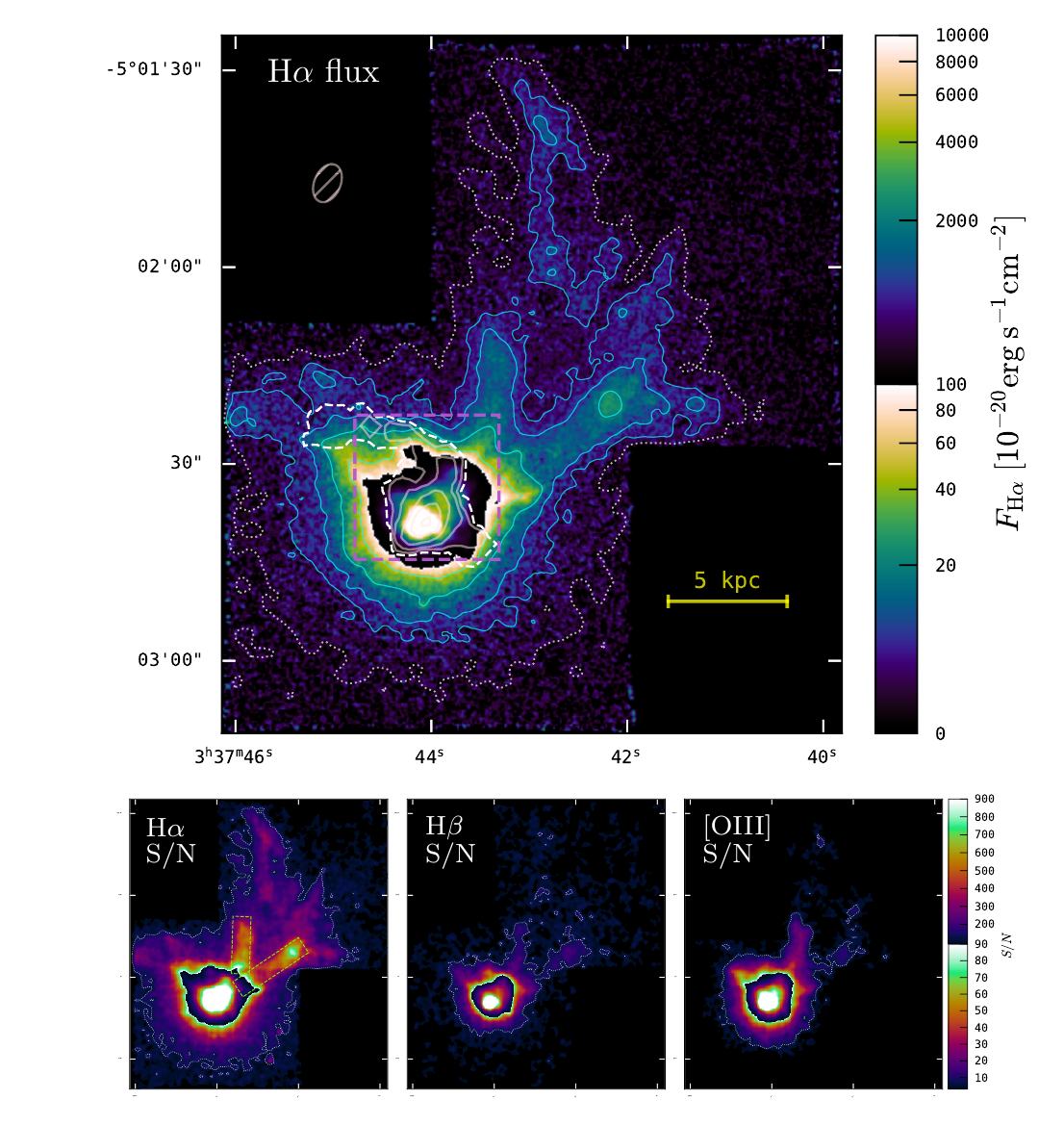}
  %\vspace{1.1em}
  \caption{\ion{H}{ii} and \ion{H}{i} emission as seen with MUSE and VLA around SBS\,0335-052E. \emph{Top panel:} Continuum-subtracted H$\alpha$ narrowband image.  Fluxes
    are encoded via an asinh-scaled cyclic colour map from $0$ to
    $10^{-18}$erg\,s$^{-1}$cm$^{-2}$ in the low surface-brightness,
    and from $10^{-18}$erg\,s$^{-1}$cm$^{-2}$ to
    $10^{-16}$erg\,s$^{-1}$cm$^{-2}$ in the central high surface-brightness
    region. The cyan solid contours demarcate H$\alpha$ isophotes at
    $\mathrm{SB}_\mathrm{H\alpha}= \{1.5, 2.5, 5, 12.5\} \times
    10^{-18}$erg\,s$^{-1}$cm$^{-2}$arcsec$^{-2}$ while the white dotted contour
    demarcates the limiting surface-brightness of
    $\mathrm{SB}_\mathrm{H\alpha}=7.5\times10^{-19}$erg\,s$^{-1}$cm$^{-2}$arcsec$^{-2}$.
    The image has been smoothed with a Gaussian of 0.4\arcsec{} FWHM.  Subdued
    grey contours indicate HI column densities
    $N_\mathrm{HI} = \{ 5, 10, 20, 30, 40 \} \times 10^{20}$cm$^{-2}$ from our
    VLA-B configuration observations; these contours are displayed more
    prominently in Fig.~\ref{fig:ha_zoom}.  The dashed white contour demarcates
    the $2.5\sigma = 2.5\times 10^{20}$cm$^{-2}$ detection limit.  The VLA-B
    configuration beam, 6.1\arcsec{}$\times$4.6\arcsec{} oriented 23.6$^\circ$
    west of north, is indicated via a grey shaded ellipse in the top left.  The
    displayed field of view is
    1\arcmin{}35\arcsec{}\,$\times$\,1\arcmin{}46\arcsec{} corresponding to
    26.2\,kpc\,$\times$\,29.2\,kpc in projection.  The dashed violet square in
    the centre indicates the viewport used in Fig.~\ref{fig:ha_zoom}.  North is
    up and east is to the left.  \emph{Bottom panels:} Signal-to-noise of the
    three principal emission lines (H$\alpha$, H$\beta$, and [\ion{O}{iii}]
    $\lambda5007$ in the \emph{left}, \emph{centre}, and \emph{right} panel,
    respectively) in which extended ionised gas is detected (see text for
    details). The maps are displayed with a linear cyclic colour map from 0 to
    90 and from 90 to 900.  White dotted contours mark out regions with
    $\mathrm{SN}>8$.  The chartreuse dashed rectangles in the H$\alpha$ panel
    outline the rectangular regions used for the line ratio analysis in
    Sect.~\ref{sec:ratio}.  }
  \label{fig:hanb}
\end{figure*}

\begin{figure}
  \centering
  \includegraphics[width=0.5\textwidth,
  trim=120 50 102 55,clip=true]{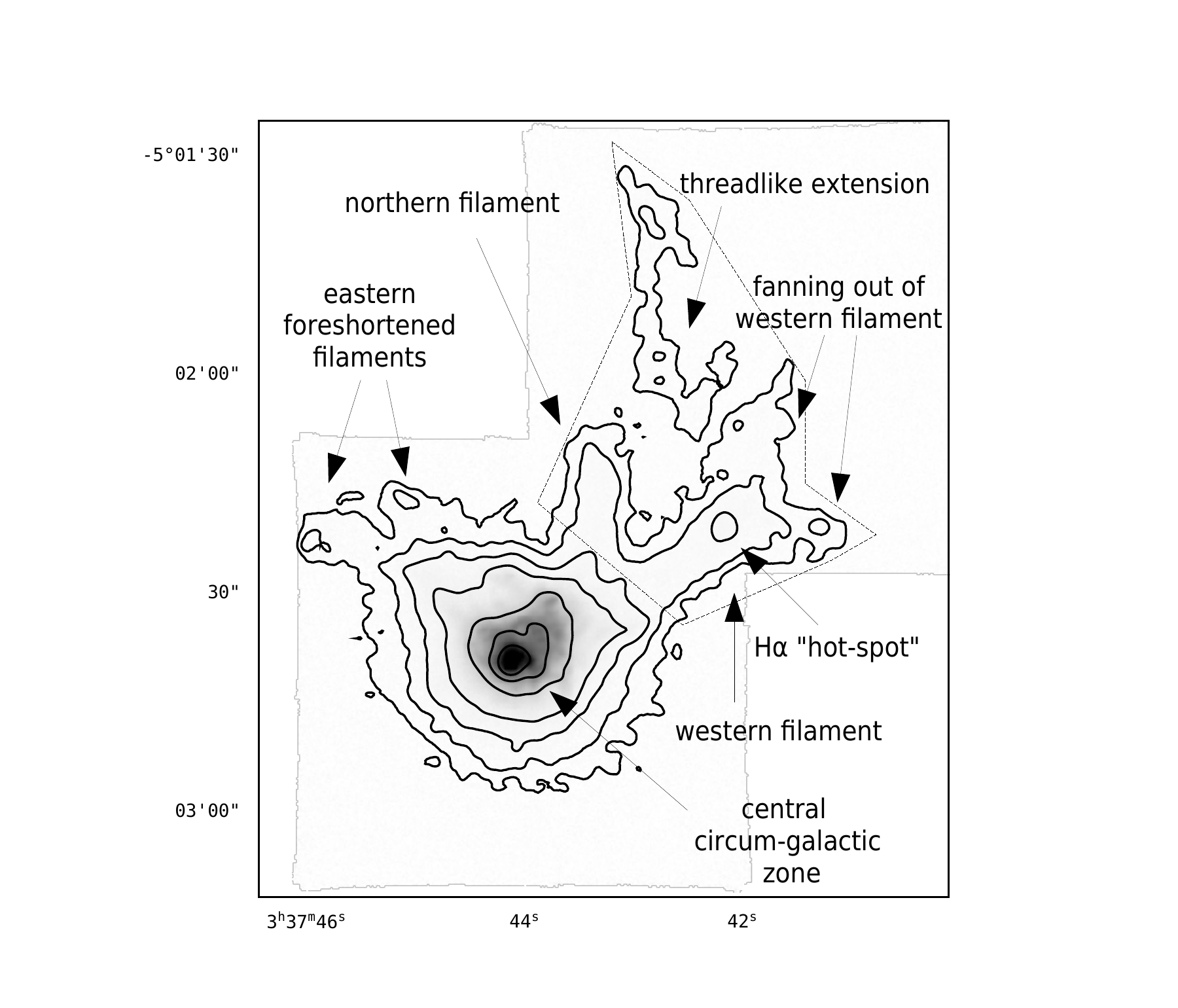}
  \caption{Annotated H$\alpha$ surface brightness contours (contour levels
    $\mathrm{SB}_\mathrm{H\alpha}= \{1.5, 2.5, 5, 12.5, 100, 700 \} \times
    10^{-18}$erg\,s$^{-1}$cm$^{-2}$arcsec$^{-2}$). In the background the
    H$\alpha$ narrowband is shown with a log-stretch to
    $5\times10^{-16}$erg\,s$^{-1}$cm$^{-2}$.  The thin-dotted line surrounding
    the filaments to the north-west demarcates the aperture that is used to
    estimate the total H$\alpha$ flux from that structure
    (Sect.~\ref{sec:disc}).}
  \label{fig:cont}
\end{figure}

\begin{figure}
  \includegraphics[width=0.45\textwidth,trim=11 20 0
  0,clip=true]{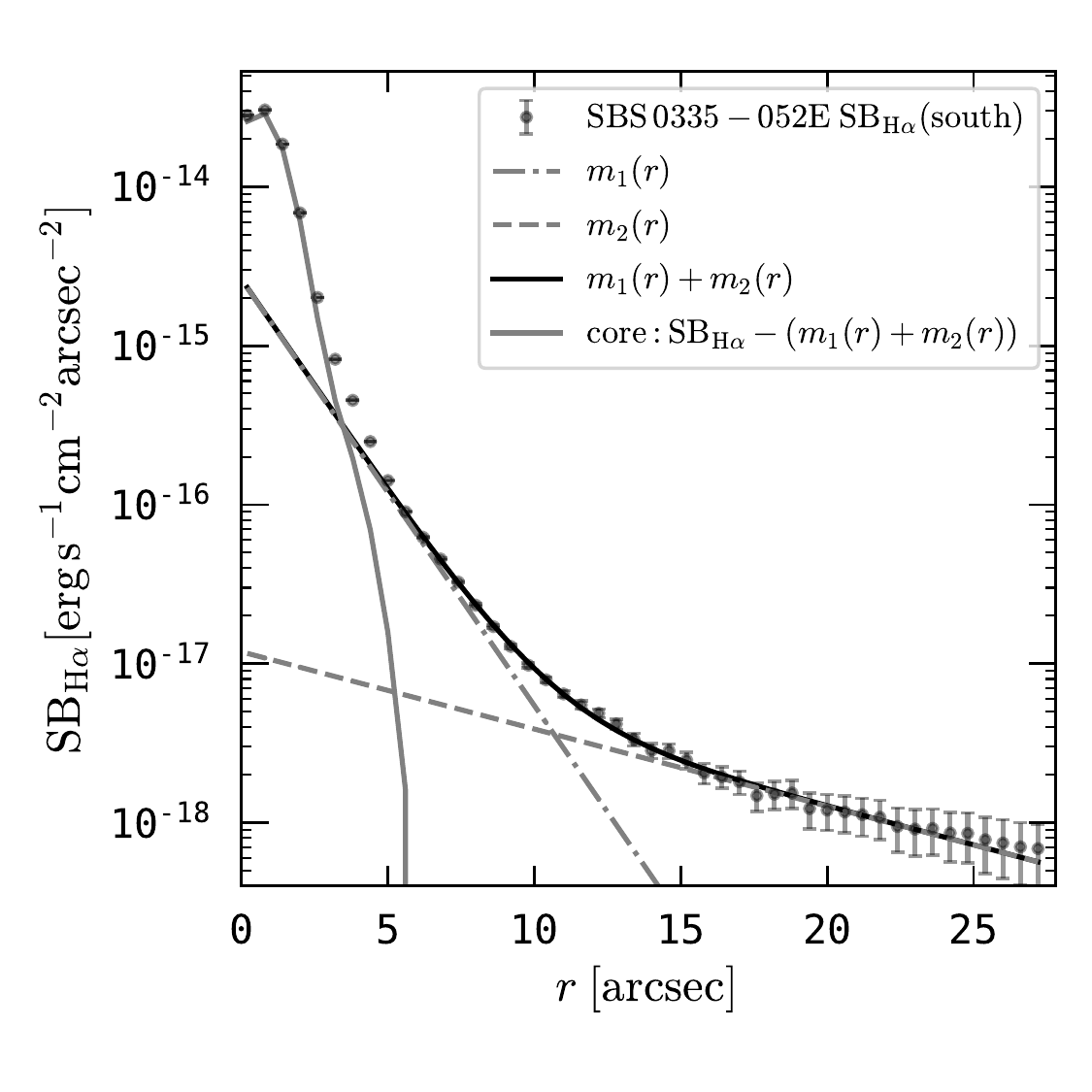}
  \caption{H$\alpha$ surface brightness profile in the south, i.e. not affected
    by the filaments in the north.  The points show the observed profile in
    circular annuli and error-bars are 2$\sigma$.  The black line indicates the
    double exponential profile fitted to the outer ($r>5\arcsec$) profile, while
    dot-dashed and dashed lines indicate the individual components with
    scale-lengths $r_1 = 1.62$\arcsec{} (447\,pc) and $r_2 = 8.94$\arcsec{}
    (2.47\,kpc).  The grey line indicates the light profile of the central
    ($r\leq5$\arcsec{}) residual component that is obtained after subtracting
    the double exponential.}
  \label{fig:sbha}
\end{figure}

\begin{figure*}
  \centering
  \includegraphics[width=0.49\textwidth,
  trim=10 25 0
  10,clip=true]{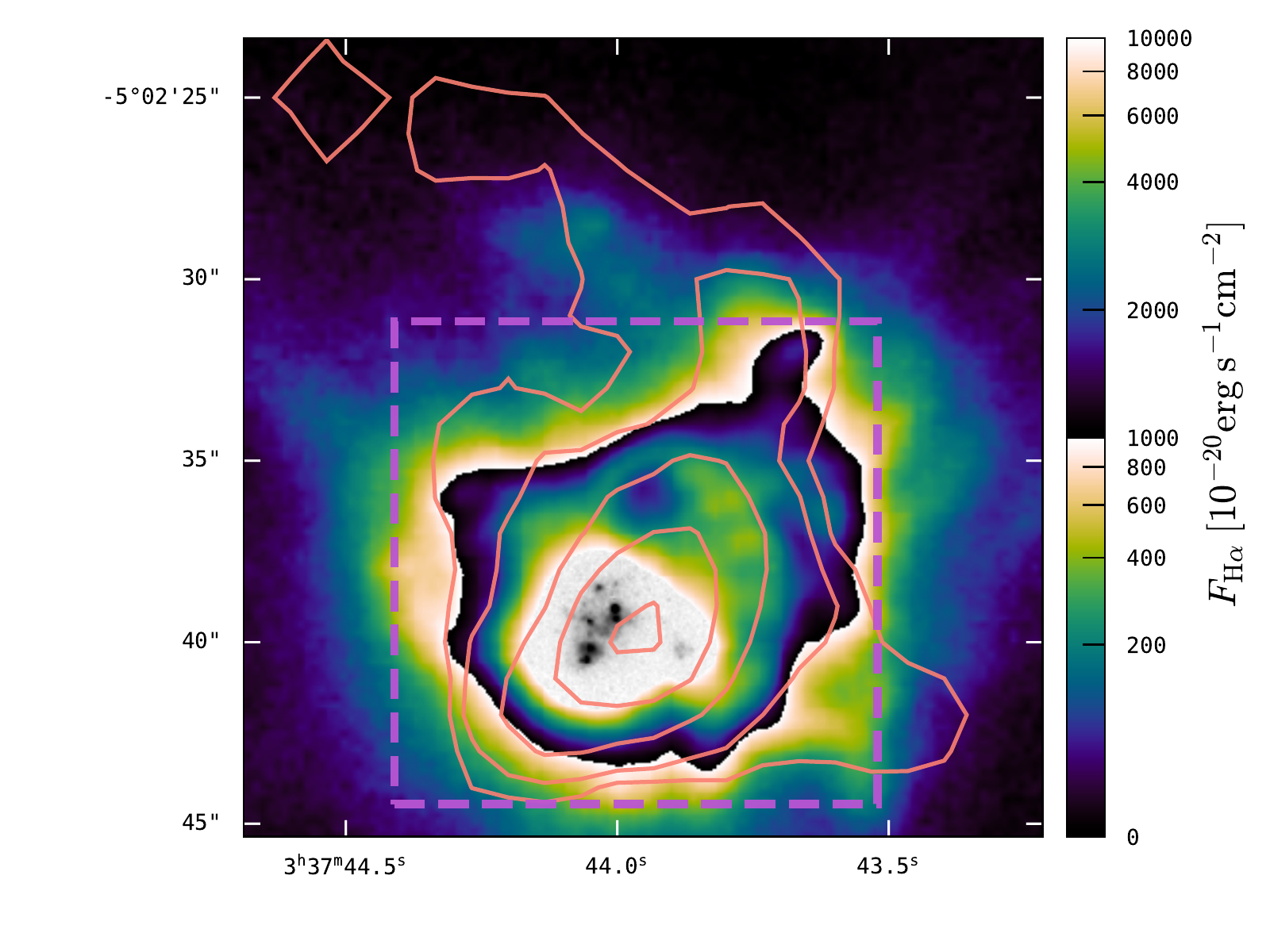} 
  \includegraphics[width=0.49\textwidth, trim=10 25 0 10,clip=True]{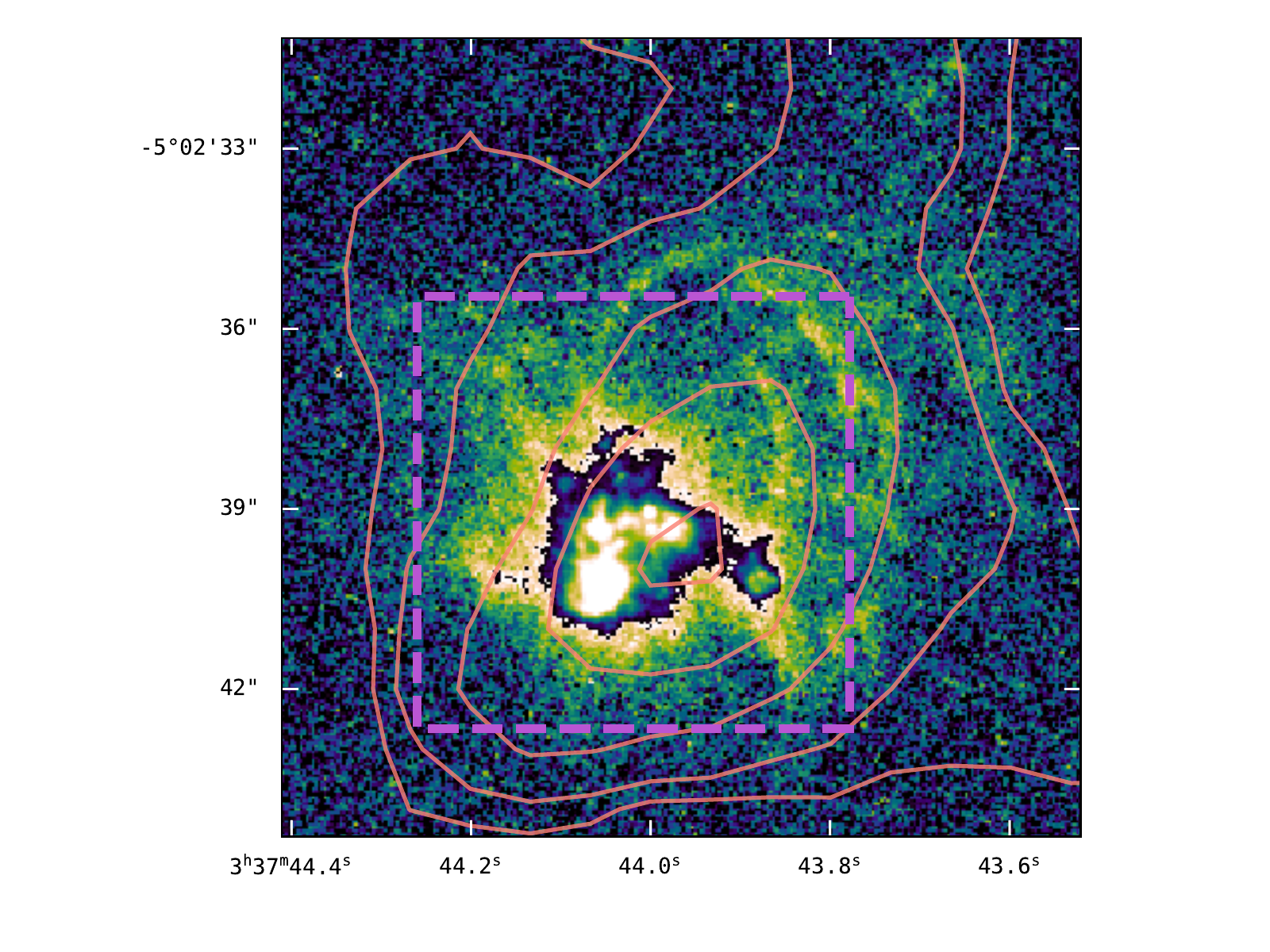}
  \caption{\emph{Left panel}: Zoomed view
    (22\arcsec{}$\times$22\arcsec{}\,/\,6\,kpc\,$\times$\,6\,kpc) of the
    H$\alpha$ narrow band image from the top panel of Fig.~\ref{fig:hanb} with
    $N_\mathrm{HI}$ contours from the VLA B-configuration 21 cm observations
    ($N_\text{HI} = \{5, 10, 20, 30, 40\}\times 10^{20}$cm$^{-2}$, rose lines);
    these contours are also shown in Fig.~\ref{fig:hanb} as subdued white lines.
    The cyclic colour map encodes H$\alpha$ flux in a asinh-scale from
    $0$ to $10^{-17}$\,erg\,s$^{-1}$cm$^{-2}$ to $10^{-15}$\,erg\,s$^{-1}$cm$^{-2}$.
    In the brightest central region the archival HST F550M image is inset, with
    an arbitrary asinh-scaling to highlight the compact super-star clusters that
    comprise the main stellar body of this system.  The violet dashed square
    indicates the region that is displayed in the right panel.  \emph{Right
      panel}: Inner high-$\mathrm{SB}_\mathrm{H\alpha}$ region as seen in the
    archival HST FR656N image
    (13\arcsec{}$\times$13\arcsec{}\,/\,3.5\,kpc\,$\times$\,3.5\,kpc) displayed
    here with a cyclic asinh-scale from 0 to
    $5\times10^{-15}$erg\,s$^{-1}$cm$^{-2}$ to $10^{13}$erg\,s$^{-1}$cm$^{-2}$.
    Here the violet dashed square indicates the region that is displayed in
    Fig.~\ref{fig:clustident}.}
  \label{fig:ha_zoom}
\end{figure*}

\begin{figure}
  \centering \includegraphics[width=0.4\textwidth,trim=20 30 0 0 ,clip=True]{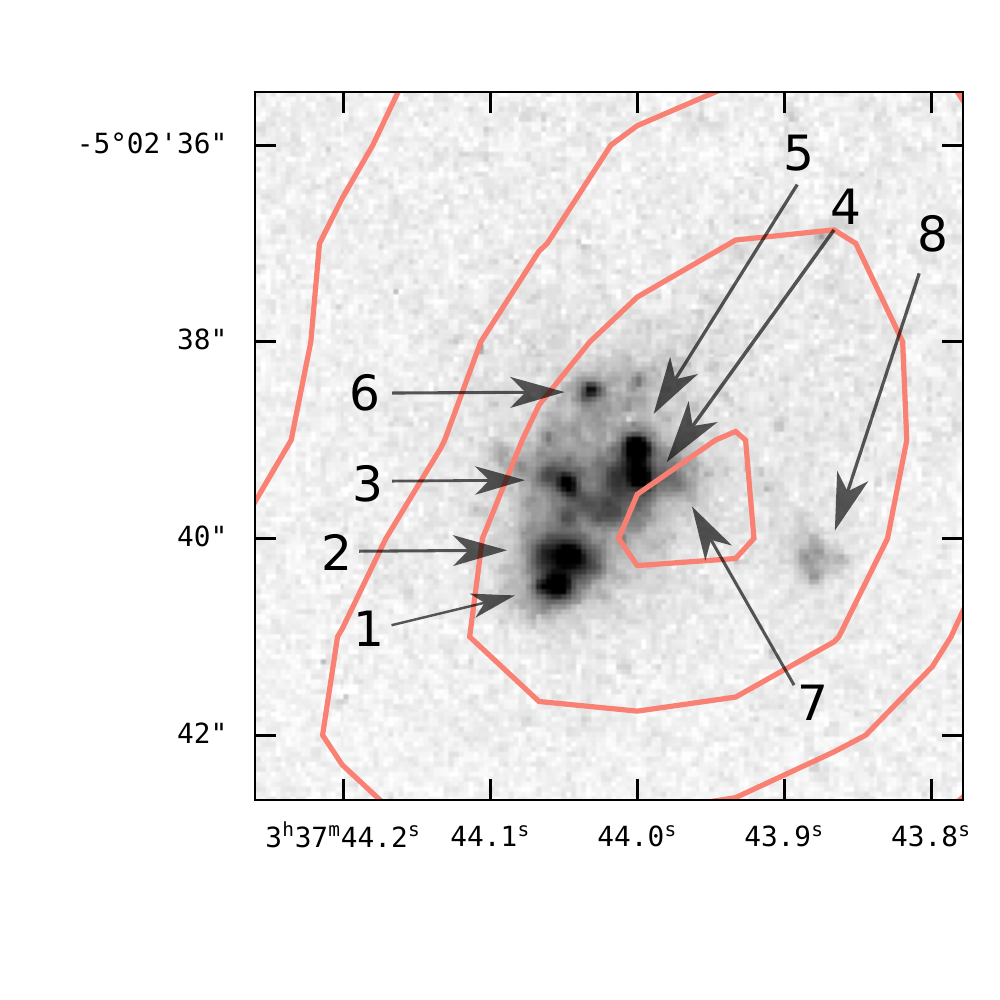}
  \caption{HST archival F550M image with super-star clusters labelled according
    to \cite{Thompson2009}; $N_\mathrm{HI}$ contours as in
    Fig.~\ref{fig:ha_zoom}.  The viewport, indicated as a violet dashed square
    in Fig.~\ref{fig:ha_zoom}, is 7\arcsec{}$\times$7\arcsec{} or
    1.88\,kpc$\times$1.88\,kpc.}
  \label{fig:clustident}
\end{figure}

\begin{figure}
  \centering \includegraphics[width=0.4\textwidth]{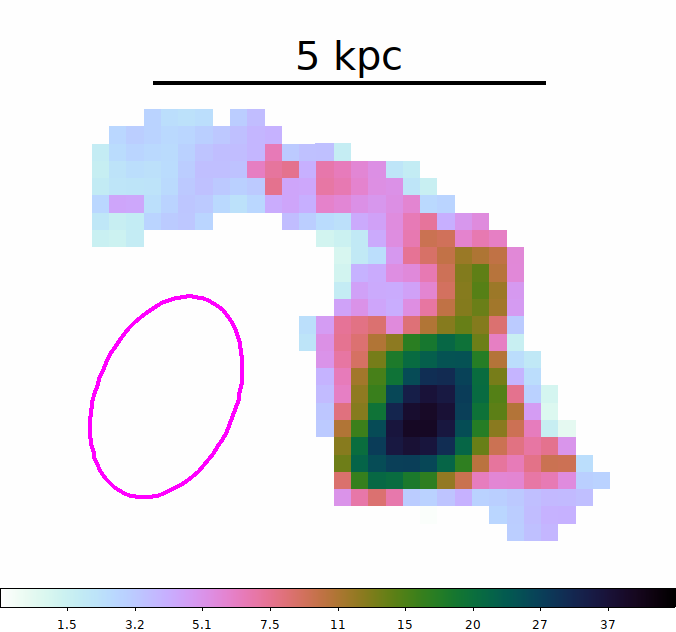}
  \caption{21\,cm moment-0 map of SBS\,0335-052E with the colour-bar encoding
    the neutral column $N_\mathrm{HI}$ in units of $10^{20}$cm$^{-2}$ using an
    asinh-scaling.  The magenta ellipse indicates the dimension and orientation
    of the VLA B-configuration beam.  This map has been used to draw the
    $N_\mathrm{HI}$ contours in Fig.~\ref{fig:hanb}, Fig.~\ref{fig:ha_zoom}, and
    Fig.~\ref{fig:clustident}.}
  \label{fig:mom0bc}
\end{figure}

\subsubsection{Ionised gas}
\label{sec:ionised-gas}

To optimally detect extended line emission in our MUSE data, we processed the
reduced datacube with routines from the \texttt{LSDCat} software
\citep{Herenz2017}.  We started by subtracting a running median in the spectral
direction.  The band-width of this filter (which effectively enables the removal
of low-frequency continua) is 151\,px (188.75\,\AA{}).  We then cross-correlated
the continuum-subtracted datacube with a 3D Gaussian template.  The result of
this procedure is a unit-less datacube, that we dub in \texttt{LSDCat} as ``S/N
cube'' in reference to the signal-to-noise ratio maximising characteristic of
matched filtering.  However, we here use \texttt{LSDCat} only as a simple linear
filter to suppresses high-frequency noise and enhance the emission line signal
in the datacube.  By experimentation we found that the parameters
$\sigma_v = 150$\,km\,s$^{-1}$ and $\sigma_G =2$\arcsec{} optimally augmented
the spectrally narrow and spatially extended emission line features in the
circumgalactic medium around SBS\,0335-052E.  We visually examined the resulting
S/N cube with the QFitsView software \citep{Ott2012} in layers with strong
nebular line detections in the integrated spectrum (cf. Fig.~3 in
\citealt{Wofford2021}).  We here report significant detections of extended
filamentary emission in H$\alpha$, [\ion{O}{iii}] $\lambda5007$, and H$\beta$
around SBS\,0335-052E (Fig.~\ref{fig:hanb}).  H$\beta$ was not reported in
\citetalias{Herenz2017b}, but the increased exposure time in the region where
the discovery data overlaps with our new observations renders the detection of
this line significant.  The filaments remain undetected in other emission lines.

The H$\alpha$ narrowband image shown in Fig.~\ref{fig:hanb} was synthesised by
summing over the three layers in the median-filter subtracted datacube
($6651.21\pm1.25$\AA{} observed frame).  Only these layers contain appreciable
H$\alpha$ emission (see also the line-width analysis in Sect.~\ref{sec:hiikin}).
The maximum S/N-images in Fig.~\ref{fig:hanb} were created by taking the maximum
values from the LSDCat cross-correlated datacube over a few layers around the
emission lines of interest.  We define
$\mathrm{SN}^\mathrm{LSD}_\mathrm{H\alpha} > 8$ as a conservative criterion for
detected emission line signal.  Spaxels in the S/N cube that fulfil this
criterion are surrounded by a white dotted contour in Fig.~\ref{fig:hanb}.

As can be appreciated from Fig.~\ref{fig:hanb}, the new MUSE observations reveal
a continuation of the filamentary structure first reported by
\citetalias{Herenz2017b}.  This H$\alpha$ emitting complex extends up to
$\gtrsim 15$\,kpc in projection towards the north-west from the galaxy's main
stellar body.  While the pronounced high-SN emission stems from the two
filamentary threads, H$\alpha$ is also detected throughout a contiguous region
in between the filaments.  The [\ion{O}{iii}] and H$\beta$ detections, on the
other hand, are confined exclusively to the filaments.  The northern
filament shows quite a strong detection in [\ion{O}{iii}]
($\mathrm{SN} \gtrsim 25$), whereas the north-western filament is significantly
weaker in [\ion{O}{iii}] ($\mathrm{SN} \gtrsim 10$).  An analysis of the
observed line ratios from the filaments is presented in Sect.~\ref{sec:ratio}.

To make the discussion of morphological features of the extended ionised
structure text more accessible, we display in Fig.~\ref{fig:cont} H$\alpha$
surface brightness contours with annotations.  The two filaments that protrude
towards the north (northern filament; $\mathrm{PA} \approx 0^\circ$) and
north-west (western filament; $\mathrm{PA} \approx 315^\circ$) are characterised
by
$\mathrm{SB}_\mathrm{H\alpha} \approx 3 \times
10^{-18}$\,erg\,s$^{-1}$cm$^{-2}$.  The northern filament fragments after
$\sim19$\arcsec{} ($\sim$5\,kpc) into a lacier prolongation.  This threadlike
extension continues outwards up to $\sim58$\arcsec{} ($\sim$16\,kpc).  The
western filament, on the other hand, brightens noticeably after
$\sim25$\arcsec{} ($\sim$7\,kpc) to
$\mathrm{SB}_\mathrm{H\alpha} \approx 5 \times 10^{-18}$\,erg\,s$^{-1}$cm$^{-2}$
(H$\alpha$ ``hot-spot'').  After this brightening the western filament fans out
into two short branches, one to the west and one to the north-east.  The
accented filamentary features are surrounded by diffuse emission at
$\mathrm{SB}_\mathrm{H\alpha} \approx 7 \times
10^{-19}$\,erg\,s$^{-1}$cm$^{-2}$.  Two shorter and fainter filaments
are found towards the north-east of the galaxy; these are labelled as ``eastern
foreshortened filaments'' in Fig.~\ref{fig:cont}.  They may trace a similar
structure as the northern and the western filaments, but appear shorter due
to a different alignment along our sight-line.

While H$\alpha$ in the northern circumgalactic halo shows complex morphological
features, the southern halo displays a rather smooth and slowly decaying light
profile.  At our sensitivity we detect diffuse H$\alpha$ emission up to
25\arcsec{} away from the main stellar body of the system.  To provide a
quantitative assessment of this emission we extract a radial surface-brightness
profile, $\mathrm{SB}_\mathrm{H\alpha}(r)$, under the assumption of circular
symmetry.  Therefore, we first determine the photometric centre
$ (\overline{x},\overline{y}) = ( \sum_{xy} I_{xy} x / \sum_{xy} I_{xy} \, , \,
\sum_{xy} I_{xy} y / \sum_{xy} I_{xy} )$, where $I_{xy}$ is the H$\alpha$ flux
value of the pixel at coordinates $(x,y)$. We find
$(\overline{x},\overline{y}) = (153.7, 162.7)$ which is located at (RA, Dec) =
(3$^\mathrm{h}$37$^\mathrm{m}$43.99$^\mathrm{s}$,
\mbox{-5}$^\circ$02\arcmin{}39.2\arcsec{}).  We then extract
$\mathrm{SB}_\mathrm{H\alpha}(r)$ in semicircular annuli of 0.6\arcsec{} width
south $\overline{y}$ and centred on $(\overline{x},\overline{y})$.  The result
of this procedure is shown in Fig.~\ref{fig:sbha}.  The measured H$\alpha$ SB
profile of SBS\,0335-052E is characterised by a central high-SB region
(``core'') which drops rapidly from $2 \times 10^{-14}$ to
$2 \times 10^{-16}$erg\,s$^{-1}$cm$^{-2}$arcsec$^{-2}$ within the first
5\arcsec{} and then kinks into a more slowly decaying (``halo'') profile.
Further out, at $\sim 13$\arcsec{} and
$\sim 10^{-17}$erg\,s$^{-1}$cm$^{-2}$arcsec$^{-2}$, another kink towards an even
flatter profile is found.

As can be seen from Fig.~\ref{fig:sbha}, the outer profile can be described
adequately by a sum of two exponentials with scale lengths $r_1 = 1.62$\arcsec{}
(447\,pc) and $r_2 = 8.94$\arcsec{} (2.47\,kpc).  We caution that given the
extreme intensity contrast -- four orders of magnitude between core and outer
halo region -- diffuse scattered light may have brightened the observed profile
at the faintest isophotes.  This observational effect of scattered excess light
was studied extensively by \cite{Sandin2014} and \cite{Sandin2015}.  These
papers caution against the physical interpretation of faint halo scale lengths around
bright compact galaxies in the absence of an accurate extended point-spread
function model for the observations.  In particular the here measured flattening
at larger radii was put forward as a tale-tale sign of diffuse scattered light
\citep[see especially Sect.~5 and Appendix~E in][]{Sandin2015}.

The ``core'' high-SB region
($\mathrm{SB}_\mathrm{H\alpha} \gtrsim 2 \times
10^{-16}$\,erg\,s$^{-1}$\,cm$^{-2}$), denoted as ``central circumgalactic zone''
in Fig.~\ref{fig:cont}, is characterised by an intricate morphology. This region
is detected in a plethora of other emission lines in the MUSE IFS data
\citep[see Figures in][]{Kehrig2018}.  To visualise the morphology of this
region adequately we display in the left panel of Fig.~\ref{fig:ha_zoom} a
zoomed-in view of the H$\alpha$ narrow band image from Fig.~\ref{fig:hanb}; we
now use a different cyclic asinh-stretch that is appropriate to trace the
brighter morphological features relevant at this magnification.  To better
visualise the size difference between circum-galactic ionised gas structures and
main stellar body of the galaxy, we inset the Hubble Space Telescope (HST) F550M
image into the brightest central
($\mathrm{SB}_\mathrm{H\alpha} > 10^{-15}$erg\,s$^{-1}$cm$^{-2}$) region.  This
band is devoid of strong emission lines and thereby traces only the morpholgy of
the stellar continuum.  We show the HST FR656N image in the right panel of
Fig.~\ref{fig:ha_zoom}.  This resolves morphological features in the central
region in greater detail and nicely demonstrates the gain in depth of our MUSE
data with respect to HST.

Figure~\ref{fig:ha_zoom} reveals that the high-SB H$\alpha$ region exhibits
multiple nested arcs or loops, as well as outward pointing filaments or spurs.
Generally loops seen in H$\alpha$ at such large scales trace the limb-brightened
edges of so-called super-bubbles or super-giant shells
\citep[e.g.][]{Chu1995,Bomans2001,Bomans2007}.  Disconnected loops are believed
to point towards regions where the shells are broken up and where the
pressurised interior can vent out.  The interaction of the hot wind ($> 10^5$ -
$10^7$\,K) with the colder material in the surroundings and warm gas
($\sim 10^4$\,K) entrained in these bubble-blow outs, as well as the merging of
super-bubbles, are believed to manifest in outward pointing filamentary spurs
\citep[e.g.][]{Cooper2008,Tanner2016}.  Here we see that the two brightest
inner arcs towards the north-west resemble closed loops, but variations in
surface brightness point at inhomogeneities that indicate cracks in the shells
caused by Rayleigh-Taylor instabilities.  Further outwards towards the
north-west, only barely detected in the FR656N image but clearly visible in the
MUSE data, we see a flaring of spur-like features.  These structures are very
extended, up to $\sim 10$\arcsec{} or 2.7\,kpc from the main star-forming
regions.  These features may be interpreted as outflowing warm gas that is
entrained in the hot wind that vents through the cracks in the shell.

\subsubsection{Neutral gas}
\label{sec:neutral-gas}

\begin{figure}
  \centering
  \includegraphics[width=0.4\textwidth]{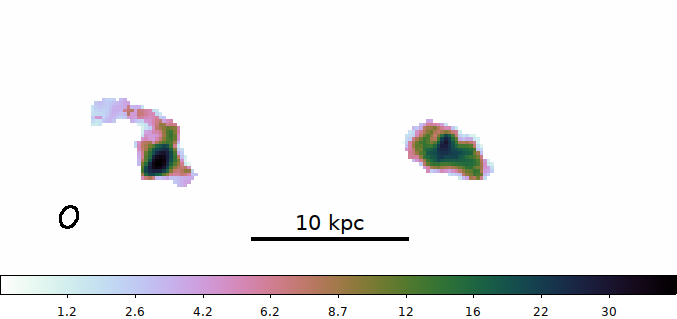} 

  \includegraphics[width=0.4\textwidth]{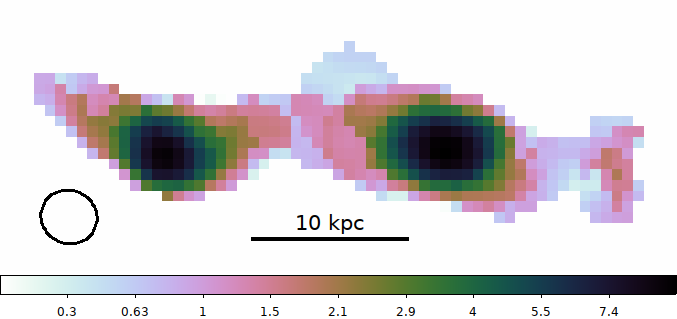} 
  \caption{Moment-0 maps from VLA B-configutration 21cm observations.  The
    colour mapping encodes $N_\mathrm{HI}$ in $10^{20}$cm$^{-2}$ and the beam for
    each dataset is indicated as a black ellipse. \emph{Top panel}: VLA-B
    configuration data. \emph{Bottm panel}: VLA-B configuration observations,
    tapered at 7k$\lambda$; this map has been used to draw the $N_\mathrm{HI}$
    contours in Fig.~\ref{fig:hI7kt}.}
  \label{fig:mom0all}
\end{figure}

\begin{figure*}
  \sidecaption
  \includegraphics[width=12cm,
  trim=2 5 3 30,clip=true]{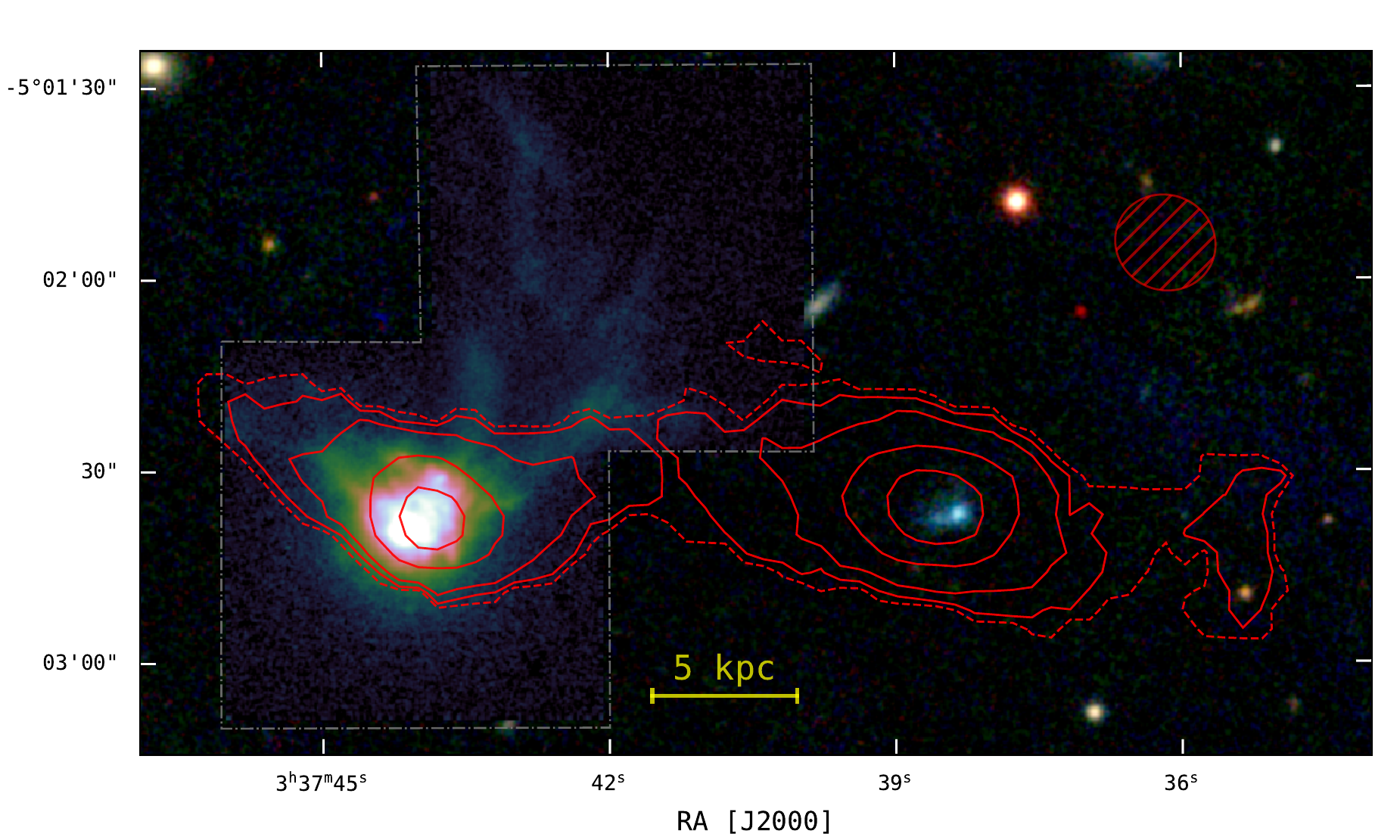}
  \caption{H$\alpha$ narrowband from MUSE (log-stretch from 0 to
    1.25$\times10^{-15}$\,erg\,s$^{-1}$cm$^{-2}$) inset into a grz colour
    composite (Pan-STARRS) with $N_\mathrm{HI}$ contours from VLA
    B-configuration data tapered at 7k$\lambda$; contours delinate
    $N_\text{HI}= \{0.5 (2.5\sigma), 1, 2, 5, 8 \} \times 10^{20}$\,cm$^{-2}$.
    The beam is 15.9\arcsec{}$\times$14.7\arcsec{} oriented at 66.4$^\circ$.
    and is indicated as a hatched ellipse.}
  \label{fig:hI7kt}
\end{figure*}

For our mapping of the signals from the neutral phase we processed the VLA
datacubes with the software SoFiA \cite[Source Finding
Application;][]{Serra2015}.  We used SoFiA's smooth+clip source detection
algorithm that is most commonly employed for searches of extragalactic HI
emission.  Here the radio datacube is smoothed with a set of top-hat kernels of
varying sizes and the final detected source mask is built from the union of the
binary masks that result from thresholding each individually smoothed cube.  We
used the recommended parameters for detection of extragalactic HI signal, i.e. a
set of 12 top-hat kernels of zero, three, seven, and 15 channels width and spatial
dimensions of 0$\times$0 (i.e. no spatial smoothing), $3\times3$, and
$6\times6$ pixels (1 pixel = 1 arcsec$^2$), and a detection threshold of 5.  The
SoFia runs provide us with moment-0 maps, $S^\nu$
[Jy\,km\,s$^{-1}$\,beam$^{-1}$], that are of interest here and moment-1 maps
that are used for the kinematic analysis in Sect.~\ref{sec:hikin}.  The moment-0
maps are converted into maps of neutral column, $N_\mathrm{HI}$ [cm$^{-2}$], via
\begin{equation}
  \label{eq:1}
  N_\mathrm{H} = 1.1 \times 10^{24} (1+z)^2 S^\nu / (ab) \; \text{,}
\end{equation}
where $a$ [\arcsec{}] and $b$ [\arcsec{}] are the FWHM of the major- and minor
axis of the synthesised beam, respectively, and $z$ is the redshift of the
galaxy \citep[Eq. 78 in][]{Meyer2017}.  The $N_\mathrm{HI}$ map of SBS\,0335-052E
from this procedure is displayed in Fig.~\ref{fig:mom0bc} and the top panel of
Fig.~\ref{fig:mom0all} shows the moment-0 map encompassing both galaxies
SBS\,0335-052E \& W.  The individual layers that contribute to the spatially
resolved \ion{H}{i} signal are assembled in Appendix \ref{sec:channel-maps} in
the form of channel maps.

Translating the detection threshold of the smooth-and-clip algorithm into an
$N_\mathrm{HI}$ sensitivity limit needs to account for the different smoothing
kernels that contributed in building up the 3D source mask.  We proceeded
empirically by inspecting the final binary-mask cube from SoFiA to determine
that the faintest pixels in the outskirts of the moment-0 map are mostly single
channel detections.  As stated in Sect.~\ref{sec:vla}, and verified by the noise
cube produced by SoFia, the typical rms-noise per channel in the B-configuration
data is $2.5 \times 10^{-4}$Jy\,beam$^{-1}$.  For a 10 km\,s$^{-1}$ wide-channel
this translates via Eq.~(\ref{eq:1}) into a 1-$\sigma$ column-density limit of
$\sigma_{N_\mathrm{HI}} = 1 \times 10^{20}$\,cm$^{-2}$.  We finally used the
contouring algorithm of ds9 \citep{Joye2003} without smoothing on the
$N_\mathrm{HI}$ map (Fig.~\ref{fig:mom0bc}) to produce contours at
$N_\mathrm{HI} = \{ 2.5 \, (2.5\sigma), 5, 10, 20, 30, 40\} \times
10^{20}$\,cm$^{-2}$.  The closing $2.5\sigma$ contour associated with
SBS\,0335-052E is shown as a white-dashed line only in Fig.~\ref{fig:hanb}; all
other contours are displayed in Fig.~\ref{fig:hanb}, Fig.~\ref{fig:ha_zoom}, and
Fig.~\ref{fig:clustident}.

From Fig.~\ref{fig:ha_zoom} it becomes apparent that the peak of $N_\mathrm{HI}$
is slightly offset to the west from the H$\alpha$ peak and from the
star-clusters.  The offset with respect to the star-clusters is made more
apparent in Fig.~\ref{fig:clustident}, where we overlay the $N_\mathrm{HI}$
contours onto the emission-line free HST/ACS F550M image.
Such offsets are observed frequently in dwarf starbursts and it is hypothesised
that that feedback from starbursting SSCs will mechanically disrupt and ionise
\ion{H}{I} in their vicinity
\citep[e.g.][]{vanEymeren2010,Cannon2016,Teich2016,Jaiswal2020}.

At lower column densities the resolved \ion{H}{I} structures appear to relate
spatially with H$\alpha$ morphological features.  Towards the north we find an
extended tail at $N_\mathrm{HI} = 10^{21}$cm$^{-2}$ that bends towards the east
at a slightly lower columns (Fig.~\ref{fig:hanb} and Fig.~\ref{fig:ha_zoom}).
The northern tail is almost co-spatial with a northern H$\alpha$ spur before it
curves around the H$\alpha$ spur towards the east.  At even lower neutral
columns, close to the detection limit of our observations
(2.5$\sigma = 2.5\times10^{20}$cm$^{-2}$; white dashed contour in
Fig.~\ref{fig:hanb}), we trace an extended tail that aligns with the
foreshortened filaments towards the east.  Towards the north-west and more
prominently towards the south-east, two ``ears'' emerge at the lowest column
density, both of which appear also co-spatial with spurs pointing in the same
direction.

Given the observed spatial correlations between HI and H$\alpha$ we suspect that
the low-column HI emission stems from cold gas that is interacting with the
ionised phase.  The entangled interactions between star-forming \ion{H}{ii}
regions and \ion{H}{i} phase are observationally well studied on small
($\lesssim 10^2$\,pc) scales in more nearby systems (e.g. \citealt{Egorov2014},
\citealt{Cannon2016}, \citealt{Egorov2018}, and review by
\citealt{Veilleux2020}).  These processes are complex, observationally
challenging to tackle, and far from being quantitatively fully understood. Here,
in the extreme environment of the starburst, we may now have resolved such
interactions on kiloparsec scales.

To study the interaction between the ionised and neutral phase on larger scales,
we use the VLA B-configuration data tapered at 7k$\lambda$
(Sect.~\ref{sec:vla}).  The $N_\mathrm{HI}$ moment-0 maps and the 1-$\sigma$
detection threshold, 2.2$\times10^{19}$\,cm$^{-2}$, were computed in the same
way as described above for the untapered dataset.  The tapered data reveal
\ion{H}{i} within and around the pair SBS\,0335-052E \& W.  We juxtapose the
moment-0 maps for the two different beams in Fig.~\ref{fig:mom0all}.  We compare the large-scale HI contours from tapered data with our MUSE H$\alpha$
image in Fig.~\ref{fig:hI7kt}.  There we use Pan-STARRS imaging
\citep{Chambers2016,Magnier2020} outside the MUSE FoV. In Pan-STARRS the fainter
companion SBS\,0335-052W is prominently visible as a blue cometary object
close to the position of the western $N_\mathrm{HI}$ peak.

The juxtaposition of the moment-0 maps in Fig.~\ref{fig:mom0all} illustrates how
different beams are sensitive to different scales of the neutral gas distribution
in and around the galaxies.  The untapered data resolves the morphology of
higher-column \ion{H}{i} within the galaxies and in their immediate vicinity.
The measured peak $N_\mathrm{HI}=4.1\times10^{21}$cm$^{-2}$ for SBS\,0335-052E
is only slightly lower than the inferred $N_\mathrm{H}^\mathrm{COS}$ from
Ly$\alpha$ absorption with HST/COS spectroscopy:
$N_\mathrm{HI}^\mathrm{COS} = (5.0 \pm 0.5) \times 10^{21}$cm$^{-2}$
\citep{James2014}.  However, this does not indicate that the gas seen by COS in
absorption is spatially resolved with the VLA.  Especially the spatial offset
between the $N_\mathrm{HI}$ peak and the star-clusters discussed above
(cf. Fig.~\ref{fig:clustident}) precludes such a conclusion.  This rather
indicates substantial substructure below our resolution limit in the neutral
phase.  The larger beam of the 7k$\lambda$-tapered data smooths out almost all
resolved features in the untapered data (Fig.~\ref{fig:mom0all}, central panel).
This is evidenced by the peak column ($9.5 \times 10^{20}$cm$^{-2}$) being only
slightly higher than the average column ($9\times 10^{20}$cm$^{-2}$) in the
untapered map.  However, the tapered data reveals the diffuse lower-column gas
that is spread over larger areas in the outskirts of the system.  Here this gas
forms a bridge between the galaxy pair and also possible tidal
features towards the east and west. The $N_\mathrm{HI}$ morphology from this
reduction appears broadly consistent with the significant detected \ion{H}{i} in
the VLA C- and D-configuration data presented in \cite{Pustilnik2001}.
We find a similar level of agreement between our maps and the GMRT
21-cm observations presented by \cite{Ekta2009}.  As described both by
\cite{Ekta2009} and \cite{Pustilnik2001} in detail, the morphological features
(and also the kinematics) of the \ion{H}{I} halo gas on larger scales strongly
indicate an early stage of a merger between both galaxies.  The onset of this
merger is hypothesised as having triggered the current star-formation episodes
in both galaxies.

Comparing the morphology of the \ion{H}{I} envelope and the extended H$\alpha$
emission, it is evident that the filaments extend into regions that are
characterised by lower neutral column densities or, at greater radial distance, are even
devoid of detected \ion{H}{I};  This is especially so for the eastern filament.
However, the western filament appears to overlap more with the low-column gas of
the bridge.  Especially in Fig.~\ref{fig:hI7kt} it appears as if the
extended H$\alpha$ structure starts to interact with the extended \ion{H}{I}
halo of the western galaxy.  Interestingly, this boundary coincides with the
fanning out of the western filament (Fig.~\ref{fig:cont}).  The
elongation of the \ion{H}{I} halo surrounding the eastern galaxy is
oriented almost perpendicular to the direction of the filaments.

\subsection{Emission line ratios along the filaments}
\label{sec:ratio}

\begin{figure}
  \centering
  \includegraphics[width=0.49\textwidth]{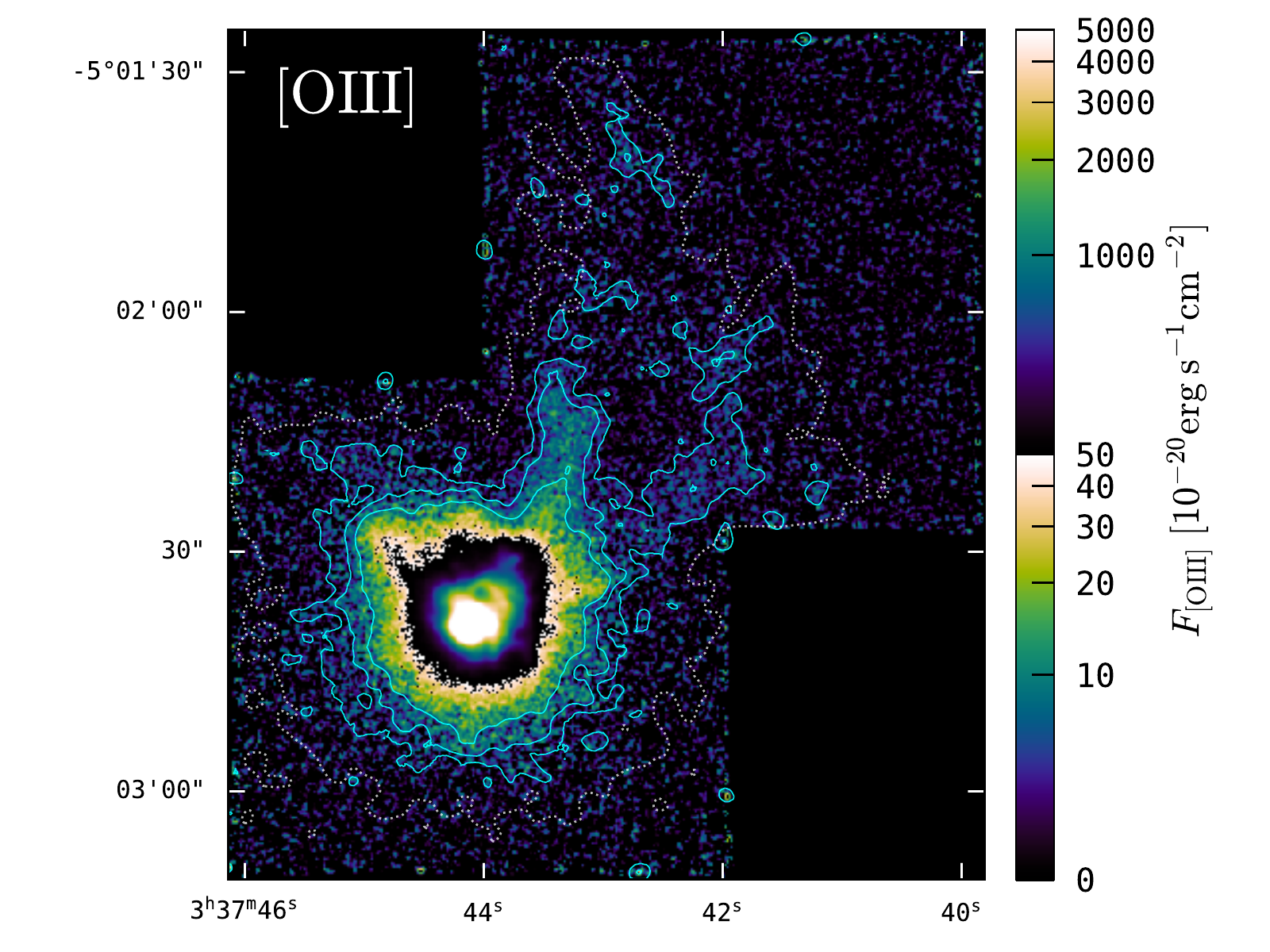} 

  \includegraphics[width=0.49\textwidth]{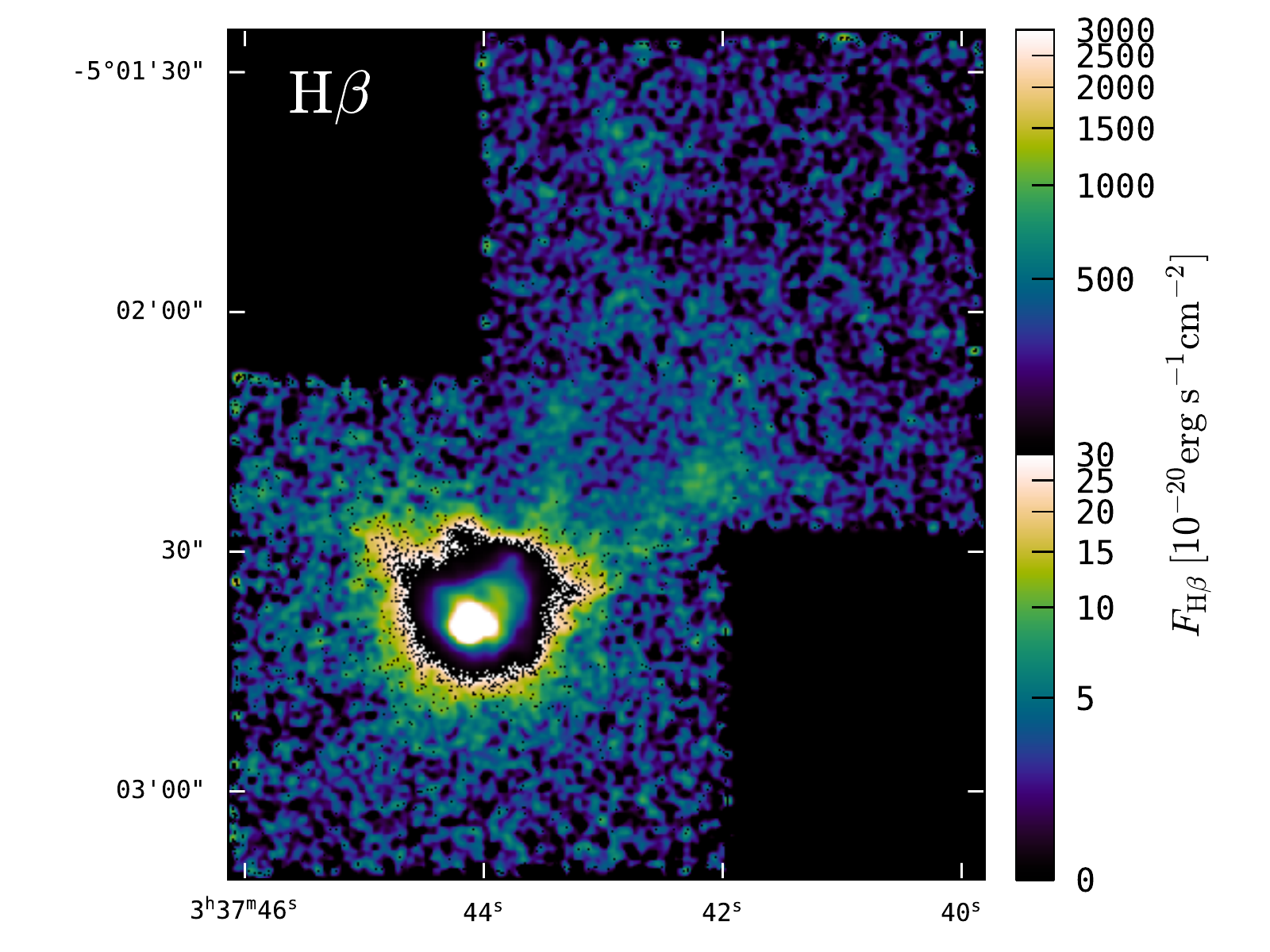} %\vspace{0.4em}
  \caption{\emph{Top panel}: Continuum subtracted [\ion{O}{iii}] $\lambda5007$
    narrowband; contours indicate
    $\mathrm{SB}_\mathrm{[OIII]}=\{5, 12.5, 25\}\times
    10^{-19}$\,erg\,s$^{-1}$cm$^{-2}$arcsec$^{-2}$. The cyclic colour map
    encodes flux densities from 0 to $5 \times 10^{-19}$\,erg\,s$^{-1}$cm$^{-2}$
    (colour bar) to $5\times10^{-19}$\,erg\,s$^{-1}$cm$^{-2}$.  The image has
    been smoothed with a Gaussian of 0.4\arcsec{} FWHM. \emph{Bottom panel}:
    Continuum subtracted H$\beta$ narrowband.  The cyclic colour map encodes
    flux from 0 to $3\times10^{-19}$\,erg\,s$^{-1}$cm$^{-2}$ to
    $3\times10^{-17}$\,erg\,s$^{-1}$cm$^{-2}$.  The image has been smoothed with
    a Gaussian of 0.95\arcsec{} FWHM to enhance the visibility of the low-SB
    H$\beta$ emission.  }
  \label{fig:oiiihb}
\end{figure}

\begin{figure*}
  \centering
  \includegraphics[width=0.43\textwidth,trim=0 40 0 0, clip=true]{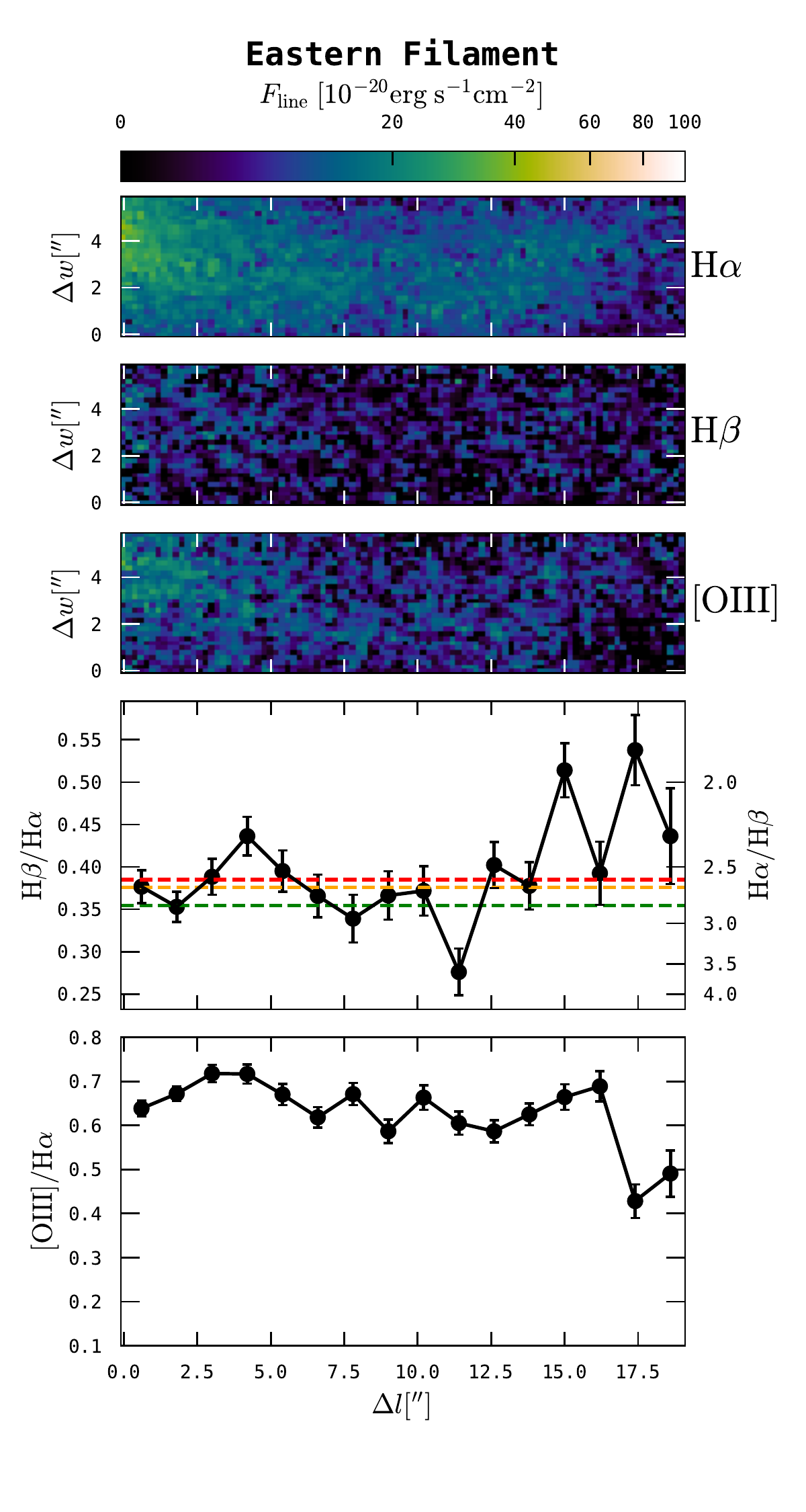}
  \includegraphics[width=0.537\textwidth,trim=0 40 0 0,clip=true]{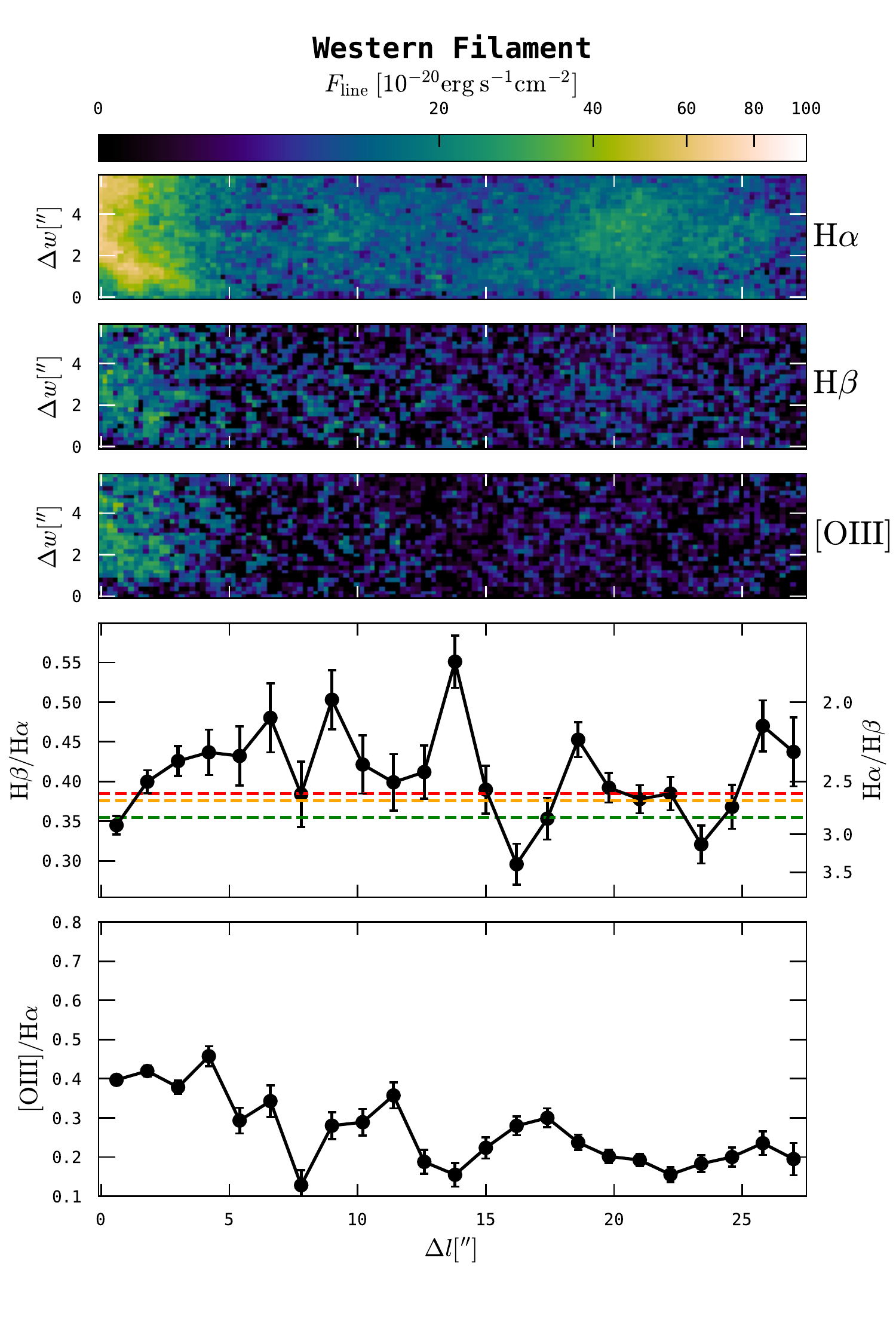}
  \caption{Analysis of emission line ratios along the eastern (\emph{left}) and
    western (\emph{right}) filament.  In the top three panels the emission in
    H$\alpha$, H$\beta$, and [\ion{O}{iii}] $\lambda$5007 is shown, now aligned
    such that the x-axis runs along the direction of the filament ($\Delta l$)
    and that the y-axis runs accros the filament ($\Delta w$); the correpsonding
    areas have also been outlined in the bottom left panel of
    Fig.~\ref{fig:hanb}.  The two bottom panels show the H$\beta$/H$\alpha$ and
    [\ion{O}{iii}]/H$\alpha$ ratio in bins of 1.2\arcsec{} along $\Delta
    l$. These ratios are obtained after integrating the emission in each line
    along $\Delta w$.  The 1-$\sigma$ errors on the ratios have been computed
    with Eq.~(\ref{eq:2}). For H$\beta$/H$\alpha$ the \mbox{Case-A}
    recombination expectations from \citet{Storey1995}, are indicated as dashed
    lines for three different temperatures ($T = \{1, 2, 3\}\times10^4$\,K in
    green, orange, and red); cf. Fig.~\ref{fig:hbha}.  For
    H$\beta$/H$\alpha$ the corresponding conventional Balmer decrement values
    are measured on the right ordinate.}
  \label{fig:ratio}
\end{figure*}

\begin{figure}
  \centering
  \includegraphics[width=0.4\textwidth,trim=20 10 10 10,clip=True]{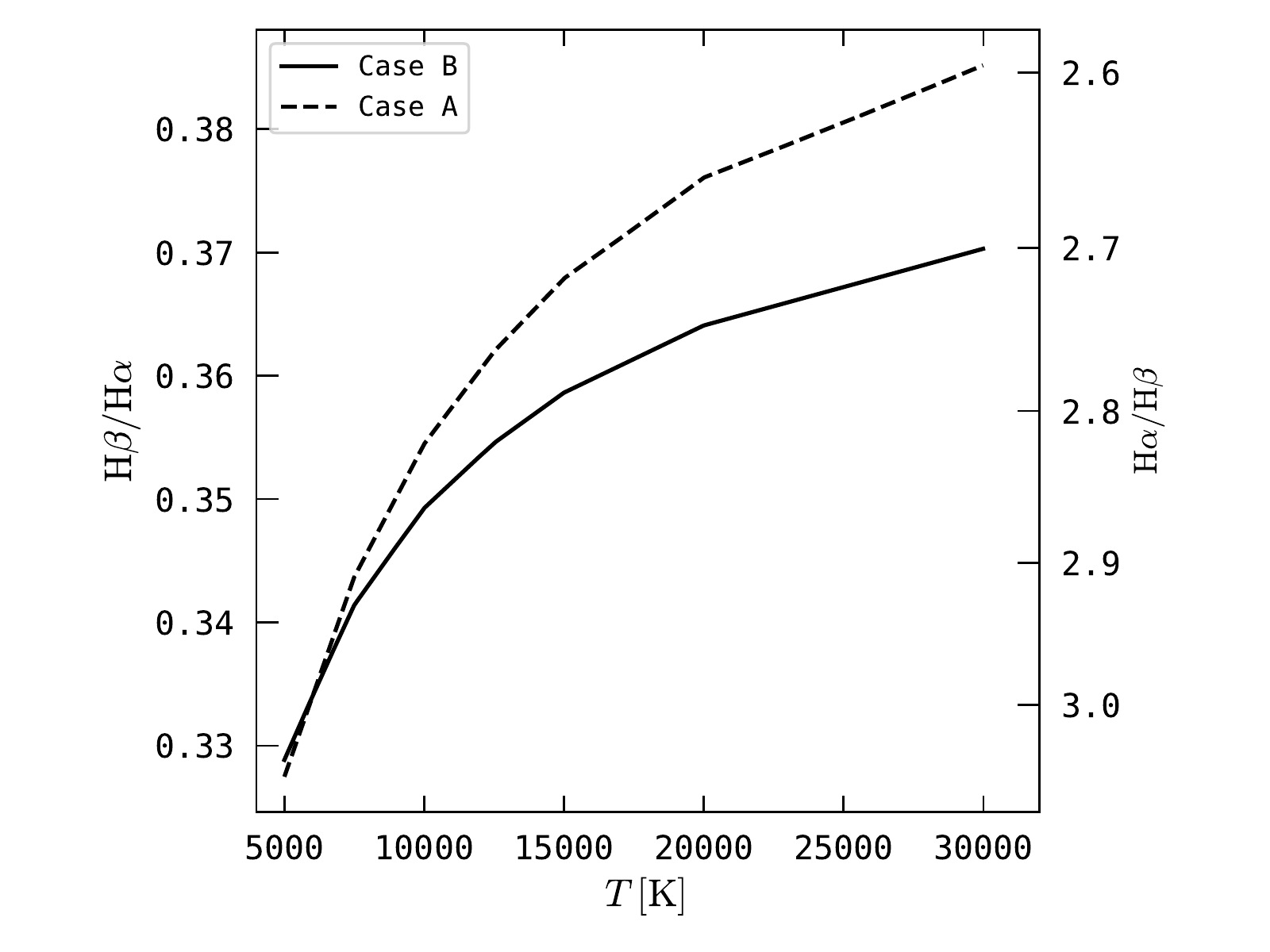}
  \caption{Inverse Balmer decrement values for Case-A (dashed line) and Case-B
    (solid line) recombination as a function of temperature as calculated by
    \cite{Storey1995}.  The right ordinate measures the conventional Balmer
    decrement values.}
  \label{fig:hbha}
\end{figure}

Our detections of H$\beta$ and [\ion{O}{iii}] $\lambda 5007$, in concert with
H$\alpha$, enable us to make limited inferences on the physical
state of the emitting plasma in the filaments.  We show in Fig.~\ref{fig:oiiihb}
the continuum-subtracted narrowband images in [\ion{O}{iii}] $\lambda5007$ and
H$\beta$.  The images were synthesised by summing over the datacube layers at
$4927.46 \pm 1.25$\AA{} and $5074.96\pm1.25$\AA{} observed frame for H$\beta$
and [\ion{O}{iii}], respectively.  These narrow windows are adequate, since the
line emission in the filaments is barely resolved (see kinematic analysis in
Sect.~\ref{sec:hiikin}).

We analyse the emission line ratios H$\beta$/H$\alpha$ (Sect.~\ref{sec:ibdc})
and [\ion{O}{iii}]/H$\alpha$ (Sect.~\ref{sec:oiiiha}) within two rectangular
regions that cover each filament of an area where all three lines are detected;
these slits are indicated in the bottom left panel of Fig.~\ref{fig:hanb}.  Both
slits cover 6\arcsec{} across each filament, and the length of the eastern and
western slits are 27.6\arcsec{} and 19.2\arcsec{}, respectively.  The flux
distribution of H$\alpha$, H$\beta$, and [\ion{O}{iii}] along ($\Delta l$) and
across ($\Delta w$) the filaments is shown in the top three panels of
Fig.~\ref{fig:ratio}.  These flux distributions were extracted directly from the
synthesised narrowband images.  Corresponding variance maps in all lines were
created by summing the respective layers in the variance datacube.  We measured
the individual emission line fluxes by summing the emission across the filaments
in bins of 1.2\arcsec{} along the direction of the slit.  We then calculate the
ratios for each bin after correcting the individual measurements for galactic
foreground emission using the \cite{Cardelli1989} extinction curve for a given
$A_V = 0.123$ from the Galactic dust map of \cite{Schlafly2011}.

\subsubsection{Inverse Balmer decrement H$\beta$/H$\alpha$}
\label{sec:ibdc}

Traditionally the Balmer decrement, H$\alpha$/H$\beta$, is used to measure
interstellar extinction.  We here use the inverse of the Balmer decrement since
the statistics for the ratios of measurements for the low surface-brightness
emission close to the detection limit are only well defined if the denominator
is chosen to have the smaller error-bar.  Then the standard 1-$\sigma$ error on
the ratio, $\Delta (\mathrm{H\beta}/\mathrm{H\alpha})$, can be calculated via
\begin{equation}
  \label{eq:2}
  \Delta (\mathrm{H}\beta/\mathrm{H}\alpha) =
  \frac{\sqrt{\mathrm{Var}(\mathrm{H}\alpha) + \mathrm{Var}(\mathrm{H}\beta) (\mathrm{H}\beta/\mathrm{H}\alpha)^2
      \,}}{\mathrm{H}\beta} \;\text{,}
\end{equation}
where $\mathrm{Var}(H\alpha)$ and $\mathrm{Var}(H\beta)$ are the
variances on the extracted fluxes H$\alpha$ and H$\beta$,
respectively.  Equation~(\ref{eq:2}) is only valid as long as
$\sqrt{\mathrm{Var}(\mathrm{H}\beta)} \lesssim 0.25 \,
\mathrm{H}\beta$ \citep[e.g.][]{Dunlap1986}, which is the case here.

To aid the interpretation of this non-standard ratio, we plot in
Fig.~\ref{fig:hbha} its behaviour as a function of temperature for a recombining
plasma under Case-A and Case-B recombination scenarios. Both in
Fig.~\ref{fig:ratio} and Fig.~\ref{fig:hbha} we also include ticks on the right
ordinate that allow reading off the conventional Balmer decrement.

From an integrated spectrum covering the star-forming sites in the galaxy
\citep[depicted in Fig.~3 in][]{Wofford2021} we find an inverse Balmer decrement
of 0.34 (or H$\alpha$/H$\beta = 2.92$).  As can be seen in Fig.~\ref{fig:ratio},
both the eastern and the western filament are predominantly characterised by
higher H$\beta$/H$\alpha$ ratios (lower Balmer decrements),
that is H$\beta$/H$\alpha \gtrsim 0.35$ (H$\alpha$/H$\beta \lesssim 2.86$).
Interesting features in the radial ratio profiles are the dips before and after
the H$\alpha$ ``hot-spot'' in the western filament, for which the higher ratios
appear closer to the galaxy; in contrast, the eastern filament shows the highest values at
its end.  The mean (median) of H$\beta$/H$\alpha$ along the eastern and western
filament are 0.4 (0.38) and 0.41 (0.39), respectively.

Some of the here observed (high) low (inverse) Balmer decrements in the
filaments are not compatible with standard Case-B or even Case-A recombination
values (Fig.~\ref{fig:hbha}).  The ansatz of the Case-B recombination scenario
is infinite optical depth in all Lyman series lines
($\tau_{\mathrm{Ly}n} = \infty$), but zero optical depth in all other
transitions, whereas for Case-A also $\tau_{\mathrm{Ly}n} = 0$.  Case-B delivers
realistic values for Hydrogen line ratios in the interstellar medium, since the
absorption cross section for Lyman series photons is very high.  Both in Case-B
and in Case-A H$\beta$/H$\alpha$ increases with increasing temperature, but the
increase is more rapid for Case-A.  However, even for $T\approx 3 \times 10^4$K,
the highest temperature provided in the calculations by \cite{Storey1995}, the
resulting H$\beta$/H$\alpha = 0.385$ for Case-A (red dashed line in
Fig.~\ref{fig:ratio}) and H$\beta$/H$\alpha = 0.370$ for Case-B are well below
the measured ratios,
$0.4 \lesssim\mathrm{H}\beta/\mathrm{H}\alpha \lesssim 0.5$, at the beginning
(end) of the western (eastern) filament.

Since the low-surface brightness filaments are observed in the low-density
outskirts of the galaxy it is qualitatively conceivable that the filaments
originate in gas where neither the $\tau_{\mathrm{Ly}n} = \infty$ nor the
$\tau_{\mathrm{Ly}n} = 0$ approximation can be justified.  Radiative transfer
effects thus may alter the population levels, and neither the Case-A or Case-B
approximation are reasonable anymore.  Calculations for this intermediate
$\tau_{\mathrm{Ly}n} \rightarrow 0$ regime were performed by
\cite{Capriotti1966} and \cite{Cox1969}; cf. Chapter 4.5 of
\cite{Osterbrock2006}.  Especially at low optical depths in Ly$\alpha$
($\tau_{\mathrm{Ly}\alpha} < 10^3$) Balmer decrements compatible with the range
of the here measured ratios in the filaments,
$2 < \mathrm{H}\alpha/\mathrm{H}\beta < 2.5$
($0.4 \leq \mathrm{H}\beta/\mathrm{H}\alpha \leq 0.5$), are possible (Fig.~4.3
in \citealt{Osterbrock2006}).  Deviations from equilibrium occur in media that
are ionised of fast-radiative shocks, but model calculations in low-metallicity
environments favour higher Balmer decrements $\approx 3$ \citep[][]{Alarie2019}.

\subsubsection{[\ion{O}{iii}]/H$\alpha$}
\label{sec:oiiiha}

The line ratio [\ion{O}{iii}]/H$\beta$ is typically used to map the excitation
of the plasma.  Given the limited coverage of detected H$\beta$ emission we here
substitute H$\beta$ with H$\alpha$ in the denominator.  The standard error on
[\ion{O}{iii}]/H$\alpha$, $\Delta (\mathrm{[\ion{O}{iii}]/H}\alpha)$, can then
be calculated in an analogous way to Eq.~(\ref{eq:2}).  In principle,
[\ion{O}{iii}]/H$\alpha$ must be understood as a lower bound to the excitation,
since extinction lowers the ratio.  This needs to be kept in mind, since the
unusual low Balmer decrements discovered in the previous section preclude us from 
making definitive claims on the presence of dust within the filaments.

From the integrated spectrum covering the central star-forming sites \citep[see
Fig.~3 in][]{Wofford2021} we measure [\ion{O}{iii}]/H$\alpha = 1.13$.  For both
filaments the ratio is significantly lower, with an mean (median) of 0.63 (0.65)
and 0.26 (0.23) for the eastern and western filament, respectively.  Such low
[\ion{O}{iii}]/H$\alpha$ ratios have been reported recently in the outskirts of
the blue compact dwarf galaxy Haro 14 \citep{Cairos2022}.  From
Fig.~\ref{fig:oiiihb} it could be already anticipated that the western filament
is significantly dimmer in [\ion{O}{iii}] (see also Fig.~1 in
\citetalias{Herenz2017b}).  The ratio in the eastern filament is relatively
flat, and only at the end ($\Delta l \gtrsim 17$\arcsec{}) of it does the [\ion{O}{iii}] flux
decrease more rapidly than H$\alpha$.  For the western filament the ratio
appears to show an overall decrease with increasing distance from the galaxy,
albeit with significant fluctuations.  There appears to be no strong correlation
between trends seen in H$\beta$/H$\alpha$ and [\ion{O}{iii}]/H$\alpha$.

Interpreting the [\ion{O}{iii}]/H$\alpha$ ratio is complicated, as it is
regulated by multiple parameters that characterise the physical conditions of
the emitting plasma.  At low-densities ($n \lesssim 1000$\,cm$^{-3}$) the
emissivity of [\ion{O}{iii}] $\lambda5007$ depends only on $T_e$
\citep[][]{Luridiana2015,Morisset2020}, but the ratio of O$^{2+}$ to H$^+$ ions
is influenced both by the oxygen abundance and the ionisation mechanism.  More
specifically, if O$^{2+}$/H$^+$ is held fixed then the important parameters
regulating [\ion{O}{iii}]/H$\alpha$ are the hardness of the ionising radiation
field and the ionisation parameter (ratio between density of ionising photons to
density of H atoms) for photo-ionisation, whereas in fast radiative shocks the
shock-velocity, density, and the strength of the magnetic field are influential
\citep[e.g.][]{Alarie2019}.  We recall that the western filament appears to
interact with the low neutral column gas that belongs to the tidal bridge that
connects the western with the eastern halo, whereas the eastern filament appears
to protrude more freely out of the neutral envelope (Sect.~\ref{sec:neutral-gas}
and Fig.~\ref{fig:hI7kt}).  Thus, a plausible interaction with the more pristine
halo gas in the western filament might lower the oxygen abundance and thereby
also attenuate the [\ion{O}{III}] line.  On the other hand, this interaction
could also lead to a higher neutral fraction within the western filament that in
turn lowers the ionisation parameter and thereby also attenuates [\ion{O}{III}]
line.

\subsection{Kinematic analysis}
\label{sec:ionh-ionhi-kinem}

\begin{figure*}
  \sidecaption
  \includegraphics[width=12cm]{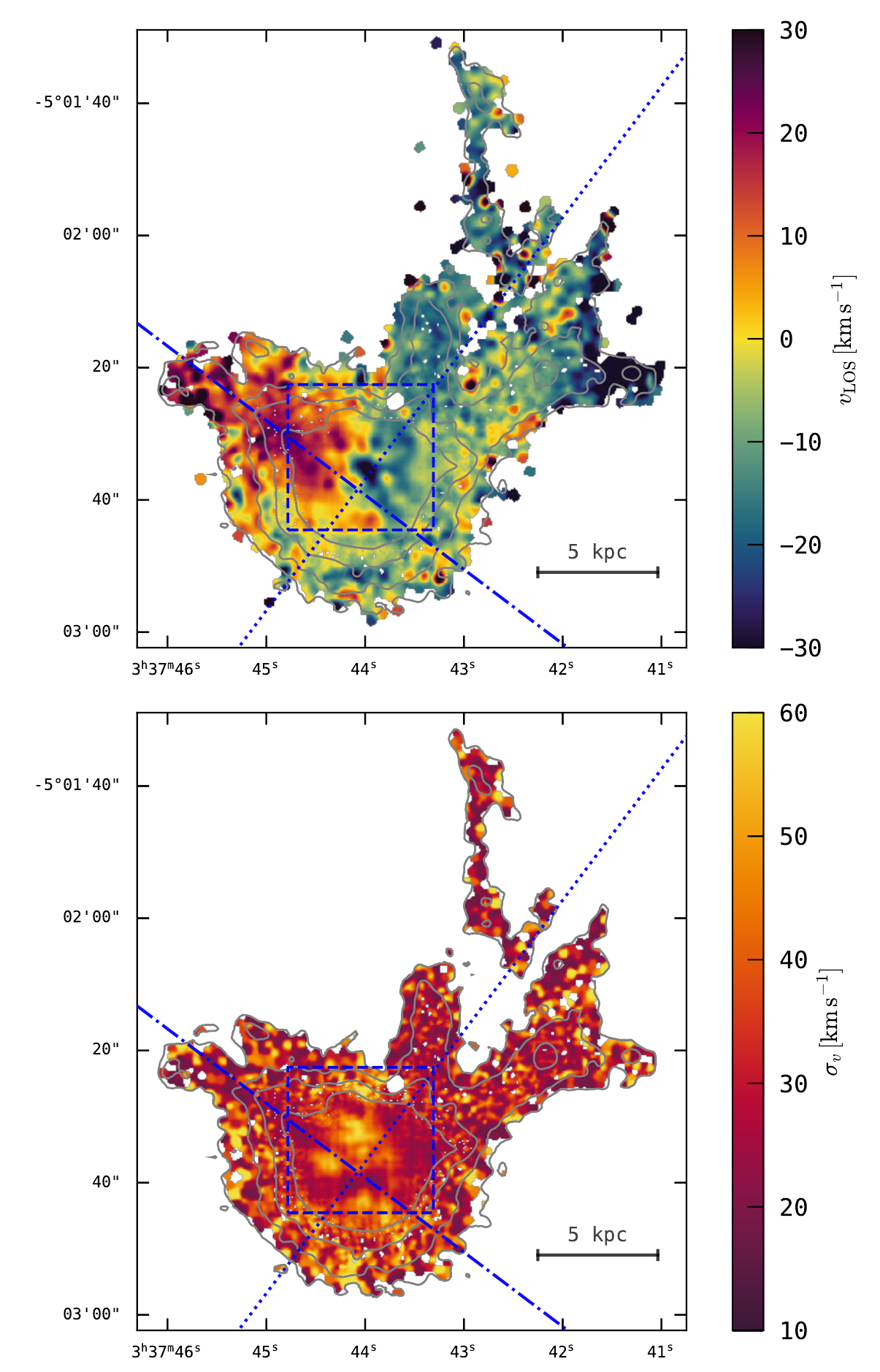}
  \caption{Line-of-sight velocity map (\emph{top}) and velocity dispersion map
    (\emph{bottom}) of the ionised gas; subdued contours mark various H$\alpha$
    surface brightness levels as in Fig.~\ref{fig:hanb}.  As zero-point velocity
    $v = c\cdot z = 4053$\,km\,s$^{-1}$ \citep{Moiseev2010} was adopted.  The
    kinematic major and minor axis after \cite{Moiseev2010} are indicated with
    a dash-dotted and dotted line, respectively.  The viewport of this figure is
    1\arcmin{}35\arcsec{}$\times$1\arcmin{}46\arcsec{} as in
    Fig.~\ref{fig:hanb}. The blue dashed square indicates the
    22\arcsec{}$\times$22\arcsec{} viewport centred on the central
    circumgalactic medium (Fig.~\ref{fig:cont}) shown in the left panel of
    Fig.~\ref{fig:ha_zoom}. }
  \label{fig:kin}
\end{figure*}

\begin{figure*}
  \sidecaption
  \includegraphics[width=12cm]{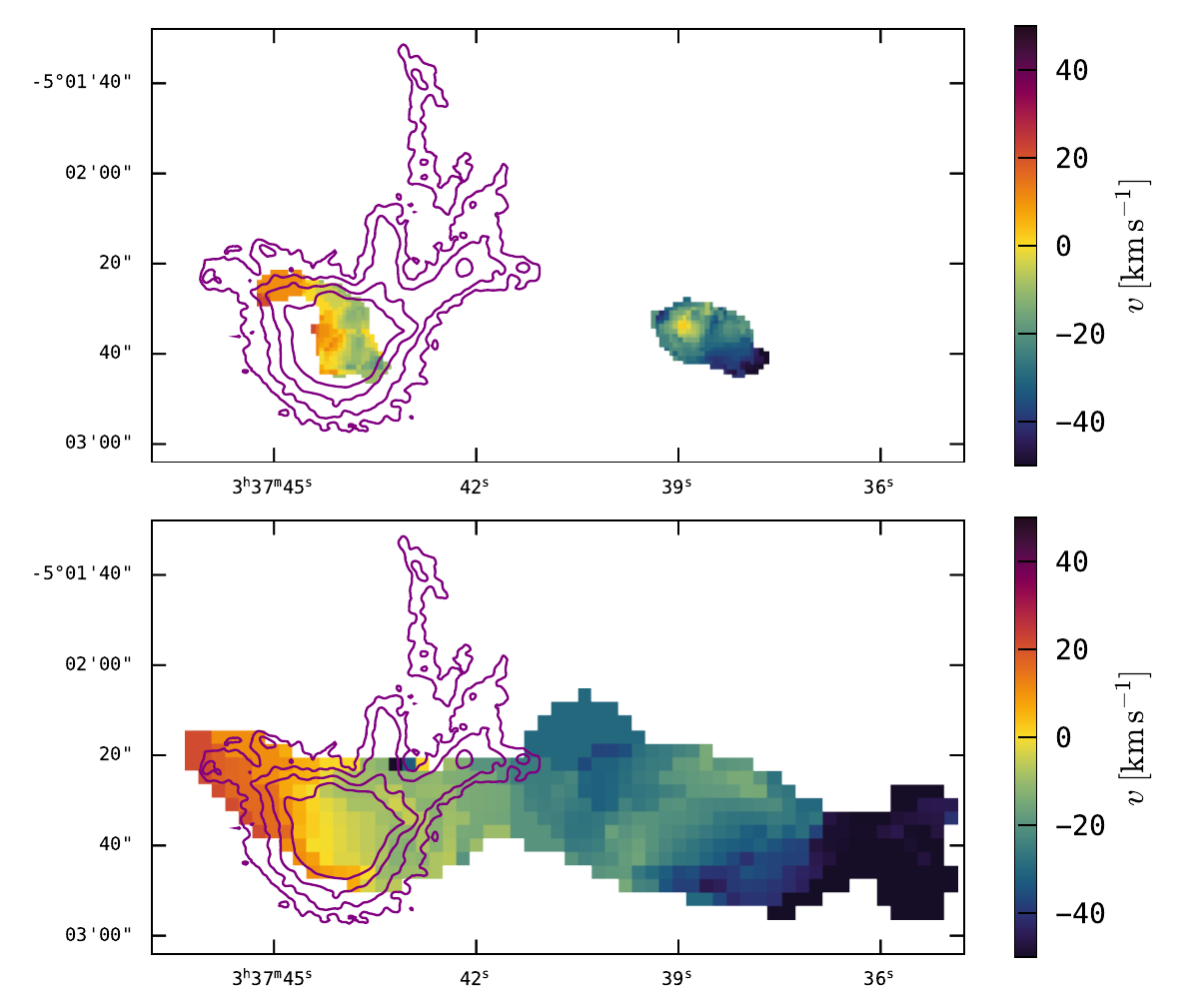}
  \caption{Moment-1 map from VLA-B configuration 21\,cm data showing the
    intensity weighted line-of-sight velocity field of \ion{H}{I} (\emph{top}:
    untapered data; \emph{bottom}: 7k$\lambda$ tapered data).  The purple
    contours mark various H$\alpha$ surface brightness levels as in
    Fig.~\ref{fig:hanb}.  For the zero-point velocity the optical recession
    velocity from the eastern galaxy, $v_0 = c \cdot z = 4053$\,km\,s$^{-1}$
    \citep{Moiseev2010}, was used.}
  \label{fig:mom1}
\end{figure*}

\begin{figure}
  \includegraphics[width=0.5\textwidth,trim=0 20 0
  30,clip=True]{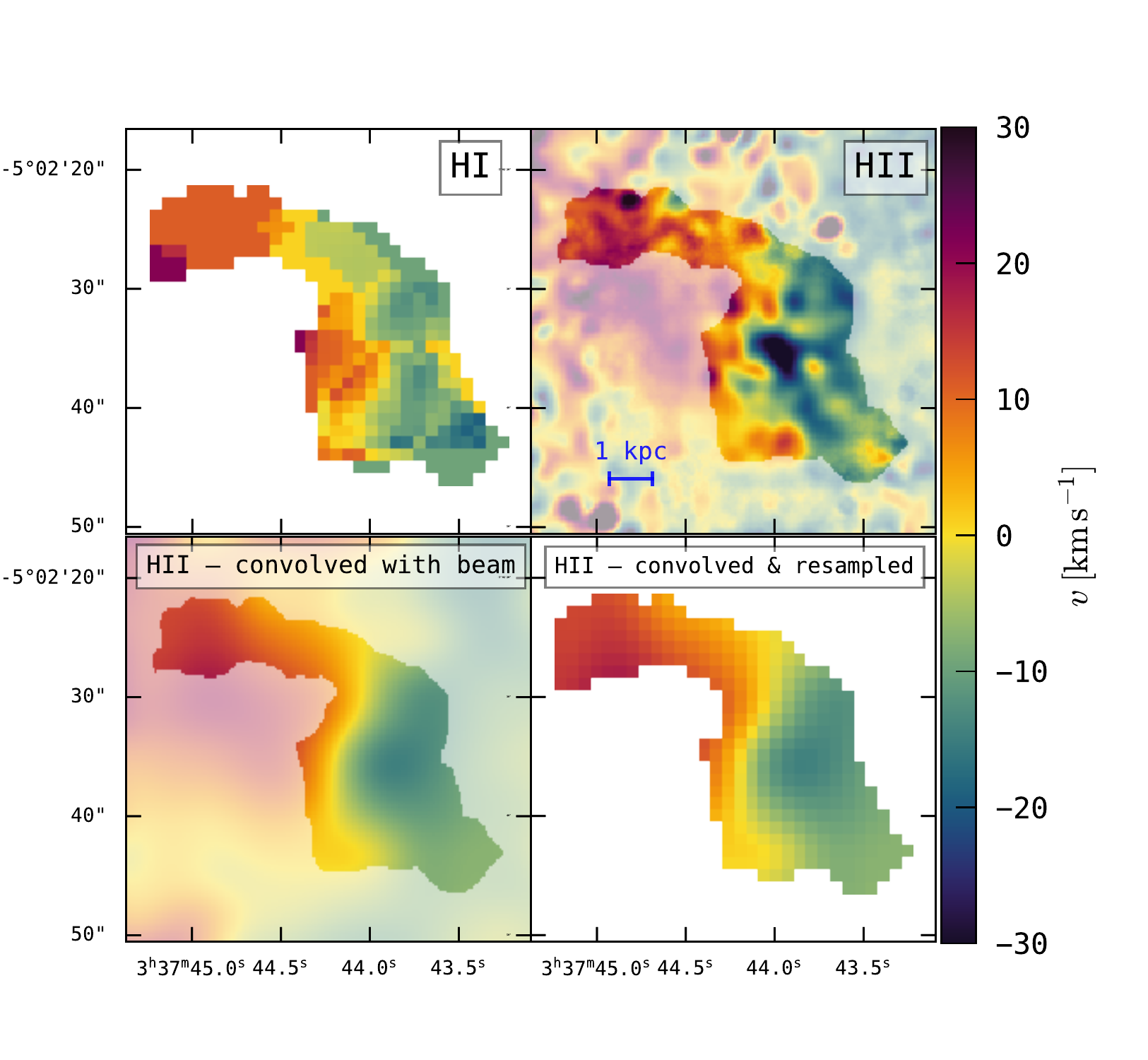}%\vspace{0.2em}
  \caption{Comparison between \ion{H}{I} and \ion{H}{II} kinematics on small
    scales.  The \ion{H}{I} velocity field (\emph{top left panel}) is a
    zoomed-in view of the velocity field shown in the top panel
    Fig.~\ref{fig:mom1}.  The \ion{H}{II} velocity field (\emph{top right
      panel}) is the same as in Fig.~\ref{fig:kin}, but regions outside of
    detected \ion{H}{I} have been subdued.  In the \emph{bottom left panel} we
    show the \ion{H}{II} velocity field convolved with a 2D Gaussian
    (6.1\arcsec{}$\times$4.6\arcsec{}, position angle -23.6$^\circ$) to mimic
    the VLA B-configuration observations beam smearing (see text). The
    \emph{bottom right panel} shows the smoothed H$\alpha$ velocity field
    resampled onto the native grid of the VLA B-configuration cube. }
  \label{fig:kinc}
\end{figure}

\begin{figure}
  \centering \includegraphics[width=0.5\textwidth,trim=30 80 0
  50,clip=True]{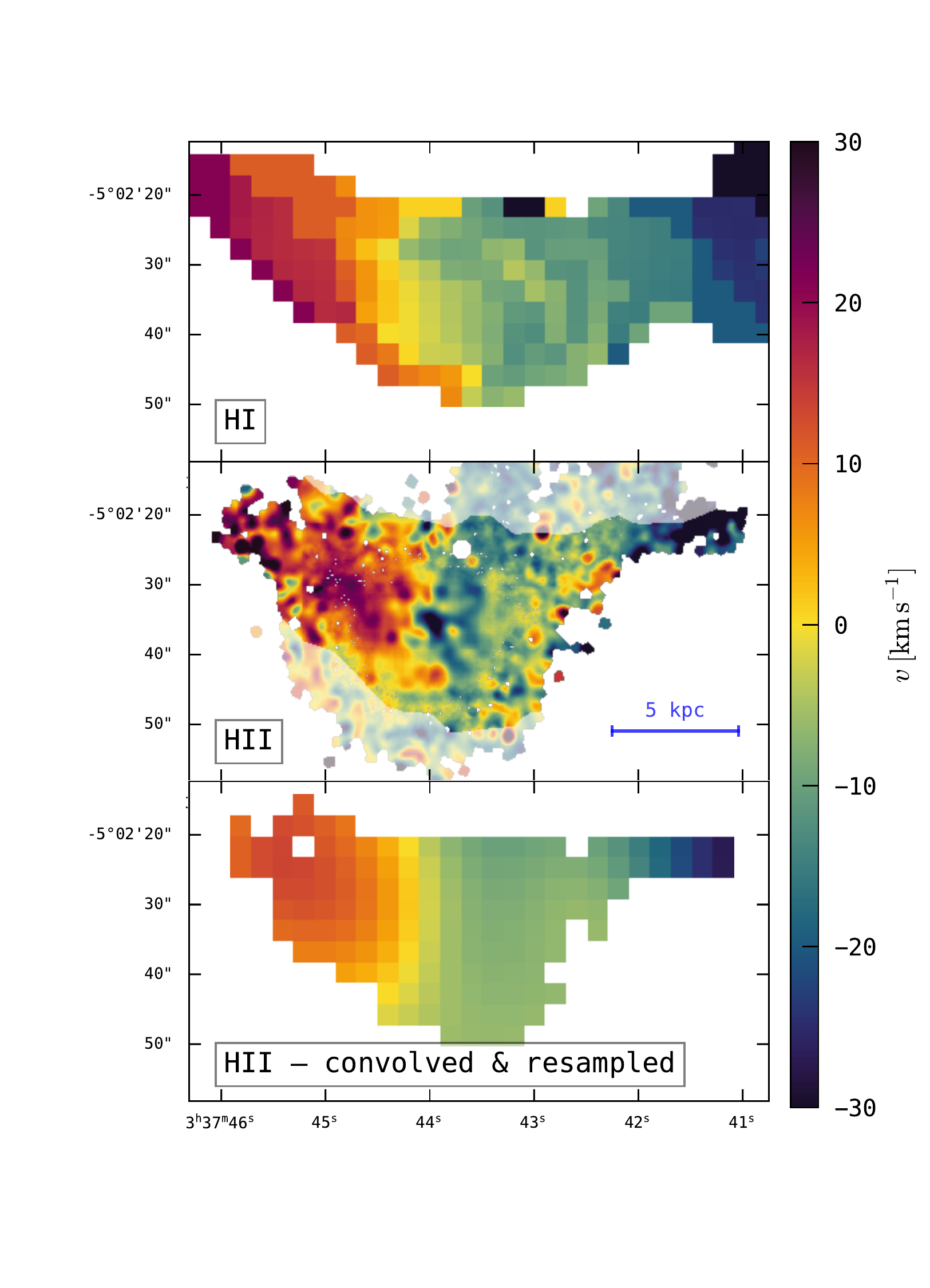}%\vspace{0.45em}
  \caption{Comparison between \ion{H}{I} and \ion{H}{II} kinematics on larger
    scales; we here display a zoomed-in view of the 7k$\lambda$ tapered 21cm
    data shown in the bottom panel of Fig.~\ref{fig:mom1}.  The \emph{top panel}
    shows the first moment from the \ion{H}{I} data, the \emph{middle panel}
    shows the ionised gas velocity field, with regions outside of \ion{H}{I}
    detections being subdued, and the \emph{bottom panel} shows the \ion{H}{II}
    velocity field convolved with a 2D Gaussian
    (15.9\arcsec{}$\times$14.7\arcsec{}, position angle 66.4$^\circ$) to mimic
    the beam smearing and resampled onto the native grid of the \ion{H}{I}
    observations.}
  \label{fig:m17kc}
\end{figure}

\begin{figure*}[t!]
  \centering
  \includegraphics[width=\textwidth]{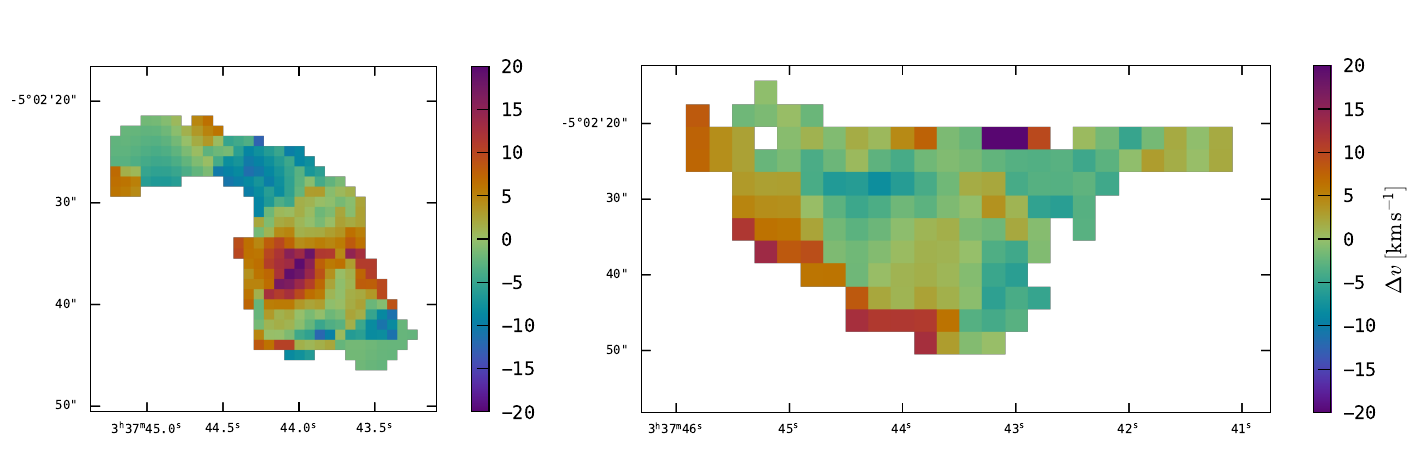}
  \caption{Maps showing
    $\Delta v = v_\mathrm{HI} - v_\mathrm{HII}^\mathrm{rep}$, where
    $v_\mathrm{HI}$ are the \ion{H}{I} line-of-sight velocities and
    $v_\mathrm{HII}^\mathrm{rep}$ are the resolution matched and re-projected
    \ion{H}{II} line-of-sight velocities.  The \emph{left panel} shows the
    comparison with the untapered VLA-B configuration data, i.e. the subtraction
    of the velocity field in the lower-right panel of Fig.~\ref{fig:kinc} from
    the upper-left panel of Fig.~\ref{fig:kinc}, while the \emph{right panel}
    shows the comparison for the 7k$\lambda$-tapered data, i.e. the subtraction
    of the velocity field in the lower panel of Fig.~\ref{fig:m17kc} from the
    upper panel of Fig.~\ref{fig:m17kc}. }
  \label{fig:diffm}
\end{figure*}

\subsubsection{\ion{H}{ii} kinematics}
\label{sec:hiikin}

In Fig.~\ref{fig:kin} we display the line-of-sight velocity field
($v_\mathrm{los}$, top panel) and the velocity dispersion ($\sigma_v$, bottom
panel) of the line emitting plasma.  We created these maps from fits of Gaussian
profiles to H$\alpha$, H$\beta$, and [\ion{O}{iii}] $\lambda\lambda$4959,5007 in
a Voronoi tessellated\footnote{The Voronoi tessellation was computed with the
  algorithm of \cite{Diehl2006}, which is an extension of the
  \cite{Cappellari2003} binning scheme.} continuum-subtracted datacube; in the
outer filaments only H$\alpha$ is contributing to the kinematic signal.  By
experimenting we found that a $S/N$ threshold of 8.5 and a maximum bin-size of 4
square arcseconds provided useful results, that is to say those parameters preserve the
morphological appearance of the filaments while also providing a large number of
high-SN bins throughout the low-SB regions; all bins in the high-SB region
consist of single spaxels.  Despite the morphological complexity, the emission
lines are all well modelled by a single Gaussian.  We show in
Appendix~\ref{sec:emiss-line-prof} the line profiles and model along the main
filaments.  Only for a few spaxels in the central regions a weaker secondary
component could be added to better fit the observed skew in some profiles.
As zero-point velocity we set $v_0 = c \cdot z = 4053$\,km\,s$^{-1}$ and
$\sigma_v$ is corrected for the instrumental line spread function using the
prescription provided in the MUSE user manual.

Low S/N $\sigma_v$ values obtained from Gaussian fits are known to be biased
high, since increasing the noise relative to the line flux artificially broadens
the fits.  Therefore we show in the bottom panel of Fig.~\ref{fig:kin} only
$\sigma_v$'s where
$\mathrm{SB}_\mathrm{H\alpha} > 1.5 \times
10^{-18}$erg\,s$^{-1}$cm$^{-2}$arcsec$^{-2}$.  Moreover, we only include bins in
the $v_\mathrm{LOS}$ map where the measurement uncertainty
$\Delta v < 15$\,km\,s$^{-1}$.  In the low-SB filaments typical values for the
1-$\sigma$ uncertainties on velocity and velocity dispersion are
$\Delta v \sim 10$\,km\,s$^{-1}$ and $\Delta \sigma_v \sim 20$\,km\,s$^{-1}$,
respectively.  Further, it is known for MUSE data, that velocity fields derived
from narrow emission lines exhibit a characteristic striping pattern on small
scales \citep[see Sect. 3.4 in][]{Weilbacher2015}; this is due to the LSF being
undersampled by the MUSE spectrograph.  For the display in Fig.~\ref{fig:kin} we
removed the most prominent effects of this striping pattern by slightly
degrading the spatial resolution of the velocity fields.
This is achieved by smoothing the computed $v_\mathrm{LOS}$ map with a circular
top-hat kernel of 0.4\arcsec{} (2 pixels) radius.

The \ion{H}{II} $v_\mathrm{LOS}$ map in the central high-SB region appears quite
chaotic at first glance.  Nevertheless, \cite{Moiseev2010}, using SAO/SCORPIO
Fabry-Perot spectroscopy of H$\alpha$ in the central region and using VLA C- and
D-configuration 21\,cm data from \cite{Pustilnik2001}, showed that the velocity
field can be characterised by a disk-like north-west to south-east gradient that
exhibits a distinct perturbation in the centre.  \citeauthor{Moiseev2010}
modelled this central disturbance, which is shaped somewhat like a hook, as the
effect of an expanding super-bubble disturbing a disk-like velocity field.  The
best-fitting disk was found to be aligned at a position angle of 53$^\circ$ east
of north.  We plot this kinematic axis and the orthogonal axis forced through
the photometric centre (Sect.~\ref{sec:ionised-gas}) in Fig.~\ref{fig:kin}.  It
can be appreciated how the orthogonal axis appears almost symmetric in between
the two filaments.

We do not observe velocity gradients along the direction of the extending filaments.
As already noted in \citetalias{Herenz2017b}, the filaments seemingly
inherit the velocity that is found at their base where they connect the
high-surface brightness region. The northern filament appears to inherit the
blue-shifts from the central perturbation, whereas the western filament appears
slightly more redshifted.  A conspicuous kinematic feature is the blue-shift
($\approx -35$\,km\,s$^{-1}$) at the western sub-branch at the tip of the
western-filament.  As analysed in Sect.~\ref{sec:neutral-gas}, this is where the
21\,cm maps indicate an overlap between H$\alpha$ emitting gas and the
\ion{H}{I} bridge that connects to the western galaxy.  Excluding this
sub-branch, the velocity difference between the average velocities in the
filament is $\sim 3 - 5$\,km\,s$^{-1}$.

The overall shear of the complete velocity field is
$v_\mathrm{shear} = 1/2 \times (v_{95} - v_{5}) = 28.5 $\,km\,s$^{-1}$, with
$v_{95}$ and $v_{5}$ being the upper and lower fifth-percentile of the velocity
map.  By limiting the calculation of $v_\mathrm{shear}$ only to the high-SB
region
($\mathrm{SB}_\mathrm{H\alpha} > 12.5 \times
10^{-18}$erg\,s$^{-1}$cm$^{-2}$arcsec$^{-2}$) we find
$v_\mathrm{shear}^{> 12.5} = 22.9$\,km\,s$^{-1}$.  This can be compared to the
disk model by \cite{Moiseev2010}, that is characterised by a maximal rotation
velocity of $v_\mathrm{max} = 28.2$\,km\,s$^{-1}$ and an inclination of
$i = 37^\circ$ \citep[see also][]{Moiseev2015}, that is the projected maximum
amplitude of this disk is $v_\mathrm{max} \cdot \sin i =
17$\,km\,s$^{-1} < v_\mathrm{shear}^{>12.5}$.  \citeauthor{Moiseev2010}
attributed this discrepancy between disk-like and observed motions in the
high-SB region to non-circular motions due to the expanding shell near the centre.  For
the large-scale \ion{H}{ii} velocity field the discrepancy is even more
pronounced.

That neither a large velocity offset between the filaments, nor a strong
velocity gradient along the filaments, nor abrupt discontinuities with respect
to the central velocity field are observed may be caused by projection effects
(i.e. because of material moving mostly perpendicular to our sightline).  However,
it may also indicate the absence of strong flows of ionised gas along the
filaments.  The eastern ``foreshortened'' filaments
(cf. Fig.~\ref{fig:cont}) are not characterised by an abrupt break in
line-of-sight velocities.  Ostensibly they just continue the velocity gradient
towards positive velocities along the north-eastern direction.  However, the
$v_\mathrm{shear} > v_\mathrm{shear}^{>12.5}$ behaviour is caused by the
redshifts of the foreshortened filaments and by the blue-shift in the western
sub-branch of the western filament. Hence, The velocities observed in those
extended regions are not compatible with an idealised disk and thus the motions
of the ionised filamentary halo gas are not predominantly driven by the
gravitational potential of the galaxy.
% if necessary, then point comparison with HI data below

The velocity dispersion map in Fig.~\ref{fig:kin} (bottom panel) shows that the
filaments are characterised by relatively narrow emission, with average values
of $\approx 30$\,km\,s$^{-1}$.  This is similar to the galaxy's luminosity
weighted average
$\sigma_0 = \sum_{x,y} \mathrm{NB}^\mathrm{H\alpha}_{x,y} \sigma_{x,y} /
\sum_{x,y} \mathrm{NB}^\mathrm{H\alpha}_{x,y} = 29$\,km\,s$^{-1}$; here
$\sigma_{x,y}$ denote the dispersion measures in each pixel of the dispersion
map and $\mathrm{NB}^\mathrm{H\alpha}_{x,y}$ are the flux values recorded in
each pixel of the H$\alpha$ narrow band image (Fig.~\ref{fig:hanb}).  This value
is in exact agreement with $\sigma_0$ derived from high-spectral resolution
($R\sim 86000$) Fabry-Perot spectroscopy \citep{Moiseev2010,Moiseev2015}.  We
note, however, that in the filaments $\sigma_v / \Delta \sigma_v \lesssim 2$,
that is there the H$\alpha$ line is barely resolved with MUSE.
 
The ratios of large-scale ordered motions, quantified by $v_\mathrm{max}$ or
$v_\mathrm{shear}$, to localised unordered motions, quantified by $\sigma_0$,
are a useful quantity to characterise the kinematical state of a galaxy
\citep[see review by][]{Glazebrook2013}.  We confirm
$v_\mathrm{max}/\sigma_0 \approx 1$ for the disk model of \cite{Moiseev2010},
and we note $v_\mathrm{shear}/\sigma_0 \sim v_\mathrm{max}/\sigma_0$.  The
$v/\sigma$-ratio for SBS\,0335-052E is lower that what is observed for typical
disk-galaxies ($v_\mathrm{max}/\sigma_0 \gtrsim 4$).  A large fraction of
star-forming galaxies at high-$z$ show also $v_\mathrm{max}/\sigma_0 \lesssim 1$
\citep[e.g.][]{Turner2017,Wisnioski2019}.  Star-formation-driven feedback is
deemed to be an important driver of those so-called dispersion dominated
kinematics.

The zone of narrow lines at the photometric centre is surrounded by zones of
elevated velocity dispersions (45\,--\,55\,km\,s$^{-1}$, towards the north and
south).  These zones appear as double-peaked H$\alpha$ profiles in high spectral
resolution ($R \sim 10000$) VLT/GIRAFFE ARGUS data \citep[][their
Fig.~7a]{Izotov2006}, but at our lower spectral resolution we here only observe
a broadening.  These double peaked profiles were interpreted as expanding
shells.  Moreover, \cite{Izotov2006} conjectured the launch of an outflow
perpendicular to the disk from the intricate line profiles seen in the small
ARGUS field of view (see especially Fig.~7 and Fig.~8).  This is consistent with
our interpretation of the filaments being large-scale effects of this outflow
(see discussion in Sect.~\ref{sec:disc}).
 %?

\subsubsection{\ion{H}{I} kinematics}
\label{sec:hikin}

In Fig.~\ref{fig:mom1} we display the intensity weighted line-of-sight velocity
field of the 21\,cm signal from SBS\,0335-052E \& W, both for the untapered and
the 7k$\lambda$-tapered data products.  The velocity maps in km\,s$^{-1}$ were
computed from the moment-1 maps produced by the source finding software SoFiA
\citep{Serra2015}. The relevant input parameters for our SoFiA runs were stated
in Sect.~\ref{sec:neutral-gas}.

The velocity map from the 7k$\lambda$-tapered datacube appears quantitatively
consistent with the maps from GMRT data \citep{Ekta2009} and with the maps from
VLA C- and D-configuration data \citep{Pustilnik2001}.  The \ion{H}{i} envelope
encompassing both systems may be characterised by an overall shear from east to
west, whose gradient is, however, not perfectly smooth.  Both \cite{Ekta2009}
and \cite{Pustilnik2001} argued for the envelope being comprised of two
disk-like systems.  Judging from our maps, but also from the maps presented in
the aforementioned publications (Fig.~6 in \citealt{Pustilnik2001} and Fig.~1 in
\citealt{Ekta2009}), this reading appears overly simplistic.  Focusing here on
the eastern galaxy, another gradient from redshifts in the south-east to
blue-shifts the north-west is perceivable.  The alignment of this gradient along
the directions of the filaments is suggestive of causal relationships between
ionised and neutral phase.  At the higher spatial resolution of the untapered
datacube the velocity fields exhibit an even higher complexity, and simple
disk-like gradients appear indiscernible. In the top-left panel of
Fig.~\ref{fig:kinc} we show a magnified view of the velocity field of the
eastern galaxy; an overall north-west to south-east shear may be envisioned, but
this velocity gradient is significantly warped.

As already noted in Sect.~\ref{sec:neutral-gas}, the faintest \ion{H}{I} signals
from the outskirts are often only single channel detections.  This can also be
appreciated from the channel maps in Appendix~\ref{sec:channel-maps}.  While our
data does not allow for robust higher-order measurements, these channel maps
supplement the view of high kinematic complexity in the \ion{H}{I} on kpc scales
in the system.  From \ion{H}{i} morphology and kinematics alone it is not
possible ascribe tidal- or feedback effects as the source of this complexity.
We thus proceed by comparing \ion{H}{I} and \ion{H}{ii} kinematics
quantitatively in the next section.

\subsubsection{Comparison between \ion{H}{I} and \ion{H}{ii} kinematics}
\label{sec:chihii}

We compare the ionised gas velocity field derived from the MUSE data
(Sect.~\ref{sec:hiikin}) to the \ion{H}{I} velocity field from the VLA
observations (Sect.~\ref{sec:hikin}).  For this comparison we need to account
for the lower spatial resolution and the coarser spatial sampling of the
\ion{H}{I} data products.  To this aim we follow the method used by
\cite{vanEymeren2009,vanEymeren2009b} and convolve the \ion{H}{II}
$v_\mathrm{los}$ map (Fig.~\ref{fig:kin}) with a 2D Gaussian, whose position
angle, major-, and minor-axis are matched to the beam of the respective 21\,cm
data.  We next re-bin this convolved velocity field to the native resolution of
the VLA datacubes.  The complete process is illustrated in Fig.~\ref{fig:kinc},
with the top right panel displaying the \ion{H}{II} velocity field as observed,
with the bottom left panel displaying it after convolution, and with the bottom
right panel displaying this convolved field after re-binning.  For the
7k$\lambda$-tapered dataset we omit the intermediate step and show only the
resulting resampled velocity field (bottom panel of Fig.~\ref{fig:m17kc}).

Qualitatively, both the velocity fields from the untapered and the
7k$\lambda$-tapered \ion{H}{I} data show a high level of congruence with the
degraded and resampled \ion{H}{II} velocity fields (Fig.~\ref{fig:kinc} and
Fig.~\ref{fig:m17kc}).    Our method ignored
the effect of intensity variations on the resulting resolution-degraded and
re-binned \ion{H}{ii} velocity fields.  If we instead convolve each MUSE
datacube layer with the beam-imitating 2D Gaussian prior to the kinematic
fitting, then the resulting degraded velocity fields show almost no spatial
variation, as they mostly trace \ion{H}{II} velocities stemming from the
brightest regions.  Hence, the unknown intrinsic 21-cm signal from the neutral
phase does not mimic the extreme intensity variations seen in the \ion{H}{ii}
emitting plasma.

The degraded \ion{H}{II} velocity fields appear more smoothly varying compared
to the \ion{H}{I} measurements, which is partly related to the low-SN of the
\ion{H}{I} data.  We analyse the quantitative differences between \ion{H}{I}
velocity fields, $v_\mathrm{HI}$, and the resolution degraded and re-sampled
\ion{H}{ii} velocity fields, $v_\mathrm{HII}^\mathrm{rep}$, in
Fig.~\ref{fig:diffm}. There we display
$\Delta v = v_\mathrm{HI} - v_\mathrm{HII}^\mathrm{rep}$ maps both for the
untapered and the 7k$\lambda$-tapered datasets.  These $\Delta v$-maps exhibit
mostly an amplitude of absolute differences $\lesssim 10$\,km\,s$^{-1}$ and thus
confirm the qualitative impression of overall congruence already anticipated
from Figs.~\ref{fig:kinc} and~\ref{fig:m17kc}.  The similarity between
both velocity fields manifests in small mean (median) absolute differences
of 0.6\,km\,s$^{-1}$ (0.2\,km\,s$^{-1}$) and 1.1\,km\,s$^{-1}$
(0.6\,km\,s$^{-1}$) for the untapered and 7k$\lambda$-tapered velocity fields,
respectively.  There are, nevertheless, some notable differences.  In the
untapered data we find $\Delta v \gtrsim 15$\,km\,s$^{-1}$ near the centre.
This $v_\mathrm{HI} > v_\mathrm{HII}^\mathrm{rep}$ zone is in close vicinity to
the most prominent ``hook-like'' disturbance of the \ion{H}{II} map.  A causal
connection between \ion{H}{I}-\ion{H}{II} velocity offsets and the expansion
thus suggests itself.  In the 7k$\lambda$-tapered data we find a prominent rim
$v_\mathrm{HI} > v_\mathrm{HII}^\mathrm{rep}$ along the south-west, parallel to
the minor axis.  The origin of this feature is not obvious.  Both
high-$\Delta v$ zones in both datasets are related to the most prominent
complexities of the \ion{H}{I} velocity fields that inhibited the identification
of simple disk-like velocity gradients in the first place
(Sect.~\ref{sec:hikin}).

\section{Discussion}
\label{sec:disc}

\begin{figure*}
  \centering
  \includegraphics[width=0.49\textwidth]{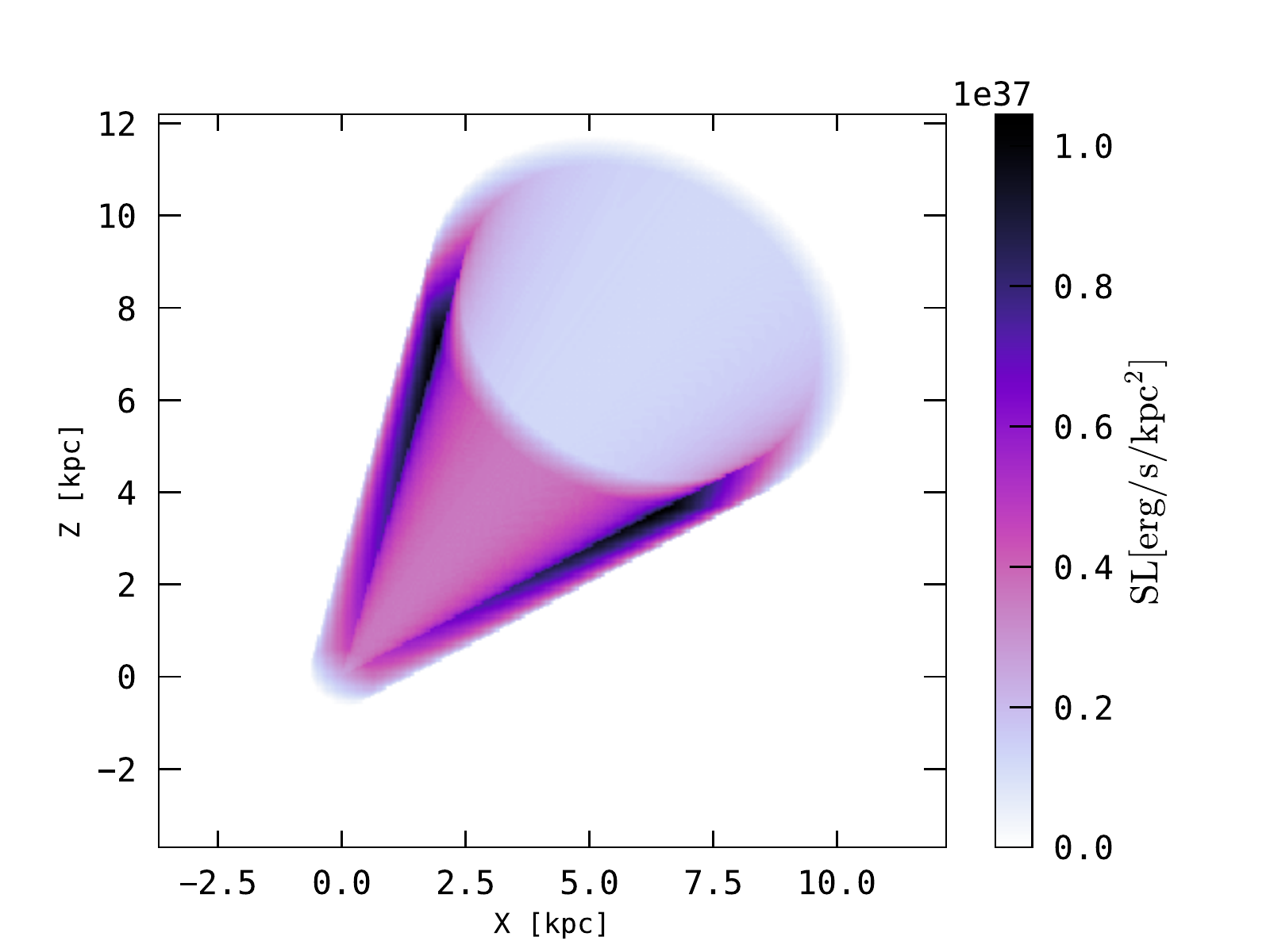}
  \includegraphics[width=0.495\textwidth]{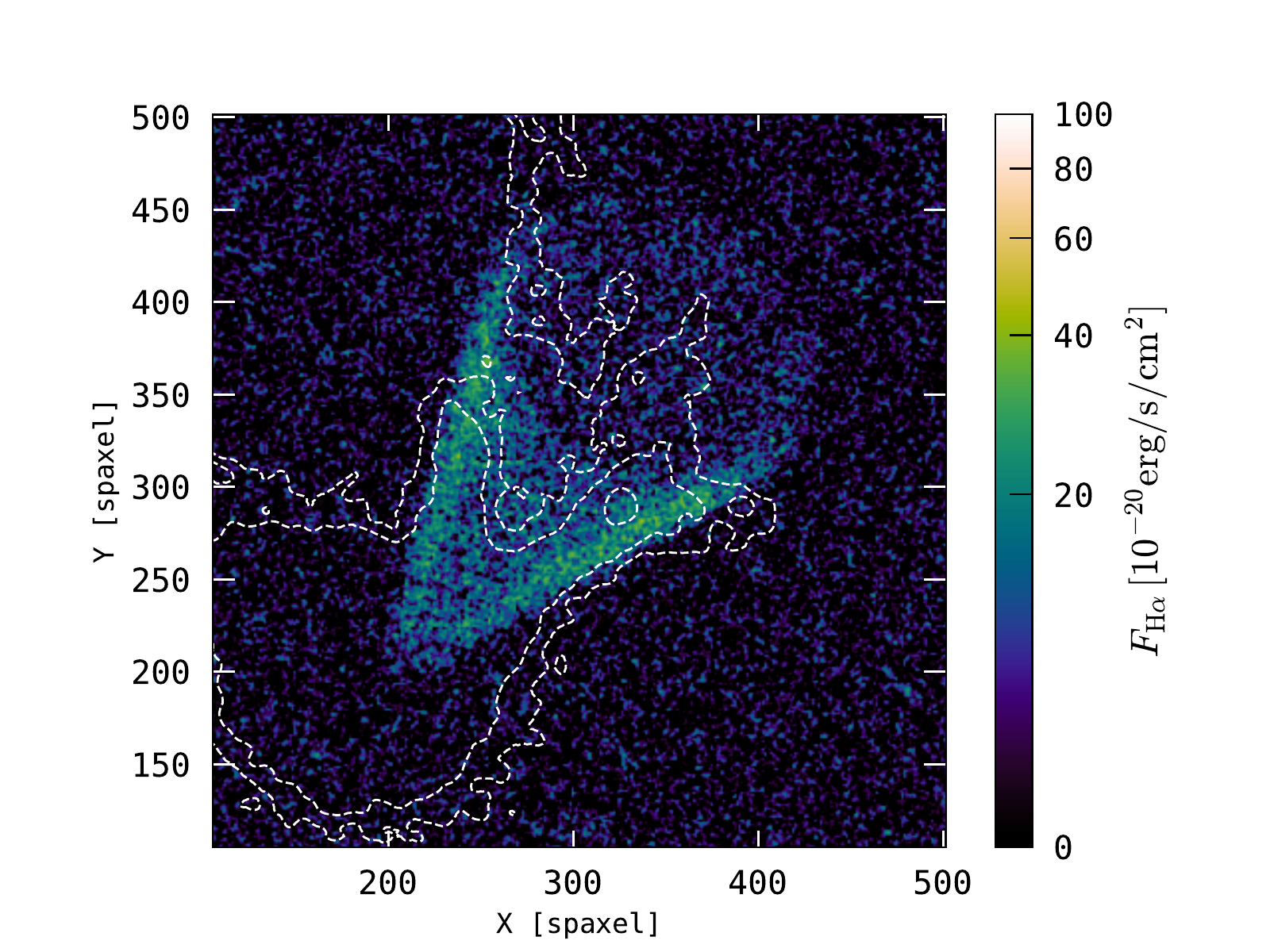}
  \caption{Toy model to explain the NW filaments as limb-brightened edges of an
    outflow cone.  We assume a cone filled with a hot wind fluid that is
    surrounded by walls filled with ionised hydrogen in recombination
    equilibrium ($T=2\times10^4$\,K, $n=0.05$\,cm$^{-3}$).  Details on the setup
    and the model parameters that are derived from the observations are given in
    the text.  The \emph{left panel} shows a projection of the resulting
    H$\alpha$ emission (as surface-luminosity in erg\,s$^{-1}$kpc$^{-1}$) from
    this model along the Y-axis of the model box; this axis is chosen to be
    parallel to the line of sight.  The \emph{right panel} shows how the H$\alpha$
    emission from this model would appear in our MUSE data under the assumption
    that the model is at the distance of SBS\,03352-052E.  Here we overlay some
    H$\alpha$ contours from Fig.~\ref{fig:hanb} (dotted white lines) to indicate
    the degree of congruence between the simulated observations of this simple toy
    model and reality.}
  \label{fig:cm}
\end{figure*}

The presented observational evidence in the previous section leads us to explain
the observed features in the circumgalactic halo of SBS\,0335-052E as
limb-brightened edges of a conical outflow phenomenon.  We briefly summarise the
main clues towards the suggested interpretation before making an attempt to
quantitatively illustrate this scenario with a didactic toy model.

VLT/MUSE reveals two faint ionised filaments of delicate morphology that extent
up to 15\,kpc to the north-west of the starburst (Fig.~\ref{fig:hanb}).  The
orientation of a perceived symmetry axis that separates the rays of the
filaments is perpendicular to the shearing direction of the large-scale velocity
field (Fig.~\ref{fig:kin}).  This large-scale shear lines up with the major axis
of the disk-model derived from SAO/SCORPIO Fabry-Perot observations in the
central high-$\mathrm{SB}_\mathrm{H\alpha}$ region
\citep{Moiseev2010,Moiseev2015}; therefore, the symmetry axis of the bifurcating
filaments is nearly parallel with the minor axis of this model
(Fig.~\ref{fig:kin}).  The ionised filaments protrude away from the
elongation axis of the \ion{H}{I} envelope (Fig.~\ref{fig:hI7kt}).  This
envelope also exhibits overall disk-like shearing motions
(Fig.~\ref{fig:m17kc}).  These observational facts are reminiscent to
prototypical conical outflows in disk-galaxies that launch such cones from their
central starburst outwards from the disk plane
\citep[][]{Veilleux2005,Bland-Hawthorn2007,Nelson2019}.

We explore the possibility of this scenario by setting up a heavily idealised
astrophysical structure with parameters suggested by the observations.  Our toy
model geometry is a simple cone of opening angle $\theta$ and height $h$.  The
apex of this cone is fixed at the centre of a disk at position angle $\vartheta$
and inclination $i$.  Thus the observed opening angle, $\theta_P$, is a
projection of the true cone opening
$\theta = 2 \times \arctan [ \sin (i) \times \tan (\theta_P / 2) ]$.  We assume
that the H$\alpha$ filaments are confining this cone.  We measure
$\theta_P = 34^\circ$ as the angle between the filaments from the H$\alpha$ NB
(Fig.~\ref{fig:hanb}) and the disk-model of \cite{Moiseev2010} which fits
$i=43^\circ$, hence $\theta = 27^\circ$.  We note, however, that conical winds
can be tilted and asymmetric with respect to the centre; our value for $\theta$
thus represents an approximate working hypothesis.  We also adopt
$\vartheta = 52^\circ$ from \cite{Moiseev2010}.  We fix $l_p = 10$\,kpc for the
projected length of the filaments from Fig.~\ref{fig:hanb}, where we ignore for
now the threadlike extension.  The height of the inclined cone is then
$h = l_p \cdot \cos(\theta/2) / \sin(i) = 16.2$\,kpc.

Our estimate of the height $h$ may be translated into a velocity requirement
that the hypothesised wind from a stellar-population of age $t_*$ must have to
blow out such a structure.  The oldest star-clusters in the north (SSC 5 \& SSC
6; Fig.~\ref{fig:clustident}) have ages $t_* \lesssim 15$\,Myr
\citep{Reines2008,Adamo2010}.  Accounting for the delay after which supernovae
start to inject kinetic energy, $t_\mathrm{SN} \approx 5 $\,Myr, we have
$v_\mathrm{wind} = h / (t_* - t_\mathrm{SN}) = 10\,\mathrm{kpc} /
10\,\mathrm{Myr} \times \cos(\theta/2) / \sin(i) \approx
1000$\,km\,s$^{-1} \times \cos(\theta/2) / \sin(i) = 1620$\,km\,s$^{-1}$.  This
rough limit, which neglects variations in energy input and flow-speed, is
broadly consistent with predictions for velocities in the tenuous hot phase
($T > 10^6 - 10^7$\,K, $n \sim 10^{-3} - 10^{-4}$\,cm$^{-3}$) of galactic winds.
This ``wind-fluid'' is believed to carry most of the momentum and energy of an
outflow, and at larger radii even most of the mass \citep{Schneider2020}.  If
the volume of the cone,
$V_\mathrm{hot} = \pi/3 \cdot \tan^2 (\theta / 2) \cdot h^3 = 256$\,kpc$^3$,
would be filled uniformly with this fluid, then its mass would comprise
$M_\mathrm{hot} \sim 6 \times 10^6 $ - $6 \times 10^5$\,M$_\odot$.  This
requires loading rates,
$\Lambda_\mathrm{hot} = (M_\mathrm{hot})/(\dot{M}_\mathrm{SFR} \cdot \Delta t)$, of
order unity if all the mass is asserted to be loaded into the wind via
star-formation over the last $\Delta t = 10$ Myr given the determined
$\dot{M}_\mathrm{SFR} \gtrsim 1$\,M$_\odot$yr$^{-1}$ in SBS\,0335-052E
\citep{Reines2008,Thompson2009}.  This rough limit for $\Lambda_\mathrm{hot}$ is higher than
the expected average mass-return from stellar winds and supernovae in stellar
populations, $\Lambda = 0.1$ \citep{Leitherer1999}, that is also used as the
injected mass fraction for the hot phase in computer simulations of wind
phenomena \citep[e.g.][]{Schneider2018,Schneider2020}.  However, a significant
fraction of the fluid will consist of shock-heated ambient ISM, hence the
required mass loaded into the interior of our cone does not appear unphysical.

The hot fluid, which does not emit in optical recombination lines, is assumed to
be surrounded by the warm ionised medium that fills the wall of the cone and
leads to the observed structure in H$\alpha$.  Denoting the thickness of the
walls as $t$ and writing $\Delta h = 2 \cdot t \cdot \cos(\theta/2)$ the volume
inhabited by the diffuse \ion{H}{ii} gas is
$V_\mathrm{HII} = \pi/3 \cdot \tan^2 (\theta/2) \cdot \left ( (h+\Delta h)^3 -
  h^3 - \Delta h^3 \right )$.  Figuratively, $V_\mathrm{HII}$ is the volume
between two identical cones pushed into each other
($\propto (h+\Delta h)^3 - h^3$), but ignoring the volume between the apexes of
the inner and the outer cone ($\propto \Delta h^3$).  From the H$\alpha$ NB we
estimate $t=1.5$\,kpc and hence $V_\mathrm{HII} = 142$\,kpc$^3$.

In the spirit of a toy-model we assume the plasma filling $V_\mathrm{HII}$ is in
recombination equilibrium, fully aware that deviations from equilibrium are
expected in shocked gas \citep[e.g.][]{Allen2008,Morisset2020} or due to
radiative transfer effects (Sect.~\ref{sec:ibdc}). 
An ionised hydrogen plasma of density $n$ ($\equiv n_e = n_p$, i.e. no electrons
from other species) and temperature $T$ filling a volume $V$ emits H$\alpha$
emission at luminosity
\begin{equation}
  \label{eq:3}
  L_\mathrm{H\alpha} = \epsilon_\mathrm{H\alpha}(T) \cdot n^2 \cdot V 
\end{equation}
in recombination equilibrium, where $\epsilon(T)$ denotes the H$\alpha$
equilibrium emissivity that follows from atomic data.  For
$T = \{1.0, 1.5, 2.0 \} \times 10^4$\,K in case-A recombination
$\epsilon^\mathrm{A}_\mathrm{H\alpha}(T) = \{2.3, 1.5, 1.1 \} \times
10^{-25}$\,erg\,s$^{-1}$cm$^3$ \citep{Storey1995}; the small density dependence
on $\epsilon$ can be neglected for our purposes.  Using a hand-drawn aperture
that encompasses the filamentary structure while not covering inner high-SB
spurs and shell-like structures (thin dotted line in Fig.~\ref{fig:cont}), we
estimate
$F_\mathrm{H\alpha} \approx 2.6 \times 10^{-15}$\,erg\,s$^{-1}$cm$^{-2}$ as the
flux from the supposed cone.  From Eq.~(\ref{eq:3}) then
$n_A = \{3,4,5\} \times 10^{-2}$\,cm$^{-3}$ follows for
$T=\{1,1.5,2\}\times10^4$\,K in case-A.  We recall from Sect.~\ref{sec:ibdc}
that Case-A recombination appears more compatible with the observed Balmer
decrement in the filaments than case-B recombination, noting here that
$\epsilon_\mathrm{H\alpha}^\mathrm{B}(T) <
\epsilon^\mathrm{A}_\mathrm{H\alpha}(T)$ and that
$\epsilon_\mathrm{H\alpha}^\mathrm{B} / \epsilon_\mathrm{H\alpha}^\mathrm{B}$ is
only mildly temperature dependent.  Therefore we can approximate
$\epsilon^\mathrm{B}_\mathrm{H\alpha} \approx 0.65
\epsilon^\mathrm{A}_\mathrm{H\alpha}$ and hence $n_B \approx 1.24 \cdot n_A$
would be required for case-B.

The required densities in the walls of the cone are only a factor of two to
three lower than the canonical $n=0.1$\,cm$^{-3}$ for diffuse warm ionised gas
in the circumgalactic medium.  They are consistent with the expectations for the
$T\sim10^4$K phase of galactic winds.  The correspondence between the derived
parameters in our idealised setup and the physical characteristics expected in
outflows appears encouraging and supportive of the idea that we indeed observe a
15\,kpc outflow cone emanating from a tiny starburst galaxy.  We remark that the
mass of ionised hydrogen in the toy-model,
$M_\mathrm{HII} = 1.5 - 1.7 \times 10^8$\,M$_\odot$, requires a mass-loading
factor $\Lambda_\mathrm{HII} \gtrsim 10$ if this mass has been loaded into the
wind over the last 10\,Myr.  Neglecting the cold phase, which is expected to
provide only a minor contribution to the total mass of outflowing material, we
arrive at a total mass loading factor of
$\Lambda = (M_\mathrm{hot} + M_\mathrm{HII})/(\dot{M}_\mathrm{SFR} \cdot \Delta
t) \gtrsim 10$.  Given the low-stellar mass of SBS\,0335-052E,
$M_\star \approx 6 \times 10^6$\,M$_\odot$ \citep{Reines2008}, the required
$\Lambda$ is in line with theoretical expectations \citep{Pandya2021} and
observed trends for the other dwarf starbursts \citep[][their
Fig.~4]{Collins2022}.  Interestingly, the ratios
$\Lambda_\mathrm{hot}/\Lambda \approx 0.1$ and
$\Lambda_\mathrm{HII}/\Lambda \approx 1$ are in line with the expectations of
the wind models by \cite{Pandya2021}.  We caution that our results have to be
taken \emph{cum grano salis}, as we assumed that the diffuse H$\alpha$ emission
in the cone is caused by pure Case-A recombination.  As explained above,
slightly higher densities would be required for the Case-B scenario, thus the
mass in the warm phase would be slightly lower.  However, we have shown in
Sect.~\ref{sec:ibdc} that the Balmer-decrement in the filaments is suggestive of
neither Case-A or Case-B being appropriate.  Ultimately photo-ionisation models
are required to obtain more stringent constraints on the observed mass in the
warm phase.

We now explore the observable H$\alpha$ morphology of our toy-model cone-wall
structure numerically.  To this aim we compute the emission on a cubic 3D
lattice with $500^3$ volume cells that cover a $25^3$\,kpc$^3$ volume, that is each
cell has a volume of $V_\mathrm{cell}=50^3$\,pc$^{3}$.  Then all grid cells $C$
are set to emit H$\alpha$ according to Eq.~(\ref{eq:3}) with $V_\mathrm{cell}$
when their centre-coordinates are within the walls of the cone.  Formally, this
is the set of cells
$C = \{ x',y',z' : f(x',y',z' | h,\theta) \leq 0 \oplus f(x',y',z' | h+\Delta h,
\theta) \leq 0\}$, where $\oplus$ denotes the exclusive disjunction and
$f(x',y',z' | h, \theta ) = x'^2 + y'^2 - z'^2 \tan^2(\theta/2)$ is the implicit
parametric equation for the surface of a single cone and $x'$, $y'$, and
$z' \in [0,h]$ are the coordinates in a rotated frame that is tilted along
$\vartheta$ and inclined along $i$ using standard intrinsic Euler rotations.  We
set up the model grid with cell-coordinates $x,y,z$ such that the $y$-axis is
parallel to the sight-line.  We project along this axis by summation after
dividing by the surface area of a cell ($6 V_\mathrm{cell}^{2/3}$).  This
provides a map of surface luminosities (erg\,s$^{-1}$kpc$^{-2}$).  This map is
transformed into observable fluxes in erg\,s$^{-1}$cm$^{-2}$ on a grid of
0.2\arcsec{} MUSE-like spaxels for the given distance to galaxy.  Prior to this
re-binning the received fluxes are convolved with a Gaussian of 1.2\arcsec{}
FWHM, which was the average DIMM seeing in our observations
(Sect.~\ref{sec:muse}).  Finally, we add noise to the re-binned image consistent
with the noise determined in the outskirts of the H$\alpha$ NB
($\sigma_\mathrm{H\alpha} = 1.8\times10^{-19}$\,erg\,s$^{-2}$\,cm$^{-2}$). We
show the result of this exercise for the observationally constrained parameters
from above ($\theta = 27^\circ$, $i=37^\circ$, $\vartheta = 52^\circ$,
$h=16.2$\,kpc, and $t=1.5$\,kpc) and
$(n,T) = (0.05\,\mathrm{cm}^{-3}, 2\times10^4\,\mathrm{K})$ in
Fig.~\ref{fig:cm}. To provide visual intuition regarding the projected
morphology for different geometrical parameters we study the effects of varying
$\theta$, $\vartheta$, and $i$ in Appendix~\ref{sec:toy-model-paramter}; we note
that the projection of the model is the same if the cone is inclined towards us
or away from us.  Furthermore we examine the effects of varying $T$ and $n$ on
the overall observability in Appendix~\ref{sec:toy-model-paramter}.  We also
verified the consistency between numerical computation and the analytical
results presented beforehand.

We overlay in Fig.~\ref{fig:cm} the two faintest H$\alpha$ SB isocontours from
Fig.~\ref{fig:hanb} to visualise morphological similarities and differences
between calculation and observation.  These contours were aligned by eye to
maximise congruity.  The resemblance between the MUSE mock data of the cone-wall
model and observed reality is readily apparent.  An interesting feature of the
model is the brightening along the walls with increasing distance from the apex.
This radial brightening is caused by more wall material being intersected
perpendicular to the sightline closer to the base of the cone.  In reality we
observe a brightening with increasing distance only in the western filament
(H$\alpha$ ``hot-spot'' in Fig.~\ref{fig:cont}); however, compared to the smooth
radial gradient in the model this feature emerges more abruptly.  In fact, the
observed filaments brighten towards the central high-SB zone.  Density or
temperature variations, as well as variations of the ionisation mechanisms, may
be invoked to explain the differences between model and observations and to
produce localised features such as the H$\alpha$ hot-spot.  

The observations reveal further morphological sub-structure that is not
reproducible with our simple purely geometrical setup.  Most prominently, the
cone-wall structure cannot explain the threadlike extension and the fanning out
of the western filament (Fig.~\ref{fig:cont}).  Parts of these filamentary
sub-structures coincide spatially with the region where only one wall of the
cone would be intersected by the sight-lines.  Zones of higher density or lower
temperature in the wall section or lumps of gas that have been advected by the
hot fluid could lead to brighter H$\alpha$ patches.  Alternatively, these
features may be interpreted as the terminating cloud-wind interface where dense
gas has been compressed and pushed up by the hot wind fluid and now cools
radiatively.

State-of-the-art numerical models of starburst-driven winds
\citep{Schneider2018,Schneider2020} produce small-scale density and temperature
variations in wind structures that all have been neglected in our didactic
scenario.  Sub-structure within the hot wind fluid arises due to interplay
between the different gas-phases; cold and warm material swept up from the ISM may
survive within the hot flow whereas density variations in the hot phase might
lead to condensation and radiative cooling.  Substructures in the HII emitting
phase produce turbulent mixing layers between the fast moving tenuous fluid and
the slower-moving denser warm-phase.  The emergence of a conical structure
appears, however, a common feature of wind phenomena.  This ``beaming'' is
believed to be a purely hydro-dynamical collimation effect due to the intrinsic
density distribution in the gaseous disk surrounding the energy and momentum
injecting star-formation sites \citep{Nelson2019}.  In this respect it appears
interesting to note that \cite{Micheva2019} advocate to interpret $\theta$ as
the opening angle of a Mach cone, that is the surfaces of the cone are created by a
shock-front from the overlapping sound waves in a supersonic flow.  The opening
angle is then related to the Mach number $\mathcal{M}$, that expresses the
velocity of the flow as multiples of the speed of sound via
$\mathcal{M} = 1 / \sin(\theta / 2)$.  For our cone we have $M \simeq 4$,
that is to say in this interpretation the wind fluid is required to be super-sonic.
Intriguingly, $M \simeq 4$ flows are again consistent with sophisticated
model expectations for the tenuous hot wind fluid.

Obviously our toy-model is static and neglects kinematics.  A kinematic model
requires many assumptions and is beyond the scope of our current analysis.
Still, some important kinematical features need to be discussed qualitatively,
especially since the absence of a strong velocity gradient and the narrow line
widths appear incompatible with the idea that the \ion{H}{II} phase is being
accelerated in an outflow.  Projection affects surely may influence the observed
line-of-sight velocities and velocity dispersions given that we look at the cone
from the side.  However, under the assumption of isotropic random motions the
combination of ordered motions and inclination can only broaden the line.  This
effect is known to as ``mixing term'' in observational studies of velocity
dispersions in disk galaxies \citep[see, e.g. Sect. 2.1 in][]{Bouche2015}.  In
this scenario then the narrowness of the lines would indeed reflect an intrinsic
narrow velocity dispersion.  A projection effect is thus only expected if the
distribution of velocities is larger parallel to the walls of the cone.
Similarly, the lack of an observed velocity gradient along the walls of the cone
may be an projection effect.  Nevertheless, the large-scale wind simulations by
\cite{Schneider2020} show that the \ion{H}{II} emitting gas at larger radii is
moving at fairly constant velocities.  Interestingly, we do not observe strong
velocity differences between \ion{H}{I} and \ion{H}{II} at the base of the cone
in the 7k$\lambda$-tapered data (Fig.~\ref{fig:m17kc} and right panel of
Fig.~\ref{fig:diffm}).  We thus speculate, that the walls of the cone are
\ion{H}{I} halo gas that is advected into the flow of the hot fluid.  During
this advection there might only be a short window of time before the dense warm
medium, which is likely in the form of clumps, is being transformed into a more
dilute wind fluid.  If this is the case, then we expect that here observed
emission of warm phase inherits mostly the velocity dispersion of the neutral
gas.  In this scenario the narrow lines appear consistent with the detection of
\ion{H}{I} emission in only 2-3 channels (see Fig.~\ref{fig:chmap7k}), which
corresponds to $\sim 20-30$\,km\,s$^{-1}$.

\section{Summary and conclusions}
\label{sec:sc}

We have presented new MUSE and VLA B-configuration 21cm observations of the
extremely low-metallicity blue compact starburst galaxy SBS\,0335-052E.  Our
observations reveal the continuation of two ionised filaments in H$\alpha$ emission
towards the north-west of the galaxy's main stellar body
(Sect. \ref{sec:ionised-gas} and Fig.~\ref{fig:hanb}).  The onset of this
structure was observed with MUSE observations centred on the galaxy and
described in \citetalias{Herenz2017b}.  The morphology of the newly detected
prolongation can be described as thread-like lacy fragments.  These somewhat
detached portions of extended H$\alpha$ emission are of lower SB
($\mathrm{SB}_\mathrm{H\alpha} \approx 1.5 \times
10^{-18}$erg\,s$^{-1}$cm$^{-2}$arcsec$^{-2}$) than the inner straight portions
of the filaments
($\mathrm{SB}_\mathrm{H\alpha} \approx 3 \times
10^{-18}$erg\,s$^{-1}$cm$^{-2}$arcsec$^{-2}$).  The rays of the inner filaments
are aligned at a projected angle of $\theta_P = 43^\circ$ relative to each
other.  Diffuse ionised gas with a smoothly declining brightness
profile is found up to $\sim 25$\arcsec{} towards the south of the galaxy.

The inner parts of the filaments are detected in H$\alpha$, [\ion{O}{iii}]
$\lambda5007$, and H$\beta$.  The flux ratios between those three lines are
different compared to typical ISM ratios in the vicinity of star formation.  The
inferred average Balmer decrement is H$\alpha /$H$\beta = 2.5$ for both
filaments, whereas [\ion{O}{iii}]/H$\alpha=0.65$ and
[\ion{O}{iii}]/H$\alpha=0.26$ for the eastern and western filament,
respectively.  The unusual low Balmer decrement, which does not agree with
standard recombination cascade expectations, is deemed to be caused by radiative
transfer effects in the low-density halo gas that alter the populations of the
contributing energy levels compared to the classical calculations
(Sect.~\ref{sec:ibdc}; \citealt{Osterbrock2006}).

Our VLA B-configuration observations enable us to resolve \ion{H}{I} at
unprecedented spatial resolution within and in the vicinity of the starburst.
The peak of the 21\,cm signal is spatially offset by $\sim 1$\arcsec{} to the
west from the main super-star-cluster complexes.  At lower $N_\mathrm{HI}$
($5 - 10 \times 10^{20}$cm$^{-2}$), we observe morphological correspondences
between neutral and ionised phases.  In particular, we observe a bent tail that
lines up in the direction of the eastern foreshortened filaments.  Our
observations are thus mapping the entanglement and interactions between the
neutral and ionised phases in feedback-driven gas.

Tapering the VLA B-configuration data at 7k$\lambda$ in the UV plane in the
imaging process enabled us to map the neutral halo around the interacting galaxy
pair SBS\,0335-052E and SBS\,0335-052W.  Our results are consistent with
the previous analysis of 21 cm data from VLA C- and D-configuration
\citep{Pustilnik2001} and GMRT \citep{Ekta2009} observations.  The H$\alpha$
emitting filaments extend outwards from the east to west elongated neutral halo
around SBS\,0335-052E.  However, there is significant overlap between halo gas
and the western filament.  Especially the coincidence of the H$\alpha$ hot spot
and the fanning-out of the western filament with the extended tidal halo gas
from the western galaxy appears intriguing.

Our kinematical analysis of the ionised gas shows that the line emission from
the filaments is very narrow, $\sigma_v \approx 20 - 30$\,km\,s$^{-1}$, and in fact
barely spectrally resolved in the MUSE data.  We do not map velocity gradients
along the filaments.  Zones nearer to the star clusters appear in the MUSE data
at high $\sigma_v$ ($45 -50$\,km\,s$^{-1}$), but they have previously been resolved
into double-peaked profiles with VLT/GIRAFFE \citep{Izotov2006}.  The seemingly
chaotic appearance of the \ion{H}{II} line-of-sight velocity map near the centre
has also been previously described as a kinematic disturbance of an expanding
shell superimposed onto a disk-like velocity field \citep{Moiseev2010}. The
position angle of the \citeauthor{Moiseev2010} model velocity field,
$\vartheta = 52^\circ$, is parallel to the shearing velocities mapped at lower
SB levels and in \ion{H}{i}.  Interestingly, this gradient is aligned
perpendicular to the symmetry axis in between the bifurcating filaments.

The line-of-sight kinematics of the neutral phase exhibit a significant degree
of complexity, both in the untapered and in the 7k$\lambda$-tapered data
products.  The complexity on large scales is likely due to tidal effects from
the encounter of the two galaxies.  However, an interaction with the
large-scale wind may also cause deviations from simple disk-like motions in and
around SBS\,0335-052E.  On smaller scales, the velocity gradient appears
significantly warped.  We find that \ion{H}{I} and \ion{H}{II} kinematics are
broadly in agreement on large and small scales.  Nevertheless, there appear to be
zones where the bulk motions of the neutral phase are redshifted by
$\sim 15 - 20$\,km\,s$^{-1}$ with respect to the velocities of the ionised
phase.  These velocity differences are deemed to be caused by different
responses of the neutral and ionised phases to injected energy and momentum from
stellar winds and supernova explosions.

The observational evidence assembled in our analysis merits an explanation as to the
unprecedented detection of more than 10\,kpc long H$\alpha$-emitting filaments
as limb-brightened edges of a conical outflow; this new hypothesis supersedes
the initial attempt at explaining the morphology in \citetalias{Herenz2017b}.  A
toy model of a cone structure, with geometrical parameters read off the MUSE
H$\alpha$ narrow band, is commensurable with the expected physical reality of a
starburst-driven wind.  Our setup is described by an inner cone that is filled
with a hot wind fluid and a surrounding wall filled with ionised plasma.  In
order to be compatible with the dimensions of the cone, the hot wind fluid would
be required to flow at radial velocities $\sim 1600$\,km\,s$^{-1}$.  This
hot wind fluid could be loaded easily with mass-loading factors
$\Lambda_\mathrm{Hot} \lesssim 0.1$.  While there is some degeneracy between
density and temperature in order to match the flux of the ionised phase, we
generally require very low densities of $n \lesssim 0.05$\,cm$^{-3}$.  If the
ionised phase is purely gas expelled from the starburst, then mass-loading
factors $\Lambda \approx \Lambda_\mathrm{HII} \gtrsim 10$ would be required.  A
mock MUSE observation of the toy model clearly shows the limb-brightening effect
at surface brightness levels similar to the observations.  Obviously, our
didactic setup fails to reproduce the morphological substructure observed in
reality, especially the thread-like lacy prolongation at the largest projected
distances.

Given the small size of SBS\,0335-052E's main stellar body, the dimensions of
its outflow structure revealed in the ionised phase appear extreme.  Similar-sized $\sim 10 - 20$\,kpc outflows are known from more massive and larger
galaxies.  The finding reported here was only possible because of the
unprecedented sensitivity of MUSE for low-SB line emission.  Currently it is
unknown whether such structures are common around other metal-poor compact
starbursts.  One unique characteristic of the SBS\,0335-052 system is its large
neutral gas reservoir, and we may speculate here that the size of the ionised
structure is connected to the size of the halo, since a significant fraction of
the H$\alpha$-emitting gas may have been ionised in situ.

In our analysis and discussion, we made no attempt to constrain the actual
ionisation mechanism.  Photo-ionisation from the starburst and the X-ray
emitting wind fluid, as well as radiative shocks between wind fluid and neutral
halo gas are expected to contribute.  While the three emission lines in this study do not
provide discriminating power with respect to these mechanisms, they can be used
in connection with photo-ionisation or shock models to design deeper
observations that may constrain the ionisation mechanisms.  We believe that this
is an intriguing avenue for future investigations.

% MAIN PAPER ENDS HERE

\begin{acknowledgements}
  We thank the anonymous referee for their valuable contribution to the manuscript.
  We express our gratitude to the ESO Office for Science in Santiago for funding
  research internships for H. Salas and C.~Moya-Sierralta to contribute to this
  project.
  J.M. Cannon and J. Inoue are supported by NSF grant
  AST-2009894.  J.M. Cannon and B.  Koenigs acknowledge support from Macalester
  College.
  M.~Hayes is fellow of the Knut \& Alice Wallenberg Foundation.
  C.~Moya-Sierralta acknowledges support from the Becas-ANID scholarship \#21211528.
  We acknowledge
  the use of the following software packages for the analysis and exploration of
  the data presented in this paper: Astropy \citep{Astropy-Collaboration2018},
  SciPy \citep{Virtanen2020}, NumPy \citep{Harris2020}, Matplotlib
  \citep{Hunter2007}, GNUastro \citep{Akhlaghi2018}, LSDCat \citep{Herenz2017},
  SoFiA \citep{Serra2015}, \texttt{cmasher} \citep{vanderWelden2020}, QFitsview
  \citep{Ott2012}, fips \citep{Kornilov2019}, pyneb
  \citep{Luridiana2015,Morisset2020}, and ds9 \citep{Joye2003}.
  Based on observations made with ESO Telescopes at the La Silla Paranal
  Observatory under programme IDs 096.B-0690 \& 0104.B-0834, and based on
  observations made with the Karl G. Jansky Very Large Array (VLA), project
  number 17B-234.
  The National Radio Astronomy Observatory is a facility of the National Science
  Foundation operated under cooperative agreement by Associated Universities,
  Inc.
  Based on observations made with the NASA/ESA Hubble Space Telescope, and
  obtained from the Hubble Legacy Archive, which is a collaboration between the
  Space Telescope Science Institute (STScI/NASA), the Space Telescope European
  Coordinating Facility (ST-ECF/ESA) and the Canadian Astronomy Data Centre
  (CADC/NRC/CSA).
\end{acknowledgements}

% References
\bibliographystyle{aa}
\bibliography{bibliography}

\begin{appendix}

\section{Emission line profiles along the filaments}
\label{sec:emiss-line-prof}

\begin{sidewaysfigure}
  \centering
  \includegraphics[width=0.4\textwidth]{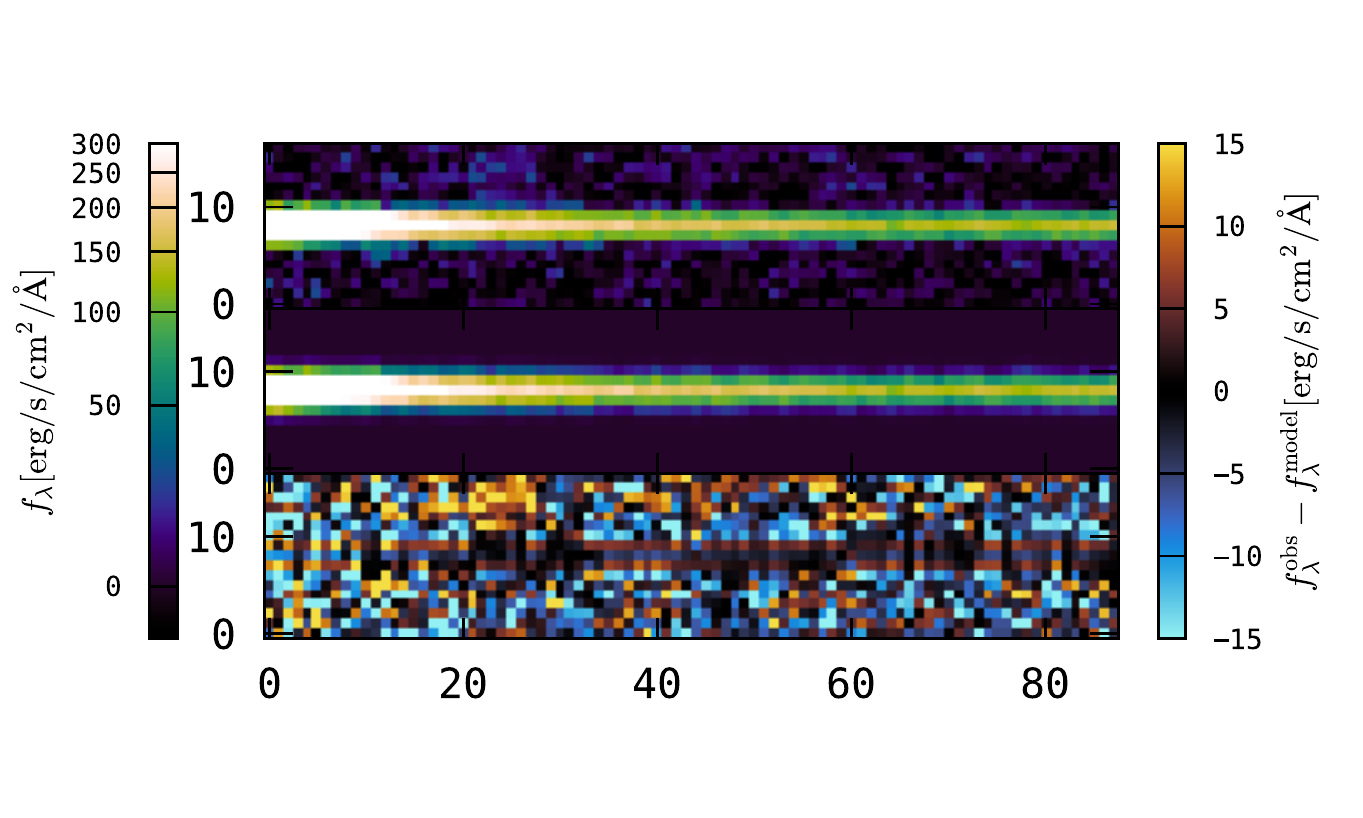}
  \includegraphics[width=0.59\textwidth]{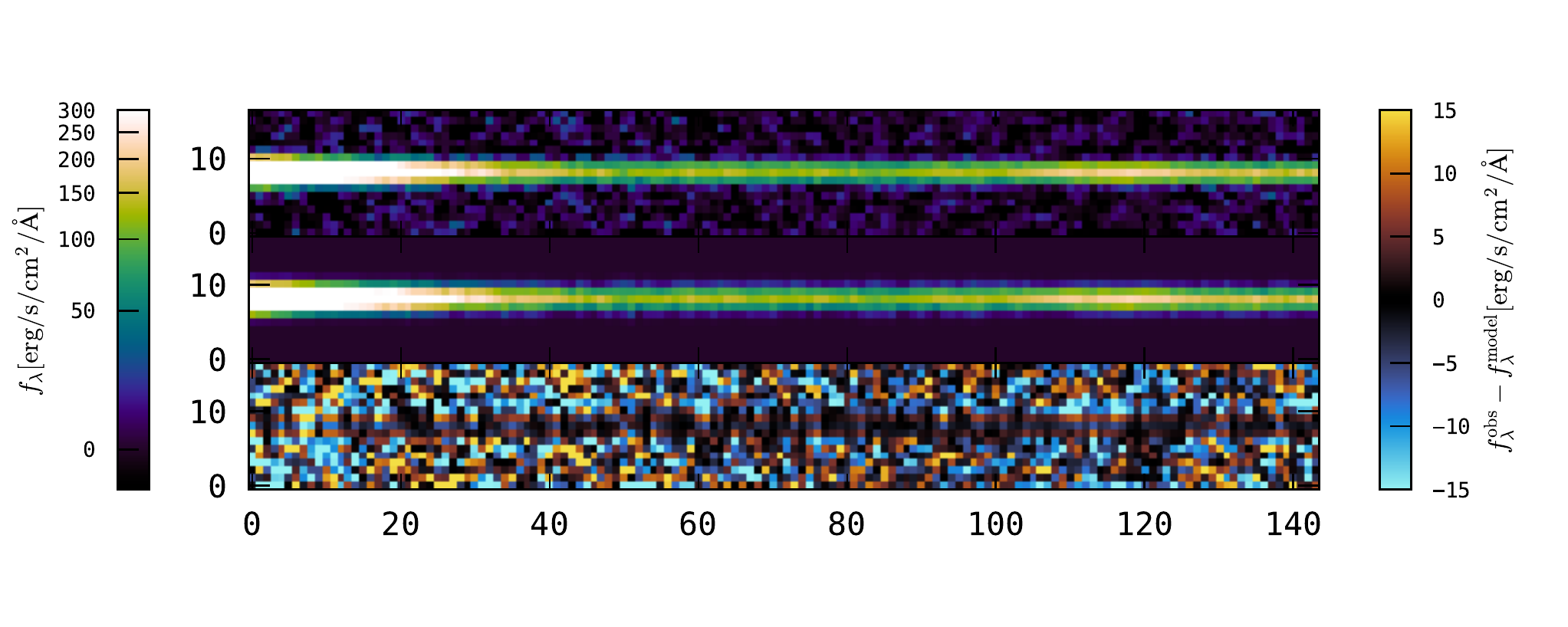}
  \vspace{-1em}
  \caption{H$\alpha$ emission line profiles extracted along the pseudo slits
    shown in the bottom left panel of Figure~\ref{fig:hanb} (\emph{top panels})
    alongside the fitted Gaussian models (\emph{centre panels}) and the
    residuals (data - model, \emph{bottom panels}).  The left panels show the
    results for the eastern filament, whereas the right panels show the result
    for the western filament and the horizontal extend is the same as in
    Fig.~\ref{fig:ratio}.  The x-axis is the cross dispersion axis in units of
    MUSE spaxels (1 spaxel = 0.2\arcsec{}) and the y-axis is the dispersion axis
    (1 pixel = 1.25\AA{}).  The colour bar on the left corresponds to the two
    top panels, whereas the colour bar on the right corresponds to the bottom
    panel. }
  \label{fig:resi}
\end{sidewaysfigure}

In Sect.~\ref{sec:hiikin} we stated that the emission lines in the outer low-SB
regions are adequately modelled by a single Gaussian profile.  To back this
claim we here visualise the line profile in the main filaments.  To this aim we
extract 2D spectra along the two pseudo-slits shown in the bottom-left panel of
Figure~\ref{fig:hanb}.  From these extractions we show in Figure~\ref{fig:resi},
the H$\alpha$ profiles alongside the fitted Gaussian models and the residuals.
It can be appreciated from this Figure in that the single Gaussian is an
adequate model and that the line is extremely narrow.  To further explore the
simple morphology and narrowness of the line profiles we encourage the
interested reader to obtain the fully reduced datacube that is released with
this publication.

\section{21cm channel maps}
\label{sec:channel-maps}

\begin{figure}[b!]
  \centering \includegraphics[width=0.5\textwidth,trim=97 0 80
  0,clip=True]{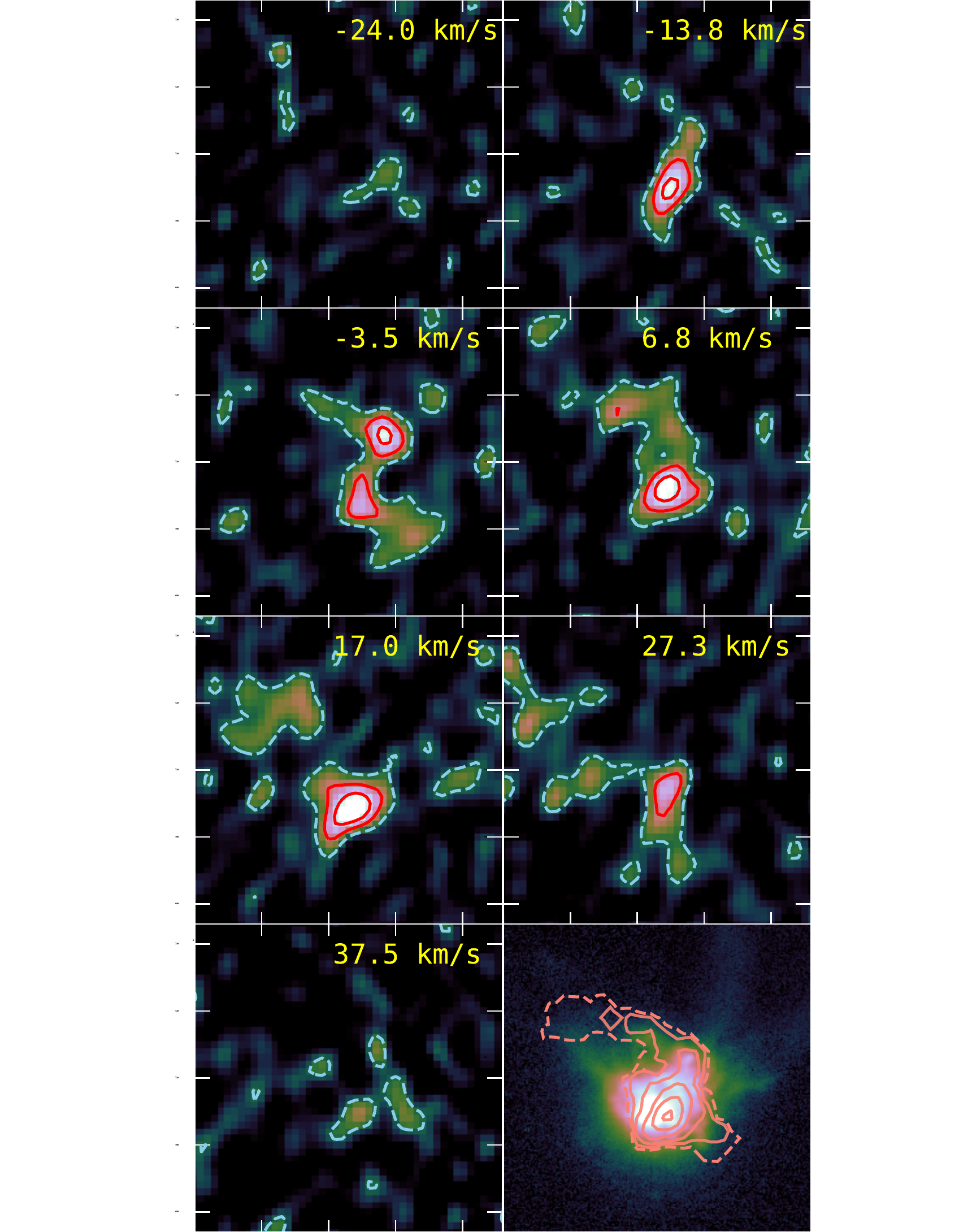} \vspace{-0.2em}
  \caption{Channel maps from the untapered VLA-B configuration datacube.  Each
    channel is shown with a linear stretch from 0 to 6$\sigma$, where
    $\sigma = 2.5\times10^{-4}$\,Jy\,beam$^{-1}$ is determined from fitting the
    negative distribution of pixels in each channel. Contours are drawn at
    2$\sigma$ (dashed), $4\sigma$, and 6$\sigma$. The bottom right panel shows
    the same viewport (46\arcsec{} / 12.7\,kpc along each axis), but with the
    H$\alpha$ NB image and the total $N_\mathrm{HI}$ contours from
    Fig.~\ref{fig:hanb}. }
  \label{fig:chmap}
\end{figure}

\begin{figure}[b!]
  \centering \includegraphics[width=0.5\textwidth,trim=0 0 0
  0,clip=False]{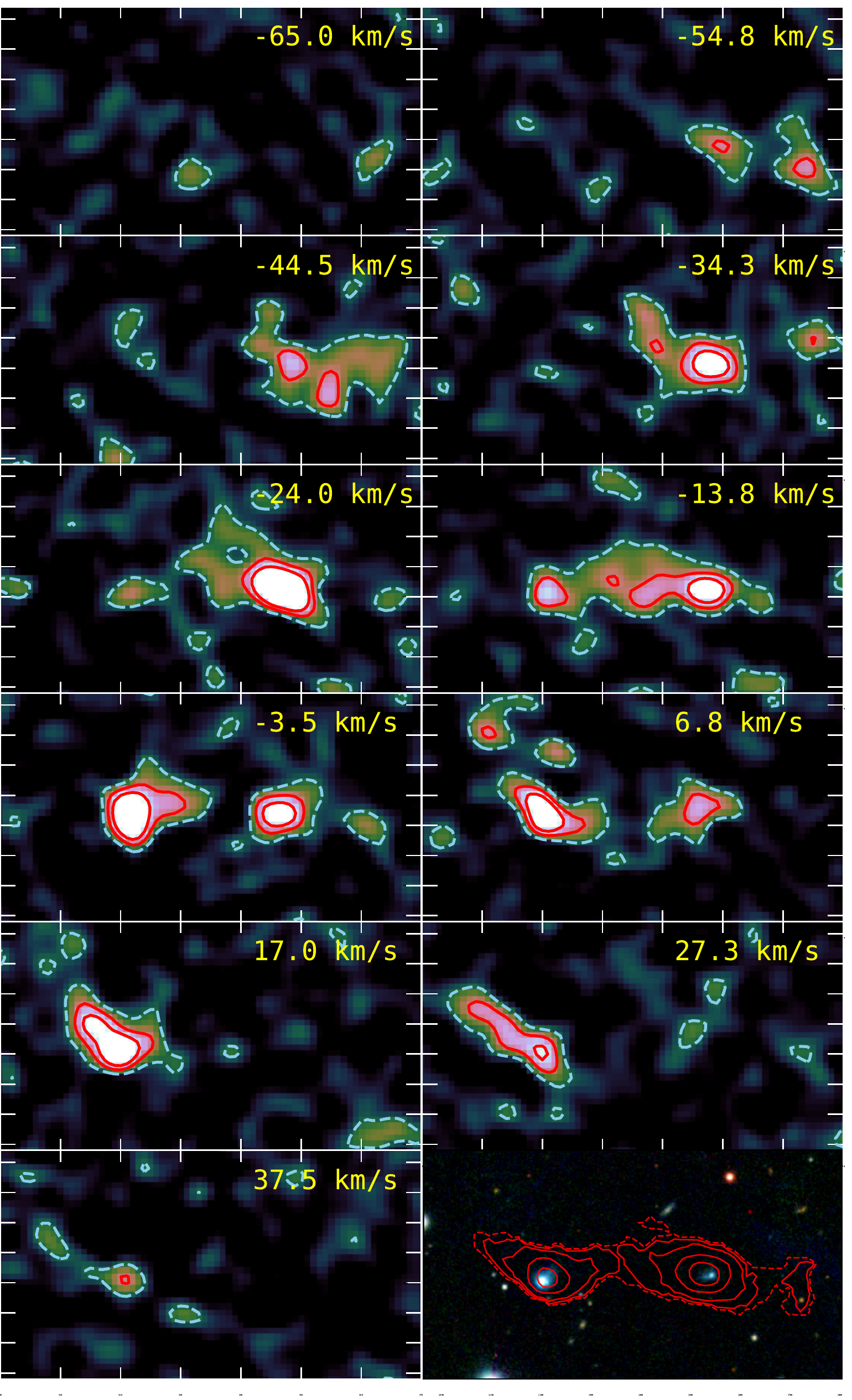} \vspace{-0.2em}
  \caption{Channel maps from the 7k$\lambda$-tapered datacube, similar to
    Fig.~\ref{fig:chmap}; here $\sigma = 4 \times 10^{-4}$\,Jy\,beam$^{-1}$.
    The bottom right panel shows the viewport (210\arcsec{}$\times$114\arcsec{}
    / 58.1\,kpc$\times$31.5\,kpc) over the Pan-STARRS false-colour image with
    the $N_\mathrm{HI}$ contours from Fig.~\ref{fig:hI7kt}.}
  \label{fig:chmap7k}
\end{figure}

We supplement our analysis with channel maps from the VLA B-configuration
datacubes.  Figure~\ref{fig:chmap} shows seven channel maps from -24\,km\,s$^{-1}$
to 37\,km\,s$^{-1}$ zoomed onto SBS\,0335-052E.  Figure~\ref{fig:chmap7k} shows 7
channel maps from -65.0\,km\,s$^{-1}$ to 37.5\,km\,s$^{-1}$ containing both
SBS\,0335-052E and SBS\,0335-052W.  Velocity offsets
in both figures are defined with respect to the optical recession velocity
$v_0 = c \cdot z = 4053$\,km\,s$^{-1}$.

%\newpage

\vspace{-1em}

\section{Cone model exploration}
\label{sec:toy-model-paramter}

\begin{figure*}[t]
  \centering
  \includegraphics[width=\textwidth]{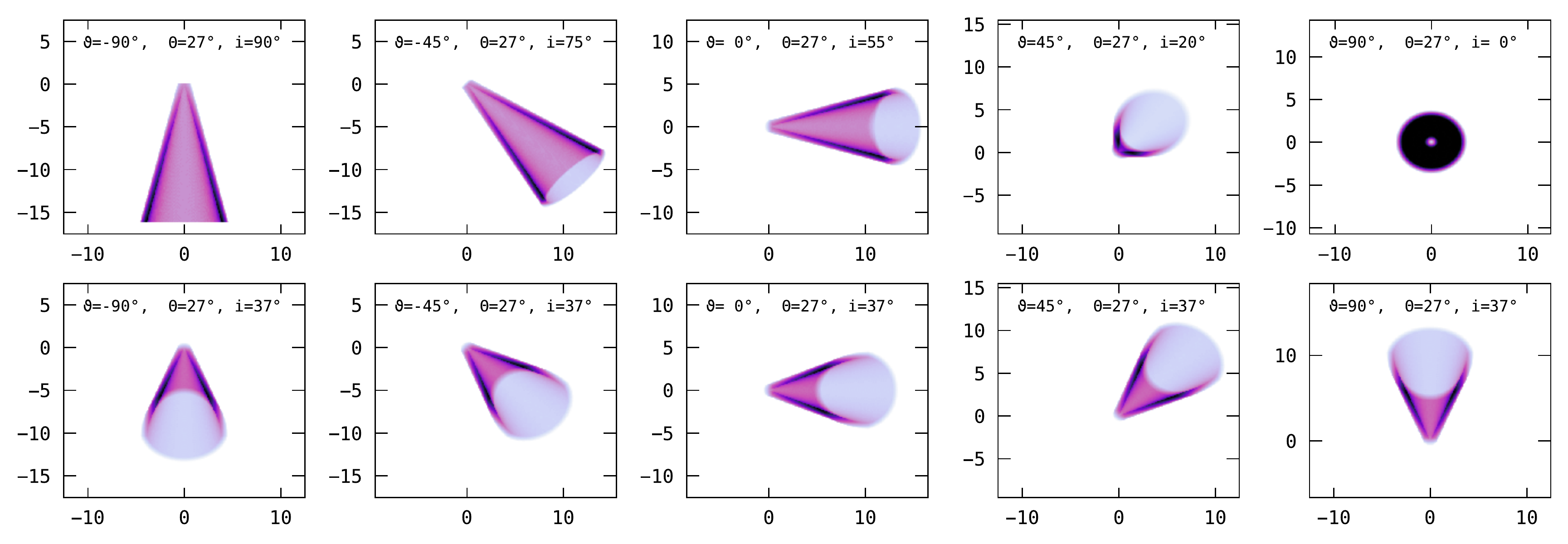}
  \caption{Surface-luminosity projections of the cone-wall structure from
    Sect.~\ref{sec:disc} for various position angles $\vartheta$ and
    inclinations $i$ as indicated in the panels. }
  \label{fig:conegeo}
\end{figure*}

\begin{figure*}[t]
  \centering
  \includegraphics[width=\textwidth]{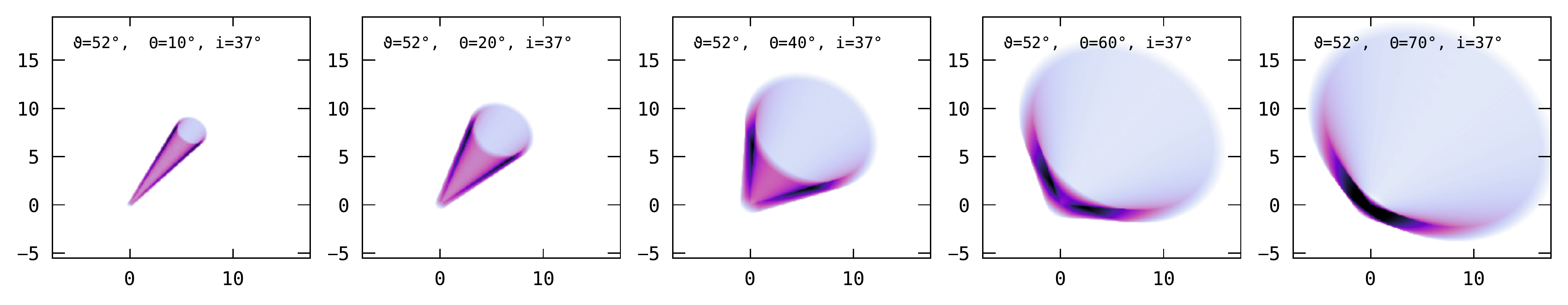}
  \caption{Surface-luminosity projections of the cone-wall structure from
    Sect.~\ref{sec:disc} for various opening angles $\theta$ as indicated in the
    panels. }
  \label{fig:conegeo2}
\end{figure*}

\begin{figure*}[t]
  \centering \includegraphics[width=\textwidth]{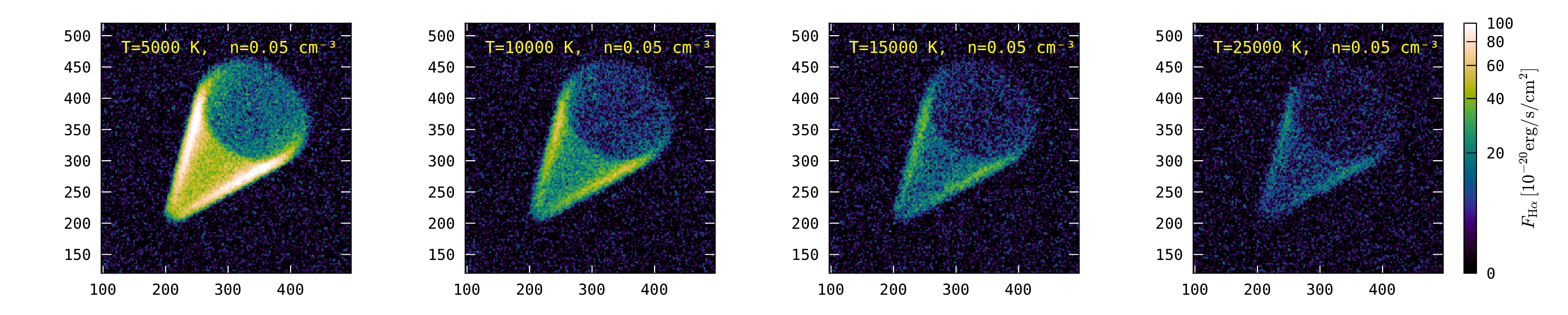}
  \caption{Simulated MUSE H$\alpha$ observation of the cone-wall structure from
    Sect. \ref{sec:disc} at the distance of SBS\,0335-052E for fixed
    $n=0.05$\,cm$^{-2}$ but varying $T$ as indicated in the panels.  }
  \label{fig:coneobs}
\end{figure*}

\begin{figure*}[t]
  \centering \includegraphics[width=\textwidth]{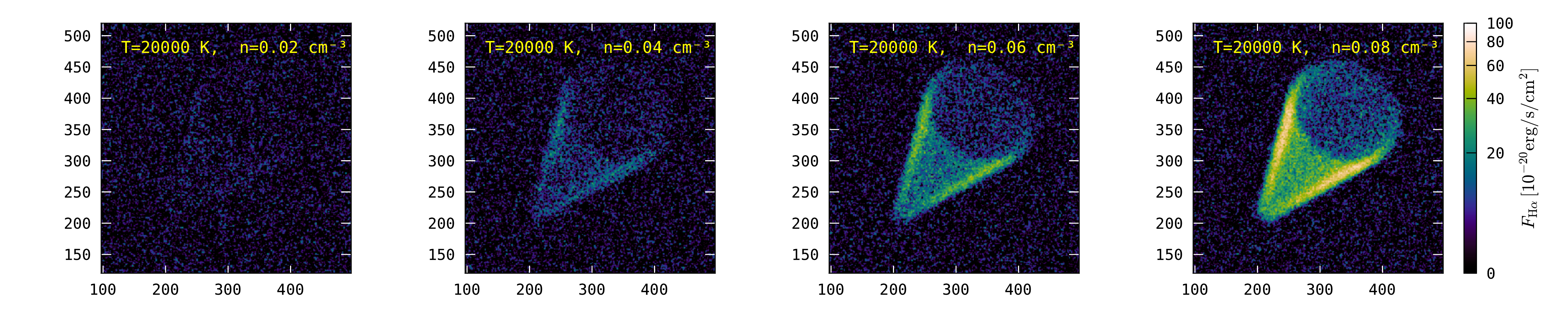}
  \caption{Simulated MUSE H$\alpha$ observation of the cone-wall structure from
    Sect. \ref{sec:disc} at the distance of SBS\,0335-052E for fixed
    $T=20000$\,K but varying $n$ as indicated in the panels.  }
  \label{fig:coneobs2}
\end{figure*}

Sect.~\ref{sec:disc} discusses the ionised cone-wall structure for parameters
that are derived from the observations.  However, it appears instructive to
visualise projections of this structures for different paramters.
In Fig.~\ref{fig:conegeo} we fix the opening angle $\theta = 27$ from
Sect.~\ref{sec:disc}, but vary $\vartheta$ and $i$.  In Fig.~\ref{fig:conegeo2}
we leave $i = 37^\circ$ and $\vartheta=52^\circ$ as in Sect.~\ref{sec:disc}, but
vary the opening angle.  While for most of the parameters the cone geometry is
apparent, projections at low inclination or large opening angles appear more
disk-like.

From Eq.~(\ref{eq:3}) it is apparent that there is a degeneracy in observable
flux between $T$ and $n$.  To obtain a feel for the effects of those parameters
on the observable morphology, we explore variations of $T$ at fixed $n$ in
Fig.~\ref{fig:coneobs} while in Fig.~\ref{fig:coneobs2} we fix $T$ but vary $n$.
As in Sect.~\ref{sec:disc} we use case-A recombination emissivities for Hydrogen
computed by \cite{Storey1995}.
\end{appendix}

\end{document}